\documentclass[12pt]{article}

\usepackage{bm}
\usepackage{amssymb}
\usepackage{enumerate}

\usepackage[russian]{babel}

\usepackage{graphicx} 
\newcommand\zm{Z_{\mu}}

\begin{document}

\title{\bf Search  for dark sector physics with NA64}

\author{S.N.Gninenko$^{1}$,  N.V.~Krasnikov$^{1,2}$ and V.A.Matveev$^{1,2}$
\\
$^{1}$ INR RAS, 117312 Moscow, Russia
\\
$^{2}$ Joint Institute for Nuclear Research, 141980 Dubna, Russia}




\date{        } 
\maketitle

\begin{abstract}
The NA64 experiment consists of two detectors which are planned to be located at the electron (NA64e) and muon (NA64$\mu$) beams  of the CERN SPS and start 
operation after the LHC long-stop 2 in 2021. Its main goals include searches for dark sector physics - particularly light dark matter (LDM), visible and invisible decays of 
dark photons ($A'$), and new  light particles  that could explain  the $^8$Be and $g_{\mu}-2$ anomalies. Here we review these physics goals,  the current status of NA64 including recent results and perspectives of further searches,  as well as other ongoing or planned experiments in this field. The main theoretical results on LDM, the problem of the origin of the $\gamma-A'$ mixing term and its connection to loop corrections, possible existence of a new light $Z'$ coupled to 
$L_\mu-L_\tau$ current are also discussed. 
\end{abstract}

$$ $$

\newpage

\section{Introduction}

At present the most striking evidence in favour of new physics beyond the Standard model (SM) is the observation of
Dark Matter (DM)  \cite{Universe00, Universe0}.  
The nature of DM is one of challenging questions in physics. 
If DM is a thermal relic from the hot early Universe then  its existence motivates 
to look for models with nongravitational interactions between dark and ordinary matter.  
There is a lot 
of candidates for the role of dark matter \cite{Universe00, Universe0}. In particular, there are  
  LDM(light dark matter) models \cite{lightdark} -\cite{Rev2018} with the mass of 
DM particles $\leq 0(1)~GeV$. 
LDM particles with masses below $0(1)~GeV$ were generally expected to be 
ruled out because 
they  overclose the Universe \cite{Lee}. However there are models \cite{lightdark} -\cite{Rev2018}
with additional 
light vector boson  and LDM  particles  that 
avoid the arguments \cite{Lee} excluding the LDM. 
The standard  assumption  that in the hot early Universe the DM particles are in equilibrium with 
ordinary matter is often used. During the Universe expansion the temperature decreases and at some 
point the thermal decoupling of the DM starts to work. Namely, at 
some freeze-out temperature the annihilation cross-section 
 of DM paricles 
$$DMparticles  ~~\rightarrow 
~~SM particles$$ becomes too small to obey the equilibrium of DM particles with 
the SM particles and the DM decouples. The experimental data are in favour of scenario          
with cold relic for which the freeze-out temperature is much lower 
than the mass of the DM particle. In other words the DM particles decouple in 
   non-relativistic regime. The value of the DM annihilation 
cross-section at the decoupling epoch determines the value of 
the current DM density in the Universe. Too big annihilation cross-section 
leads to small DM density and vise versa too small annihilation 
cross section leads to DM overproduction. The observed 
value of the DM density fraction $\frac{\rho_{d}}{\rho_{c}} \approx 0.23$ \cite{particledata} allows 
to estimate the  DM annihilation cross-section 
into the SM particles and hence to estimate the discovery potential of the LDM
 both in direct underground and accelerator experiments. 
Namely, the annihilation cross-section leading to the correct DM density is 
estimated to be $\sigma_{an} \sim 1~pbn$ and the  value of
 the cross-section depends rather weakly on the 
DM   mass \cite{Universe00, Universe0}.  
Models with the LDM ($m_{\chi} \leq 1~GeV$) can be classified by the 
spins and masses of the DM particles  and mediator. The scalar DM 
mediator models are severely restricted   \cite{Toro, Krnjaic}  but not completely excluded  by  
rare  $K$- and $B$-meson decays. 
Models with light vector bosons \cite{lightdark1,Okun,Holdom} (vector portal) are
 rather popular now. In these models light vector boson  $A'$    mediates between our world and the dark sector \cite{lightdark1}. 
Another possible hint in favour of new physics is the   muon $g_{\mu}-2$ anomaly which is 
the 3.6 $\sigma$ discrepancy between the 
experimental values \cite{exp1,particledata}  and the SM  predictions \cite{th1,th2,th3,th4}  for  the anomalous magnetic moment of the muon. 
Among several  extensions of the SM explaining the   $g_{\mu}-2$  anomaly, the   models predicting the existence  
of  a weak leptonic force  mediated by a sub-GeV gauge boson $Z'$ that couples predominantly  to the difference between the muon and tau lepton currents, $L_{\mu} - L_{\tau}$,  
are of general interest. The abelian symmetry $L_{\mu} - L_{\tau}$ is an anomaly-free global symmetry
within the SM \cite{foot1, foot2, foot3}.  The $L_{\mu} - L_{\tau}$ gauge symmetry breaking 
 is  crucial  for the appearance  of a new relatively light, with 
a mass $m_{Z'} \leq 1~GeV$,  vector boson ($Z'$) which couples very weakly to  muon and tau-lepton with the 
coupling constant $\alpha_\mu \sim O(10^{-8})$ \cite{vecmuon1}- \cite{vecmuon8} and explain muon $g_{\mu}-2$ anomaly.  
Recent claim \cite{17mev} 
of the discovery of  $~17~MeV$ vector particle observed 
as a peak in $e^+e^-$ invariant mass distribution in nuclear transitions makes the question of possible light vector boson  existence  
extremely interesting and important and enhance motivation for the experimental searches at 
low energy intensity frontier. 

At present the most popular vector mediator model is 
the model with additional light vector boson  $A'$ (dark photon) \cite{lightdark1, Holdom} which couples 
to the SM electromagnetic current.
However other light vector boson models, in particular, 
model with $L_{\mu} - L_{\tau}$ interaction  \cite{Krasnikov, gkm, Koz, Esk}, 
are possible as messenger candidates beetween our world and DM world.

The aim of this paper  is review of the search for LDM at the NA64 fixed target experiment \cite{NA64}-
\cite{NA64explast4} at CERN and related 
current and future experiments on the search for LDM. 
 Also we review essential part of the phenomenology related with the LDM models. 
The paper is organized as follows. In section 2 we describe phenomenology of the dark photon model. 
In particular, we discuss the bound on low energy effective coupling constant 
$\bar{\alpha}_d(m_{A'}) \equiv \alpha_d$ derived from the requirement of the absence of Landau pole singularity 
up to some scale $\Lambda_{pole}$.  We present the main formulae for the $A'$ electroproduction reaction
$e Z \rightarrow e Z A'$ on nuclei.
We review muon $g_{\mu} - 2$ anomaly and the possibility to 
explain it due to existence of new light vector boson interacting with muons. Also we discuss the 
problem of the origin of  photon-dark photon  mixing term $\frac{\epsilon}{2}F^{\mu\nu}F'_{\mu\nu}$ and 
its connection with loop corrections.  
In sections 3  we review 
current  accelerator and nonaccelerator bounds including experiments on direct LDM detection.  
In section 4 we describe the NA64 experiment  on the search for 
both invisible and visible $A'$ boson decay.
In section 5 we  review  the last NA64 results and 
discuss future NA64 perspectives on  the search for LDM and, in particular,  
we discuss the NA64 LDM discovery potential  with the use of muon beam. 
In section 6 we outline  some other future experiments 
related with the search for dark photon and LDM at NA64.     
Section 7  contains the main conclusions. In Appendix A we collect the main formulae used for 
the approximate DM density calculations. In Appendix B we discuss the discovery potential 
of NA64 for the case of visible dark photon $A'$ decays 
$A' \rightarrow \chi_1\chi_2 \rightarrow e^+e^-\chi_1\chi_1$
with large missing energy.

\section{A little bit of theory}

\subsection{Model with dark photon}


In  model with ``dark photon'' \cite{lightdark1, Holdom} new light vector boson (dark photon) $A'$ 
interacts with the Standard $SU_c(3) \otimes SU_L(2) \otimes U(1)$ gauge model only 
due to kinetic mixing with $U'(1)$ gauge field $A'_{\mu}$. Dark photon interacts also with 
LDM. In renormalizable models DM particles  have spin 0 or 1/2. 
The Lagrangian of the model has the form  
\begin{equation}
L = L_{SM} + L_{SM,dark} + L_{dark} \,,
\label{1}
\end{equation}
where $L_{SM}$ is the SM Lagrangian, 
\begin{equation}
L_{SM,dark} = -\frac{\epsilon}{2\cos\theta_w}B^{\mu\nu}F'_{\mu\nu} \,,
\label{2}
\end{equation}
$B^{\mu\nu} = \partial^{\mu}B^{\nu} - \partial^{\nu}B^{\nu}$,   
$F'_{\mu\nu} = \partial_{\mu}A'_{\nu} - \partial_{\nu}A'_{\mu}$ and the $L_{dark}$ 
is the DM Lagrangian\footnote{Here $B_{\mu}$ is the SM $U(1)$ gauge field.}. For Dirac LDM $\chi$ the DM Lagrangian is 
\begin{equation}
L_{dark} = -\frac{1}{4}F'_{\mu\nu}F'^{\mu\nu}   +i\bar{\chi}\gamma^{\mu}\partial_{\mu}\chi -
m_{\chi}\bar{\chi}\chi + e_d \bar{\chi} \gamma^{\mu} \chi A'_{\mu}  + \frac{m^2_{A'}}{2}A'_{\mu}A'^{\mu} \,,
\label{3}
\end{equation}
The abelian gauge symmetry 
\begin{equation}
A'_{\mu} \rightarrow  A'_{\mu} + \partial_{\mu} \alpha \,, 
\label{4}
\end{equation}
\begin{equation}
 \chi \rightarrow  exp(ie_d\alpha)\chi \, 
\label{5}
\end{equation}
is explicitly broken due to the mass term $\frac{m^2_{A'}}{2}A'_{\mu}A'^{\mu}$ in the Lagrangian (\ref{3}). 
However we can use the Higgs mechanism for dark photon $A'_{\mu}$ mass creation, namely we can 
use the Lagrangian 
\begin{equation}
L_{\phi} = (\partial_{\mu}\phi - ie_dA'_{\mu}\phi) (\partial^{\mu}\phi - ie_dA'^{\mu}\phi)^* 
- \lambda(\phi^*\phi - c^2)^2 \,.
\label{6}
\end{equation}
Here $\phi$ is scalar field. The spontaneous breaking of the  gauge symmetry (\ref{4}, \ref{5})
due to $<\phi> \neq 0$ leads to nonzero 
dark photon mass. 
As a consequence of the mixing term $ L_{SM,dark} = -\frac{\epsilon}{2\cos \theta_w}B^{\mu\nu}F'{\mu\nu}  $
the low energy interaction between dark photon $A'_{\mu}$ and the SM fermions is described by the effective Lagrangian 
\begin{equation}
L_{A',SM} = \epsilon eA'_{\mu} J^{\mu}_{em} \,,
\label{8}
\end{equation} 

where $ J^{\mu}_{em}$ is the SM electromagnetic current.
The invisible and visible  decay rates of  $A'$ for fermion DM particles $\chi$ are given by 
\begin{equation}
\Gamma(A' \rightarrow \chi\bar{\chi}  ) = 
\frac{\alpha_D}{3}m_{A'}(1 + \frac{2m^2_{\chi}}{m^2_{A'}})
\sqrt{1 - \frac{4m^2_{\chi}}{m^2_{A'}}} \,,
\label{9}
\end{equation}
\begin{equation}
\Gamma(A' \rightarrow e^+e^-) = 
\frac{\epsilon^2 \alpha}{3}m_{A'}(1 + \frac{2m^2_{e}}{m^2_{A'}})
\sqrt{1 - \frac{4m^2_{e}}{m^2_{A'}}} \,.
\label{10}
\end{equation}
Here $\alpha = \frac{e^2}{4\pi} = 1/137 $  and  $\alpha_D = \frac{e^2_d}{4\pi} $ is the analog of
the electromagnetic  fine coupling constant for dark photon. 
For scalar DM particles $\chi $ the invisible decay width is  
\begin{equation}
\Gamma(A' \rightarrow \chi\chi^*) = 
\frac{\alpha_D}{12} m_{A'}(1 - 4 \frac{m^2_{\chi}}{m^2_{A'}})
\sqrt{1 - \frac{4 m^2_{\chi}}{m^2_{A'}}}  \,.
\label{11}
\end{equation}

\subsection{Upper bound and range of  $\alpha_D$ }
One can  obtain upper bound on $\alpha_D$ by the requirement of  the absence of Landau 
pole singularity for the effective coupling constant 
$\bar{\alpha}_D(\mu)$ up to some scale  $\Lambda $ \cite{Marciano}.
One loop $\beta$-function for $\bar{\alpha}_D(\mu)$ is
\begin{equation}
\beta(\bar{\alpha}_D) = \frac{\bar{\alpha}_D^2}{2\pi}[\frac{4}{3}(Q^2_Fn_F +Q^2_S\frac{n_S}{4}) ] \,.
\label{12}
\end{equation} 
Here $\beta(\bar{\alpha}_D) \equiv \mu\frac{d\bar{\alpha}_D}{d\mu}$ and $n_F$ ($n_s$) is the number of fermions
 (scalars) with the $U^{'}(1)$ charge $ Q_F(Q_S)$. For the model with 
pseudo-Dirac fermion \cite{asymferm} we have to introduce an additional scalar with $Q_S = 2$ 
to realize nonzero  splitting between fermion masses,  so 
 one loop $\beta$-function is $ \beta(\bar{\alpha}_D) =\frac{4\bar{\alpha}_D^2}{3\pi}$.
For the model with Majorana fermion  we also have to introduce an additional scalar field 
with the charge $Q_S = 2$ and additional 
Majorana field to cancel $\gamma_5$-anomalies, so the $\beta$-function coincides with 
the  $\beta$-function for the model with pseudo-Dirac fermions. 
For the model with charged scalar DM   to create  nonzero dark photon mass  
in a gauge invariant way we have to introduce additional scalar field  with 
$Q_S = 1$,  so  one loop $\beta$-function is $\beta =  \alpha^2/3\pi$. 
From  the requirement that $\Lambda \geq 1$~TeV \cite{Marciano}   we find that  $\alpha_D \leq 0.2$
for pseudo-Dirac and Majorana fermions and  $\alpha_D \leq 0.8$
for charged scalars \footnote{For smaller values of 
$\Lambda$ we shall have  charged particles in the specrtrum  with masses $\leq 1$~TeV \cite{Marciano} that contradicts to the LHC bounds.}.
Here $\alpha_D $ is an effective low energy coupling constant at scale $\mu \sim m_{A'}$, i.e.
 $\alpha_D = \bar{\alpha}_D(m_{A'})$. In our calculations as a reper point we used the value $m_{A'} = 10 $~MeV.  
In the assumption  that dark photon model is valid up to Planck 
scale, i.e.   $\Lambda = M_{PL} = 1.2 \times 10^{19}$~GeV, we   find 
that for pseudo-Dirac and Majorana fermions  $\alpha_D \leq 0.05$ 
while for scalars $\alpha_D \leq 0.2$. 
In  the SM the $SU_c(3)$, $SU_L(2)$ and $U(1)$ gauge coupling constants are equal to $\sim (1/30 - 1/50) $ at the Planck scale. 
It is natural to assume that  the   effective gauge coupling $\bar{\alpha}_D(\mu = M_{PL})$ is of the order of 
 $SU_c(3)$, $SU_L(2)$ and $U(1)$ gauge coupling constants, i.e.
$   \bar{\alpha}_D(\mu = M_{PL})     \sim (1/30 - 1/50)$.  As a result of this assumption  we find that the values of
the low energy coupling  $\alpha_D$ in the range 
 $ \alpha_D  \sim (0.015 - 0.02) $ are  the  most natural.

\subsection{Some comments on the origin of the mixing parameter $\epsilon$}

In  Holdom paper \cite{Holdom}\footnote{Recent discussion of the $\epsilon$ parameter origin is contained in ref.\cite{Pospelov2019}}
 the origin of the mixing  $\epsilon$ parameter was assumed to be  related 
with  radiative corrections. To clarify this statement consider the simplest model 
 with two free $U(1)\otimes U'(1)$ gauge fields $A_{\mu}$ and $A'_{\mu}$. The Lagrangian of the model is 
\begin{equation}
L_o = - \frac{1}{4}F_{\mu\nu}F^{\mu\nu}  - \frac{1}{4}F'_{\mu\nu}F'^{\mu\nu} + \frac{m^2_{0,A'}}{2} A'^{\mu}A'_{\mu} 
-\frac{1}{2}\epsilon_{0l} F'_{\mu\nu}F^{\mu\nu} \,,
\label{13}
\end{equation} 
where $F_{\mu\nu}= \partial_{\mu}A_{\nu} - \partial_{\nu}A_{\mu}$ and $F'_{\mu\nu}= \partial_{\mu}A'_{\nu} - \partial_{\nu}A'_{\mu}$.
For $\epsilon_{0l} = 0$ the Lagrangian (\ref{13}) is invariant under two independent discrete symmetries 
$A_{\mu} \rightarrow  -A_{\mu} $ and $A'_{\mu} \rightarrow  -A'_{\mu} $.
After diagonalization we find that the spectrum of the model for $|\epsilon_{0l}| \ll 1$ consists of 
massless vector particle(photon) and massive vector particle(dark photon) with a mass
$m^2_{A'} = m^2_{0,A'}(1  + \epsilon_{0l}^2)$. Let us add to 
the model  massive fermion field $\psi_M$ with a mass $M$ which interacts both 
with $A_{\mu} $ and $A'_{\mu}$ with the interaction Lagrangian 
\begin{equation}
\Delta L = e\bar{\psi}\gamma^{\mu}\psi A_{\mu} +   e'\bar{\psi}\gamma^{\mu}\psi A'_{\mu} \,.
\label{14}
\end{equation}
At one-loop level 
the  propagator  
  $\int e^{ipx}  
<T(A_{\mu}(x)A_{\nu}(0))>d^4x$
depends on virtual momentum $p^2$. It means that one-loop correction  $\epsilon_{1l}$ 
depends  on virtual momentum $p^2$, namely  
\begin{equation}
\epsilon_{1l}(p^2) = \frac{e e'}{16\pi^2}
\int^1_{-1}(1 - \eta^2) ln[\frac{4 M^2 - p^2(1-\eta^2)}{\mu^2}]d\eta \,.
\label{16}
\end{equation}
Here $\mu$ is some renormalization point, so  
one-loop contribution to the tree level $\epsilon_{0l}$ parameter depends  on the renormalization scheme. 
To our mind the most natural choice of the renormalization point $\mu$ is to require that radiative corrections 
to the tree level  $\epsilon_{0l} $ parameter vanish at the  $A'$ mass shell
\begin{equation}
\epsilon_{1l}(p^2 = m^2_{0,A'}) = 0 \,.
\label{17}
\end{equation} 
The renormalization condition (\ref{17}) guarantees us that radiative corrections don't modify  the tree level formula
 $m^2_{A'} = m^2_{0,A'}(1  + \epsilon_{0l}^2)$ for the pole dark photon mass.
The renormalization condition (\ref{17}) leads to well defined value of the $\epsilon$ parameter at one-loop level
\begin{equation}
\epsilon_{0+1l}(p^2) = \epsilon_{0l} + \frac{e e'}{16\pi^2}
\int^1_{-1}(1 - \eta^2) ln[\frac{4 M^2 - p^2(1-\eta^2)}{4 M^2 - m^2_{0,A'}(1-\eta^2)          }]d\eta \,.
\label{18}
\end{equation}
For the normalization condition  (\ref{17}) one-loop contribution to 
the $\epsilon_{0l}$ parameter vanishes as  $\epsilon_{1l} \sim \frac{1}{M^2}$ for large fermion masses $M \gg m_{0,A'}$ 
 that agrees with   the decoupling expectations. For the model with two massive fermions 
$\psi_1$, $\psi_2$ with masses $M_1$, $M_2$, the charges $e, e'$ and $e, -e'$ one-loop 
correction to the $\epsilon_{0l}$ parameter is ultraviolet finite and it does not depend on 
the renormalization point $\mu$  
\begin{equation}
\epsilon^{naive}_{1l}(p^2) =   \frac{e e'}{16\pi^2}
\int^1_{-1}(1 - \eta^2) ln[\frac{4 M_1^2 - p^2(1-\eta^2)}{4 M_2^2 - p^2(1-\eta^2)           }]d\eta \,.
\label{19}
\end{equation}
However the $\epsilon^{naive}_{1l}(0) = \frac{ee^`}{12\pi^2}ln[\frac{M^2_1}{M^2_2}]$ does not vanish 
for $M_1 \rightarrow \infty$,  $M_2 \rightarrow \infty $ in contradiction with naive  decoupling expectations. 
To cure this situation we can add one-loop finite counter-term $-  \frac{\Delta_{1l}}{2} F'_{\mu\nu}F^{\mu\nu}$
to the Lagrangian (\ref{13})  with $ \Delta_{1l} = - \epsilon_{1l} (p^2 = m^2_{A'}) $, 
so  one-loop expression for $\epsilon_{1l}(p^2)$ reads  
\begin{equation}
\epsilon_{1l}(p^2) =  \epsilon^{naive}_{1l}(p^2) - \epsilon^{naive}(p^2 = m^2_{0,A'}) \,.
\label{20}
\end{equation} 
One can find that $\epsilon_{1l}(0) \rightarrow 0$ for $M_1 \rightarrow \infty$,  $M_2 \rightarrow \infty $ 
in accordance with decoupling expectations. 
Let us formulate our  main conclusion - within the abelian
$U(1)\otimes U'(1)$ gauge model we can't predict the value of the mixing parameter $\epsilon$ 
and to our mind the most natural renormalization scheme is based on the use of the condition 
that loop corrections to the $\epsilon(p^2)$ vanish at the $A'$ mass shell, so    $\epsilon_{0l}$ 
is free arbitrary parameter of the model. 
 
The situation with the $\epsilon$ prediction changes drastically if we assume that one of the $U(1)$ 
abelian gauge groups  arises due to gauge symmetry breaking of nonabelian gauge group. 
As a simplest example consider  the model where dark photon originates from $SU'(2)$ gauge 
symmetry breaking $SU'(2) \rightarrow U'(1)$. The unbroken $U(1) \otimes SU'(2)$ gauge symmetry 
prohibits  the mixing term $-\frac{\epsilon}{2}F^{\mu\nu}F'_{\mu\nu}$. 
Suppose $SU'(2)$ gauge symmetry is broken to
$ U'(1)$ due to the Higgs field $\Phi_b$ $(b= 1,2,3)$  
 in adjoint representation. The $U(1)\otimes U'(1)$  mixing term 
 arises as a result of  $SU'(2)$ breaking due to the 
effective term $\frac{\Phi_a}{\Lambda}F'^{a}_{\mu\nu}F^{\mu\nu}$. Suppose we have doublet(under $SU'(2)$) of 
vector-like   fermions $\psi_a$ $(a =1,2)$ with the mass $M $  and  the $U(1)$  charge $e $. 
The Yukawa interaction of  vector-like fermions with scalar triplet $\Phi_b$  is 
$L_{Yuk} = - h\Phi_a\bar{\psi}\sigma_a\psi $. 
Nonzero vacuum expectation value $<\Phi_3> \neq 0$ leads to $SU'(2) \rightarrow U'(1)$ gauge symmetry breaking and to the 
splitting of fermion masses for fermion doublet $\psi_a$, namely $M_{1,2} = M \pm h<\Phi_3>$. As 
a consequence of fermion doublet mass splitting we find nonzero one-loop contribution to 
the $\epsilon$ parameter, namely
\begin{equation}
\epsilon_{1l} =   \frac{eg}{6\pi^2} ln[\frac{M+h<\Phi_3>}{M-h<\Phi_3>}] \,.
\label{21}
\end{equation}
Here $g$ is the $SU(2)$ gauge coupling.
The expresssion (\ref{21}) vanishes for $<\Phi_3> = 0$ and for $ M \rightarrow \infty $. For 
$M \gg ~h<\Phi_3> $ the $\epsilon$ parameter is 
\begin{equation}
\epsilon_{1l} =   \frac{eg}{6\pi^2}\frac{2h<\Phi_3>}{M} \,.
\label{22}
\end{equation}
So we find that for the  model with nonabelian extension of one of the $U(1)$ 
gauge groups the $\epsilon$ parameter arises as a result of nonabelian 
 gauge symmetry breaking and in principle but not in practise  we can predict it as a function of 
the parameters of the model. To conclude   we can say that at present state of art we 
can't predict reliably the value of the $\epsilon$ parameter.

\subsection{Dark photon production}

There are several $A'$  production mechanisms \cite{lightdark1}. In proton nucleus collisions the 
$A'$ are produced mainly  in $\pi^0/\eta $ decays
$\pi^0/\eta  \rightarrow \gamma A'$. 
The use of visible $A' \rightarrow e^+e^-$ decay allows to detect dark photon $A'$  as a peak in the $e^+e^-$ 
invariant mass distribution. Also direct $A'$ production in proton nucleus collisions is 
possible in full analogy with the photoproduction in proton nucleus collisions. 

Other perspective way is the $A'$ production in electron nucleus interactions, 
namely the use of  
the reaction
\begin{equation}
e^-(p) ~+~ Z(P_i) \rightarrow e^-(p') ~+~Z(P_f) ~+~ A'(k) \,.
\label{23}
\end{equation}
Here $p = (E_0, \vec{p})$ is the 4-momentum of incoming electron, $P_i = (M,0)$ denotes the $Z$ nucleus 
4-momentum in the initial state, final state $Z$ nucleus momentum is defined by $P_f = (P^0_f, \vec{P}_f)$, 
the $A'$-boson momentum is $k = (k_0, \vec{k})$ and $p' =(e', \vec{p}')$ is the momentum of electron 
recoil. In the improved Weizsacker-Williams (IWW) approximation  the differential and total cross-sections 
for the reaction (\ref{23}) for $m_{A'} \gg m_e$ can be written \footnote{Exact tree level calculations for the 
$e^-Z \rightarrow e^-ZA'$ reaction have been performed in refs.\cite{Exact, Exactkirpich}. For a certain kinematic 
region of the 
parameters $m_{A'}, ~E_{A'}$ the $A'$ yeld derived in the IWW approximation could differ significantly 
from the exact tree level calculations \cite{Exact, Exactkirpich}}\cite{Bjorken} as

\begin{equation}
\frac{d\sigma^{A'}_{WW}}{dx}   = (4\alpha^3\epsilon^2 \chi_{eff})(1-x +x^2/3)(m^2_{A'}\frac{1-x}{x} +m^2_ex)^{-1} \,, 
\label{24}
\end{equation}  
\begin{equation} 
\sigma^{A'}_{WW} = \frac{4}{3} \frac{\epsilon^2\alpha^3}{m^2_{A'}} \cdot log( \delta^{-1}_{A'}) \,,
\label{25}
\end{equation}
\begin{equation}
\delta_{A'} = max[ \frac{m^2_e}{m^2_{A'}},~\frac{m^2_{A'}}{E^2_0}] \,,
\label{26}
\end{equation}
where $\chi_{eff}$ is an effective flux of photons 
\begin{equation}
\chi_{eff} = \int^{t_{max}}_{t_{min}} dt \frac{t - t_{min}}{t^2}[G^{el}_2(t) ~+~ G^{inel}_2(t)] \,,
\label{27}
\end{equation}
and $x = \frac{E_{A'`}}{E_o}$. Here $t_{min} = m^4_{A'}/4E^2_0$, $t_{max} = m^2_{A'}+m^2_e$ and 
$G^{el}_2(t)$, $G^{inel}_2(t)$ are elastic and inelastic form-factors respectively. For NA64  energies $E \leq 100~GeV$  the 
elastic form-factor dominates. The elastic form-factor can be represented in the form \cite{Bjorken} 
\begin{equation}
G^{el}_2 = (\frac{a^2t}{1 +a^2t})^2 (\frac{1}{1 +t/d})^2 Z^2 \,,
\label{28}
\end{equation}
where $a =111Z^{-1/3}/m_e$, $d = 0.164~GeV^2 A^{-2/3}$ 
and $A$ is atomic number of nuclei. We consider the quasielastic reaction (\ref{23}) so
the inelastic nuclear formfactor is not taken into account. Numerically, 
$ \chi_{eff} = Z^2 \cdot Log$, where the function $Log \sim (5~-~10) $ and it  depends weakly on atomic screening, nuclear size 
effects and kinematics.

\subsection{Muon  $g_{\mu}-2$ anomaly and the light vector boson $Z'$}

Recent precise measurement of the  anomalous magnetic 
moment of the positive muon $a _{\mu}= (g_{\mu}-2)/2$ from Brookhaven AGS experiment 821 \cite{exp1} gives result which is about
$3.6 \sigma$ higher \cite{dorokhov,   muonmoment} than the 
SM prediction 
\begin{equation}
a^{exp}_{\mu} -a^{SM}_{\mu} = 288(80)\times 10^{-11}  \,.
\label{29}
\end{equation}
This result may signal the existence of new physics beyond the 
SM. New light  (with a mass $m_{Z'} \leq O(1)~GeV$)
vector boson (dark photon) which couples very weakly with muon with $\alpha_{Z'} \sim O(10^{-8})$ 
can explain $g_{\mu}-2$ anomaly \cite{vecmuon1} - \cite{vecmuon8}.  Vector-like interaction of $Z'$ boson with muon  
\begin{equation}
L_{Z'} = g'\bar{\mu}\gamma^{\nu}\mu Z'_{\nu}\,
\label{30}
\end{equation}
leads to  additional contribution to  muon 
anomalous magnetic moment \cite{muonmoment}     
\begin{equation}
\Delta a  = \frac{\alpha'}{2\pi} F(\frac{m_{Z'}}{m_{\mu}}) \,,
\label{31}
\end{equation}
where
\begin{equation}
F(x) = \int^1_0 dz \frac{[2z(1-z)^2]}{[(1-z)^2 + x^2z]} \,
\label{32}
\end{equation}
and $\alpha' = \frac{(g')^2}{4\pi}$. The relations (\ref{31}, \ref{32}) allow to determine the coupling constant $\alpha'$ which 
explains the value (\ref{29}) of muon anomaly. For $ m_{Z'} \ll m_{\mu}$ one can  find that
\begin{equation}
\alpha' = (1.8 \pm 0.5) \times 10^{-8}\\.
\label{33}
\end{equation}
For another limiting case  $ m_{Z'} \gg m_{\mu}$ the $\alpha'$ is 
\begin{equation}
\alpha' = (2.7 \pm 0.7) \times 10^{-8} \times \frac{m^2_{Z'}}{m^2_{\mu}}   \\.
\label{34}
\end{equation}

However  the postulation of the interaction (\ref{30})   is not the end of the story. 
The main question: what about the interaction of the  $Z'$  boson with other quarks and leptons? 
The  renormalizable $Z'$ interaction with the SM fermions $\psi_{k}$ $(\psi_k = e, \nu_e, u, d, ...)$ 
has the form
\begin{equation}
L_{Z'} = g'Z'_{\mu}J_{Z'}^{\mu}\\,
\label{35}
\end{equation}
\begin{equation}
J_{Z'}^{\mu} = \sum_{k}[q_{Lk}\bar{\psi}_{Lk}\gamma^{\mu}\psi_{Lk} + q_{Rk}\bar{\psi}_{Rk}\gamma^{\mu}\psi_{Rk}] \\,
\label{36}
\end{equation}
where  $\psi_{Lk,Rk} = \frac{1}{2}(1 \mp \gamma_5)\psi_k$ and 
$q_{Lk}, q_{Rk}$ are the $Z'$ charges of the $\psi_{Lk}, \psi_{Rk}$ fermions.
The $Z'$ can interact with   new hypothetical particles beyond the SM, for instance, with  DM fermions $\chi$
\begin{equation}
L_{Z'\chi} = g_D Z'_{\mu}\bar{\chi}\gamma^{\mu}\chi \\.
\label{37}
\end{equation}
There are several models of the current $J_{Z'}^{\mu}$. In a model with dark photon \cite{Holdom} $Z'$ boson interacts with
photon $A_{\mu}$  due to 
kinetic mixing term\footnote{Here $F_{\mu\nu} = \partial_{\mu}A_{\nu} - \partial_{\nu}A_{\mu}$ and 
$Z'_{\mu\nu} = \partial_{\mu}Z'_{\nu} - \partial_{\nu}Z'_{\mu}$.} 
\begin{equation}
L_{mix} = -\frac{\epsilon}{2}  F^{\mu \nu} Z'_{\mu \nu} \\.
\label{38}
\end{equation}
 As a result of the mixing (\ref{38}) the field $Z'$ interacts with the SM electromagnetic field 
$J^{\mu}_{EM} = \frac{2}{3}\bar{u}\gamma^{\mu}u - \frac{1}{3}\bar{d}\gamma^{\mu}d - \bar{e}\gamma^{\mu}e + ...$
with the  coupling constant  
$g' = \epsilon e$ ($\alpha = \frac{e^2}{4\pi} = \frac{1}{137}$). 
However experimental data exclude dark photon model  as 
an explanation of muon $g_{\mu} - 2$ anomaly.        Other interesting scenario is the model
\cite{LEE} where $Z'$ 
(the dark leptonic gauge boson)  
interacts with the SM leptonic current, namely 
\begin{equation}
L_{Z'} = g'[\bar{e}\gamma^{\nu}e + \bar{\nu}_{eL} \gamma^{\nu}\nu_{eL} + \bar{\mu}\gamma^{\nu}\mu +  
\bar{\nu}_{\mu L} \gamma^{\nu}\nu_{\mu L} \nonumber \\
+ \bar{\tau}\gamma^{\nu}\tau +  \bar{\nu}_{\tau L} \gamma^{\nu}\nu_{\tau L}]Z'_\nu  \,. 
\label{39}
\end{equation}

In refs. \cite{vecmuon1} - \cite{vecmuon3} for an explanation of $g_{\mu} - 2$ muon anomaly a model where 
$Z'$ interacts predominantly with the second and third generations through the  
$L_{\mu} - L_{\tau}$ current 
\begin{equation}
L_{Z'} = g'[\bar{\mu}\gamma^{\nu}\mu +  \bar{\nu}_{\mu L} \gamma^{\nu}\nu_{\mu L}
- \bar{\tau}\gamma^{\nu}\tau  -  \bar{\nu}_{\tau L} \gamma^{\nu}\nu_{\tau L}]Z'_{\nu}
\label{40}
\end{equation}
has been proposed.
The interaction (\ref{40}) is $\gamma_5$-anomaly free, 
it commutes with the SM gauge group  and moreover it  escapes (see next section) from the most restrictive current 
experimental bounds because the interaction (\ref{40}) does not contain quarks and first generation leptons $\nu_e$, $e$.
In ref.\cite{NKSG} a model where $Z'$ couples with a right-handed current of the first 
and second generation  SM fermions including the right-handed neutrinos has been suggested. The model is 
able to explain the muon $g_{\mu}-2$ anomaly due to existence of light scalar and it can be tested in future experiments. 

The Yukawa interaction of the scalar field with muon
\begin{equation}
L_{Yuk, \phi} = -g_{\mu \phi} \phi \bar{\mu}\mu  \,.
\label{41}
\end{equation}
leads to  additional one loop contribution to  muon 
anomalous magnetic moment \cite{muonmoment}     
\begin{equation}
\Delta a_{\mu} = \frac{{g}^2_{\mu \phi}}{8\pi^2}\frac{m^2_{\mu}}{m^2_{\phi}}
\int^1_0\frac{x^2(2-x)dx}{(1-x)(1-\lambda^2x) + \lambda^2 x} \,,
\label{42}
\end{equation}
where $\lambda =  \frac{m_{\mu}}{m_{\phi}}$. 
For heavy scalar $m_{\phi} >> m_{\mu}$ 
\begin{equation}
\Delta a_{\mu} = \frac{{g}^2_{\mu \phi}}{4\pi^2}\frac{m^2_{\mu}}{m^2_{\phi}} [ln(\frac{m_{\phi}}{m_{\mu}}) 
- \frac{7}{12}] \,
\label{43}
\end{equation} 
and for light scalar $m_{\mu} \gg m_{\phi}$  
\begin{equation}
\Delta a_{\mu} = \frac{3{g}_{\mu \phi}^2}{16\pi^2}  \,.
\label{44}
\end{equation}

\subsubsection{LDM and $Z'$ boson interacting with $L_{\mu} - L_{\tau}$ current \cite{Krasnikov, gkm, Koz, Esk}.    }

It is interesting that  an extension of the $L_{\mu} - L_{\tau}$ model is able to explain today DM density in the Universe. 
 Consider as an example the simplest  extension with  complex scalar 
LDM $\chi$\footnote{The annihilation cross-section for scalar DM 
has $p$-wave suppression that allows to escape CMB bound \cite{Planck}.}.  
The interaction of the DM $\chi$ with the $Z'$ boson is described by  the Lagrangian
\begin{equation}
L_{\chi Z'} = (\partial^{\mu}\chi - ie_dZ'^{\mu}\chi)^{*}(\partial_{\mu}\chi - 
ie_dZ'_{\mu}\chi) - m^2_{\chi}\chi^{*}\chi -  \lambda_{\chi}  (\chi^{*} \chi )^2  \,.
\label{45} 
\end{equation} 
The nonrelativistic annihilation cross section  $\chi \bar{\chi} \rightarrow \nu_{\mu}\bar{\nu}_{\mu}, \nu_{\tau}\bar{\nu}_{\tau}$  
 for $s \approx 4 m^2_{\chi}$
has the form\footnote{Here we consider the case $m_{Z'} > 2 m_{\chi}$.}
\begin{equation}
\sigma v_{rel} = \frac{8\pi}{3} \frac{\epsilon^2\alpha\alpha_D m^2_{\chi}v^2_{rel}}
{(m^2_{Z'} - 4 m^2_{\chi})^2 } \,.
\label{46}
\end{equation} 
We use standard assumption  that in the hot early Universe DM is in equilibrium with 
ordinary matter. Using the formulae of Appendix A one can find that 
\begin{equation}
\epsilon^2 \alpha_{D}  = 
    k(m_{\chi}) \cdot 10^{-6}\cdot(\frac{m_{\chi}}{GeV})^2 \cdot\Bigl[\frac{m^2_{Z'}}{m^2_{\chi}} - 4\Bigr]^2  \\.
\label{47}
\end{equation}
Here the coefficient $k(m_{\chi})$ depends logarithmically on DM  mass $m_{\chi}$ and 
$k_{DM} \sim O(1)$ for $1~MeV \leq m_{\chi} \leq 300~MeV$.

As a consequence of (\ref{47}) we find that 
for $m_{Z'} \ll m_{\mu}$ the values   $\epsilon^2 = (2.5 \pm 0.7) \cdot 10^{-6} $ and 
\begin{equation}
\alpha_{D} \sim  0.4   k(m_{\chi}) \cdot(\frac{m_{\chi}}{GeV})^2 
\cdot\Bigl[\frac{m^2_{A'}}{m^2_{\chi}} - 4\Bigr]^2  \,
\label{48}
\end{equation}
explain both 
the  $g_{\mu} - 2$ muon anomaly and today DM density.




\section{Current experimental bounds}

\subsection{The reactions used for the search for LDM} 

 Here we briefly  describe the most interesting reactions used(or will be used) for the search 
for both visible and invisible $A'$ decays at  accelerators.  

\subsubsection{Visible $A'$  decays searches}
There are a lot of dark photon searches  based on the use of visible $A'$ decays $A' \rightarrow e^+e^-,~\mu^+\mu^-$.
The production mechanisms are $e^+e^- \rightarrow \gamma A'$,   $eZ \rightarrow eZA'$ reactions,  neutral 
meson decays $pZ \rightarrow (\pi^0/\eta^0 \rightarrow A'\gamma) + ~... $ in proton nuclei collisions 
or direct $A'$ production in 
proton nuclei reactions \cite{lightdark1}. The $A'$ boson is reconstructed as a narrow resonance.
Also  vertex detection for  $A' \rightarrow l^+l^-$ decay can be used.
Really,  the $A'$ decay length is proportional to $(\epsilon^2 m_{A'})^{-1}$ implying that searches for displaced vertices 
probe low values of the $\epsilon$-parameter. Typical example  is NA64 experiment. 

\subsubsection{Invisible $A'$ decays}

The DM is produced in the  reactions like $eZ \rightarrow eZ(A' \rightarrow \chi \bar{\chi})$ or 
$e^+e^- \rightarrow \gamma(A' \rightarrow \chi\bar{\chi})$ and identified through the missing energy 
carried away by the escaping DM particles. The hermeticity of the detector is crusial for 
background rejection. Resonance  hunt in missing mass distribution is very effective for the search for 
$A'$ invisible decays. For instance, 
BaBar collaboration \cite{BABAR} used the reaction $e^+e^- \rightarrow \gamma(A' \rightarrow \chi\bar{\chi})$. 
The $e^+$, $e^-$ and $\gamma$ momenta are measured with good accuracy $O(10^{-2})$ that allows to restore  
the  missing mass $m_{mis} = \sqrt{(p_{e^+} +p_{e^-} - p_{\gamma})^2}$. The $A'$ is searched for as a peak 
in distribution of the missing mass $m_{mis}$. 

However there are  experiments where the exact measurement of the initial and final particle 
momenta is impossible. For instance, the NA64 experiment \cite{NA64} uses the 
reaction  $eZ \rightarrow eZ(A' \rightarrow \chi \bar{\chi})$ for the search for $A'$ invisible decays 
and measures only initial and final electron energies. 
The typical signature for the LDM detection is missing energy in electromagnetic calorimeter 
without essential activity in hadronic calorimeter. 
Good hermeticity of the detector allows to suppress the background at the level $O(10^{-11})$ or even less 
that is crusial for the $A'$ detection. The 
number of signal events at NA64 is proportional to $\epsilon^2$.

\subsubsection{Electron and proton beam dump experiments}

In beam dump experiments DM is produced in decays $\pi^0/\eta^{(`)} \rightarrow \gamma(A' \rightarrow \chi \bar{\chi})$ 
or in the reactions $pZ \rightarrow pZ(A' \rightarrow \chi{\bar{\chi}})$, $eZ \rightarrow eZ(A' \rightarrow \chi{\bar{\chi}})$ 
and it is detected via  reactions $e\chi \rightarrow e\chi $, $N\chi \rightarrow N\chi $ in downstream detectors \cite{lightdark1}. These 
experiments  probe LDM twice and they are  sensitive to LDM  coupling constant $\alpha_D = \frac{e^2_D}{4\pi}$ 
with dark mediator $A'$. 
The number of events is proportional to $\epsilon^4 \alpha_D$. Therefore a large proton(electron) flux is required.

\subsection{Bound from electron magnetic moment}

The experimental and theoretical values for electron magnetic moment coincide at the $2.4 \sigma $ level \cite{ge}
\begin{equation}
\Delta a_e \equiv a^{exp}_e - a^{SM}_e = -(0.87  \pm 0.36) \times 10^{-12} \,.
\label{49}
\end{equation}
The $A'$ boson contributes to the $\Delta a_e$ at one loop level, see formulae (\ref{42} - \ref{44}). From the value (\ref{49}) of $\Delta a_e$ 
it is possible to restrict the couplng constants $g_{Ve}$ and $g_{Ae}$. For the model 
with equal muon and electron vector couplings $g_{Ve} = g_{V \mu }$ and  $g_{Ae} = g_{A \mu} = 0$ 
   the $g_{\mu} - 2$ muon anomaly explanation  is excluded for 
      $M_{A'} \leq 20 ~MeV$ \cite{gegmu}.

\subsection{Visible $A'$ decays}

\subsubsection{Fixed target electron  experiments}

Fixed target experiments APEX \cite{APEX} and A1 at MAMI(Mainz Microtron) \cite{MAMI} searched for $A'$ in electron-nucleus 
scatterings using the $A'$ bremsstrahlung production 
$e^- Z \rightarrow e^- ZA'$ and subsequent $A'$ decay into electron-positron pair $A' \rightarrow 
e^+e^- $. 
The absence of the resonance peak in the invariant $e^+e^-$ mass spectrum allows to obtain 
upper limits on the $A'$ boson  coupling constants $g_{Ve}$, $g_{Ae}$ of the $A'$ with electron, see Fig.1. 
The A1 collaboration 
excluded the masses 
$50~MeV < M_{A'} < 300~MeV$ \cite{MAMI} for $g_{\mu}-2$ muon anomaly explanation in the model with equal muon and electon couplings 
    of the $A'$ boson with  a sensitivity to the mixing parameter up to $ \epsilon^2 = 8 \times 10^{-7}$. 
APEX collaboration used $\sim 2~GeV$ electron beam at Jefferson Laboratory and excluded masses 
$175~MeV < M_{A'} < 250~MeV$ for $g_{\mu}-2$ muon anomaly explanation in the model with equal muon and electon couplings 
 of the $A'$ boson.  Recently NA64 collaboration studied long lived $A' \rightarrow e^+e^-$ decays and 
obtained new bounds on mixing parameter $\epsilon$, see sect. 4.

\subsubsection{$e^+ e^-$  experiments}

BaBar collaboration \cite{BABARvis} looked for visible decays of light $A'$ bosons  in the reaction $e^+e^- \rightarrow \gamma A', 
~A' \rightarrow l^+l^-(l = e, \mu)$  as resonances in the $l^+l^-$ 
spectrum. For the model with  the $A'$ dark photon  the mixing strength values 
$10^{-2} - 10^{-3}$ are 
excluded for $0.212~GeV < m_{A'} < 10~GeV$ \cite{BABARvis}
in the assumption that visible $A'$ decays into the SM particles dominate, see Fig.1.
The KLOE  experiment  at the DA$\Phi$NE $\Phi$-factory  in Fraskati      searched for $A'$ in 
decays $\Phi \rightarrow \eta A' \rightarrow \eta e^+e^-$ and   $\Phi \rightarrow \gamma (A' \rightarrow 
\mu^+\mu^-)$ \cite{KLOE}. 
The obtained bounds  are weaker than those from  
NA48/2 \cite{NA-48} and MAMI \cite{MAMI} bounds.

Recently BaBar collaboration used  the reaction $e^+e^- \rightarrow Z'\mu^+ \mu^-, ~Z' \rightarrow \mu^+\mu^-$ to search 
for the $Z'$ boson coupled with muon. The use of this process allows to restrict directly the muon coupling $g_{V\mu}$ of 
the $Z'$ boson. The obtained results exclude  the model with $L_{\mu} - L_{\tau}$ interaction as 
possible explanation of  $g_{\mu}-2 $ muon anomaly  for $m_{Z'} > 214 ~MeV$ \cite{BABAR0}.

\subsubsection{Fixed target proton   experiments}

The NA-48/2 experiment used simultaneous $K^+$ and $K^-$ secondary beams produced by $400~GeV$ primary 
CERN SPS protons for the search for light $A'$ boson in 
$\pi^0$ decays \cite{NA-48}.  The decays $K^{\pm} \rightarrow \pi^{\pm} \pi^{0}$ and $K^{\pm} \rightarrow \pi^0 \mu^{\pm} \nu$ 
have been used to obtain tagged $\pi^{0}$ mesons. The decays $\pi^0 \rightarrow \gamma A^{`}$, 
$A' \rightarrow e^+e^-$ have been  used for the search for
$A'$ boson. The  $A'$ boson manifests itself as a narrow peak in the distribution of the 
$e^+e^-$ invariant mass  spectrum. For the model with dark photon  
the obtained bounds  exclude the $g_{\mu}-2$ muon anomaly explanation for $A'$ boson masses 
$9 ~MeV < m_{A'} < 70~MeV$ \cite{NA-48}, see Fig.1. 
It should be noted that the decay width $\pi^0 \rightarrow \gamma A'$ is proportional to $(g_{Vu}q_u -g_{Vd}q_d)^2 =
(2g_{Vu} + g_{Vd})^2/9$
and for the models with  nonuniversal $A'$-boson couplings\footnote{In ref.\cite{FENG} models with  $2g_{Vu} + g_{Vd} \approx 0 $ have 
been suggested for an explanation of recent discovery claim \cite{17mev} of   $~17~MeV$  narrow resonance  observed 
as a peak in $e^+e^-$ invariant mass distribution in nuclear transitions.}, for instance, for the model 
with $L_{\mu} - L_{\tau}$ interaction current the NA-48/2 bound \cite{NA-48} is not applicable.

\begin{figure}[tbh!]
\begin{center}
\includegraphics[width =0.45\textwidth]{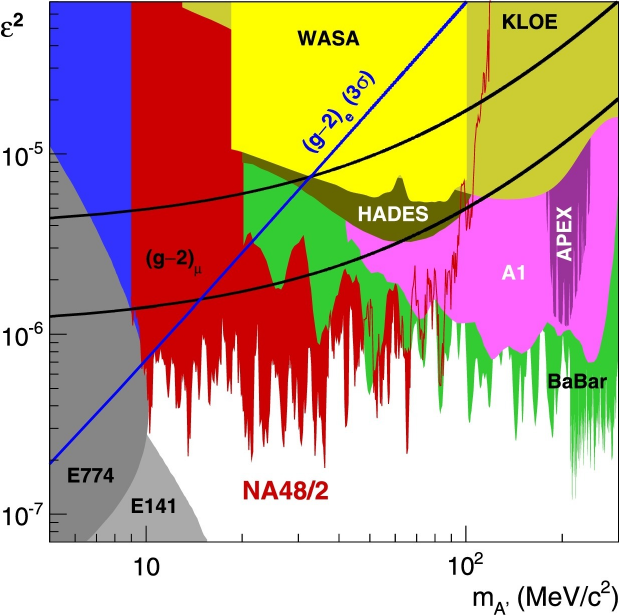}
\caption{Current limits at 90 \% CL on the mixing parameter $\epsilon^2$ versus the $A'$ mass 
for visible $A'$ decays, taken from ref.\cite{NA-48}}
\label{Fig1}
\end{center}
\end{figure}

\subsubsection{ATLAS and CMS bounds on light particles  in Higgs boson decays}

ATLAS collaboration searched for new light particles $\gamma_d$ in Higgs boson decays 
$h \rightarrow 2 \gamma_d + X$, 
 $h \rightarrow 4 \gamma_d + X$ \cite{ATLAS}. In the assumption that new boson 
$\gamma_d$ decays mainly into muon pair bounds on  $Br(h \rightarrow 2 \gamma_d + X)$ 
and $Br(h \rightarrow 4 \gamma_d + X)$ have been otained \cite{ATLAS}. It should be stressed that 
for the model with dark photon the bound on $\epsilon $ parameter is rather  weak.  

CMS collaboration also searched for new particles \cite{CMS} in the Higgs boson decay
  $h \rightarrow 2 a + X  \rightarrow 4 \mu + X$. Bounds similar to the ATLAS bounds 
have been obtained.

\subsubsection{LHCb bound on  $A' \rightarrow \mu^+\mu^- $ decays}

Recently LHCb collaboration performed the search for $A'$ bosons on the base of visible 
$A' \rightarrow \mu^+\mu^- $ decay. In the assumption that the $A'$ production arises 
as a result of $\gamma A'$ mixing the bound on mixing parameter $\epsilon$ has 
been derived for wide range of $A'$ masses from $214~MeV$ up to $70~GeV$ for prompt decays and 
for $214~MeV < m_{A'} < 350~MeV$ for long lived $A'$ \cite{LHCb}. No evidence for signal has been found and 
upper bound on $\epsilon$ parameter has been derived. The obtained bounds are  the most stringent to date 
for the  masses $10.6~GeV < m_{A'} < 70~GeV$.

\subsection{Invisible $A'$ decays}

\subsubsection{Constraints from $K \rightarrow \pi + nothing$ decay}

Light vector boson $A'$ can be produced in the  $K \rightarrow \pi A'$ decay in the  analogy with the SM 
decay $K \rightarrow \pi \gamma^*$ of K-meson  into pion and virtual photon. For the model with the 
dominant $A'$ decay into invisible modes nontrivial bound on the  $A'$ boson mass and 
the coupling constant arises. Namely,  the results of 
BNL E949         and  E787 experiments \cite{E949}
on the measurement of the  $K^{+}  \rightarrow \pi^{+}\nu\bar{\nu} $ decay width  were used to obtain 
an  upper bound on the $Br(K^+ \rightarrow \pi^+ A')$ decay as a function of the $A'$ mass in 
 the assumption that  $A' \rightarrow invisible $ decay dominates. In the model where the $A'$ is 
dark photon,
  the explanation of muon $g_{\mu}-2$ anomaly  due to the $A'$ existence is excluded for $M_{A'} > 50~MeV $ except the 
narrow region around     $m_{A'} =  m_{\pi} $ \cite{LM1} - \cite{LM3}. Note that in  models with 
non-electromagnetic current interactions of $A'$ with quarks and leptons, for instance, 
in the model where  the $A'$ interacts with 
the $L_{\mu} - L_{\tau}$ current only, the bound from   $K \rightarrow \pi + nothing $  decay does 
not work or it is  rather weak \cite{LM2}. 


\subsubsection{The use of the reaction $eZ \rightarrow eZA'$, $A' \rightarrow invisible$}

\begin{figure}[tbh!]
\begin{center}
\includegraphics[width =0.6\textwidth]{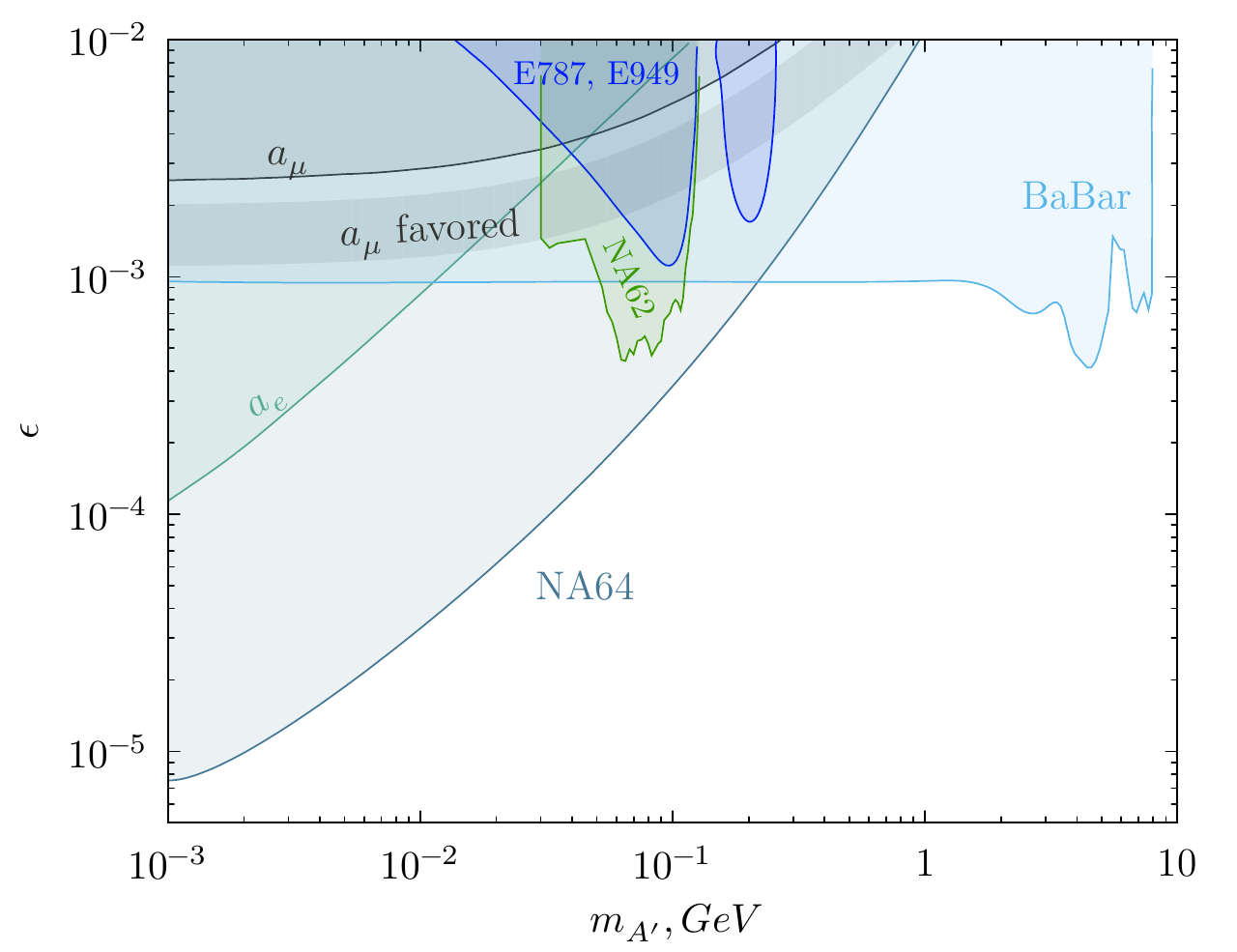}
\caption{Limits at 90 \% C.L. 
on the mixing parameter $\epsilon$ versus the $A'$ mass for invisible $A'$ decays, taken from ref.\cite{NA64explast3}} 
\label{Fig2}
\end{center}
\end{figure}

The NA64 collaboration  \cite{NA64explast1, NA64explast2} 
used the reaction $eZ \rightarrow eZA'$, $A' \rightarrow invisible$ for the search for 
invisible dark photon  decays  into LDM  particles. The obtained bounds exclude the 
dark photon model as an explanation of muon $g_{\mu}-2$, see Fig.\ref{Fig2}.

\subsubsection{$e^+ e^-$  experiments}

Recently BaBar collaboration \cite{BABAR1} used the reaction $e^+e^- \rightarrow \gamma A'$, 
$A' \rightarrow invisible$ for the search for invisible decays of $A'$. In the assumption that 
$A'$ invisible decays dominate the bound $\epsilon \leq 10^{-3}$ has been obtained for $m_{A'} \leq 9.5~GeV$, 
see Fig.2.



\subsubsection{Electron beam dump experimemts}

In electron beam dump experiments the reaction $eZ \rightarrow eZA'$ is used for 
the $A'$ production. After some shield the $A'$ bosons are manifested  as visible 
decays $A' \rightarrow e^+e^-, ~\mu^+\mu^-$. If $A'$ decays mainly 
into LDM particles $A' \rightarrow \chi\bar{\chi}$  
the use of elastic scattering $\chi e \rightarrow \chi e$,  $\chi N \rightarrow \chi N$ 
in the far  detector allows to detect LDM particles.  The results of electron beam dump experiments  \cite{E137, E774}
  at SLAC and   FNAL have been  used \cite{Essig0} 
to constrain the couplings of light gauge boson $A'$. For the case of dominant $A'$ decays into 
visible particles electron beam dump experiments  exclude $10^{-7} \leq \epsilon \leq 10^{-6}$ for 
$m_{A'} \leq 20~MeV$. For the case where the $A'$ decays dominantly into LDM  particles the 
experiment E137 gives the most stringent bounds and it excludes  the 
parameter $y \equiv \epsilon^2\alpha_D(\frac{m_{\chi}}{m_{A'}})^4 \geq 10^{-11}(10^{-9}) $ 
for $m_{A'} \leq 1(100)~MeV$.

\subsubsection{Proton beam dump experiments}

In proton beam dump experiments the main source of the $A'$ arises as a result of $\pi^0(\eta)$ 
production $ pZ \rightarrow \pi^0(\eta) + ...$ with the subsequent $\pi^0(\eta) \rightarrow \gamma A'; A' \to e^+e^-$ 
decays, see e.g \cite{sng-nomad, sng-charm}. In the case of  dominant $A'$ decay into LDM particles 
$A' \rightarrow \chi \bar{\chi} $ the reactions  $ \chi e \rightarrow  \chi e $ and  $ \chi N \rightarrow  \chi N $ 
are used for  dark matter identification. 

The LSND (Liquid Scintillarion Neutrino Detector) \cite{lsnd} at Los Alamos  was constructed to 
detect neutrino. Neutrino arise mainly 
from the reaction $pZ \rightarrow \pi^+ + ...$ with the subsequent  $\pi^+ \rightarrow \mu^+ \nu_{\mu}$ decays. 
LSND data with $N =10^{24}~$POT also allow to restrict  the  dark photon couplings. 
Dark photons $A'$ are produced mainly in the reaction $pZ \rightarrow (\pi^0 \rightarrow \gamma A')  + ...$. 
The LSND bound on the   parameter $y \equiv \epsilon^2\alpha_D(\frac{m_{\chi}}{m_{A'}})^4 $  is  by factor $O(10)$ 
more strong that the corresponding bound from electron beam dump experiment E137.  
The MiniBoone experiment at FNAL is also proton beam dump experiment which uses the FNAL $8~GeV$ Booster proton beam.  
As in LSND dark photons are produced mainly in $\pi^0$ decays and detected in a 800 tonn mineral oil 
Cherenkov detector situated $\sim 500~m$ downstream of the beam dump.  
Recently  MiniBoone experiment
 has obtained bound \cite{MINIboon} on  
$y  \leq 10^{-8}$ for $\alpha_D = 0.5$ and for DM masses 
$0.01 < m_{\chi } < 0.3~GeV$ in a dedicated run  with $1.86 \times 10^{20}$ protons delivered to a steel beam dump.   

\subsubsection{COHERENT at ORNL} 

The primary goal of the COHERENT experiment \cite{COHERENT} at Oak Ridge National Laboratory(USA) is to measure coherent  
elastic neutrino scattering ($CE\nu NS$) process and to check the $N^2$ dependence of the cross section. 
Recently the COHERENT experiment measured the $CE\nu NS$ process \cite{COHERENTres} and the results are in agreement with 
the SM expectations.  
The COHERENT  is  beam-dump experiment
and LDM can be  produced mainly in $\pi^0 \rightarrow 
\gamma A' \rightarrow \gamma \chi \bar{\chi} $  decays.
DM particles scatter in scintillating cristals and liquid argon detectors 
at the Apallation Neutron Source at ORNL. The DM particles(if they exist) are  
produced via $\pi^0/\eta \rightarrow \gamma A'$ decays and they can be identified through coherent scattering leading to detectable 
nuclear recoil. In ref. \cite{COHERENTdark} recent COHERENT data \cite{COHERENTres} have been used for 
the derivation of the bounds for  LDM. For $1 < m_{\chi} < 90 MeV$ the bound on
$\epsilon e_d^{1/2}$ is between $10^{-5}$ and $10^{-4}$.

\subsection{Bound from the neutrino trident process $\nu_{\mu}N \rightarrow \nu_{\mu} N \mu^+\mu^- $}

The neutrino trident $\nu_{\mu}N \rightarrow 
\nu_{\mu}N  \mu^+ \mu^- $ events 
allow to restrict a model where $Z'$ boson interacts with $L_{\mu} - L_{\tau}$ current. 
The  data of the CHARM  and the CCFR  experiments  exclude the $g_{\mu} -2$  
  muon anomaly explanation for  $m_{Z'} \geq 400~MeV$  \cite{POSPELOV}.

\subsection{Nonaccelerator bounds}

\subsubsection{CMB bound} 

The residual annihilation of DM particles after equilibrium annihilation and before recombination can still reionize hydrogen and 
hence modify  the CMB (cosmic microwave background) power spectrum. The Planck experiment constraint \cite{Planck} 
 rules out thermal DM  below 10~GeV if the annihilation 
is s-wave (velocity independent). The p-wave annihilation is allowed since at recombination epoch the temperature is $T \sim eV$ 
and the p-wave annihilation is suppressed by factor $T/m_{\chi}$. Also models with pseudo-Dirac 
LDM \cite{lightdark1, asymferm}   escape the CMB bound.

\subsubsection{Constraints from stars} 

Light $A'$ boson can be produced in stars. The energy loss of the stars through the $A'$ places 
strong limits $\epsilon \leq O(10^{-14})$  on the $A'$ couplings for $m_{A'} \leq 0.01~MeV$ \cite{Sun}  -  \cite{Jeong}. 
The constraints on the $A'$ couplings 
result from the requirement that the energy loss by the $A'$ emission has to be less than  10 percent of the solar energy in photons 
\cite{10percent}. Also for  $m_{A'} \leq 0.3~MeV$  similar but more weak  limit on $\epsilon$   
can be deived from  horizontal branch stars and red giants where the temperatures are higher 
than in the Sun \cite{10percent}.

\subsubsection{Supernnova 1987A bounds}

Bounds from Supenova 1987A are based on the fact that if dark photons are produced in sufficient quantity, they reduce the 
amount of energy emitted in the form of neutrinos, in conflict with observations. In ref.\cite{Supernovae} 
bounds on $\epsilon$ parameter were obtained for the model with dark photon. Bounds on $\epsilon$ parameter 
exist for $m_{A'} \leq 120~MeV$  \cite{Supernovae}. For the most interesting case    $m_{A'} \geq 2m_e$ 
the value  $  \epsilon  \geq   O(10^{-7})$     does not contradict to data drom Supernova 1987A   \cite{Supernovae}. It 
means  that the bounds from Supernova 1987A  don't restrict severely the LDM hypothesis.

\subsubsection{Constraints from BBN}

Big Bang nucleosynthesis (BBN) can also provide the constraints on $A'$ coupling constants. During the first 
several minutes after the Big Bang, the temperature of the Universe rapidly decreased as a 
consequence of the Universe expansion. During the Universe expansion some light elements are produced and the 
predictions of their abundance from BBN agree with experimental data \cite{BBN}. The constraints on 
new interactions are based on the fact that  new relativistic particle increases the expansion rate of the 
Universe through an additional degree of freedom which usually expressed in terms of extra neutrinos $\Delta N_{\nu}$. 
The larger Universe expansion rate increases the freeze-out temperature, therefore the $n/p$ ratio and 
as a consequence the $^4He$ abundance is increased.  The observed value of the $^4He$ abundance 
leads to the bound on   $\Delta N_{\nu}$ that is equivalent  to the bounds on coupling constants of new 
relativistic particle. For dark photon model BBN constraints  have been obtained in ref.\cite{BBNPosp}.  
The   $A'$ dark photon  model with  $m_{A'} \leq O(1)~MeV$ is excluded \cite{BBN1} as 
a mediator explaining current DM abundance. 
Note that in ref. \cite{boehm} lower bound   $m_{\chi} \geq O(1)~MeV$ on the mass of the LDM particle was obtained 
from the  experimental bound on effective number of neutrinos. 

\subsection{Direct LDM detection}
The main problem of the LDM detection   via elactic LDM scattering at nuclei is the size 
  of the nuclear recoil energy \cite{lightdark1}. 
The velocity of DM is $v_{\chi} \sim 10^{-3}c $ and the maximum possible energy transfer is proportional to the square of the reduced mass 
$ \mu_{red} = \frac{m_{nuclei} m_{\chi}}{m_{nuclei}+ m_{\chi}}$. The nuclear recoil energy is \cite{lightdark1}
\begin{equation}
E_{NR} = \frac{q^2}{2m_{nuclei}} \leq \frac{2\mu_{red}^2v^2_{\chi}}{m_{nuclei}} \leq 190 ~eV \cdot (\frac{m_{\chi}}{500~MeV})^2
\cdot (\frac{16~GeV}{m_{nuclei}}) \,
\label{50}
\end{equation}
that makes the detection of LDM with masses $m_{\chi} \leq O(1)~GeV$ at nuclei extremely difficult. 
The remaining possibility is the use of electron LDM elastic scattering \cite{lightdark1}.
For electron LDM scattering the maximum energy transfer to electron is
\begin{equation}
E_e \leq \frac{1}{2}m_{\chi}v^2_{\chi} \leq 3~eV(\frac{m_{\chi}}{MeV}) \,.
\label{51}
\end{equation}
Bound  electrons with binding energy $\Delta E_{B}$ can produce measurable signal 
at  \cite{lightdark1}
\begin{equation}
m_{\chi} \geq 0.3~MeV \times (\frac{\Delta E_{B}}{1~eV}) \,.
\label{52}
\end{equation}
The elasic nonrelativistic cross-section of scalar or fermion LDM in  dark photon model at $m_{\chi} \gg m_e$ is \cite{lightdark1, Essig}
\begin{equation}
\sigma(e\chi \rightarrow e\chi) = \frac{16\pi m^2_{e}\alpha\epsilon^2 \alpha_D }{(m^4_{A'})} \,,
\label{53}
\end{equation}
while the elastic Majorana cross-section  is suppressed by factor $k_M = \frac{2 m^2_e}{m^2_{\chi}}v^2_{\chi}$
\begin{equation}
\sigma(e\chi_{Majorana} \rightarrow e\chi_{Majorana}) = \frac{16\pi m^2_{e}\alpha\epsilon^2 \alpha_D }{(m^4_{A'})}\cdot k_{M} \,
\label{54}
\end{equation}
that makes the direct detection of Majorana LDM in dark photon model extremely difficult or even hopeless.

\begin{figure}[tbh!]
\begin{center}
\includegraphics[width =0.6\textwidth]{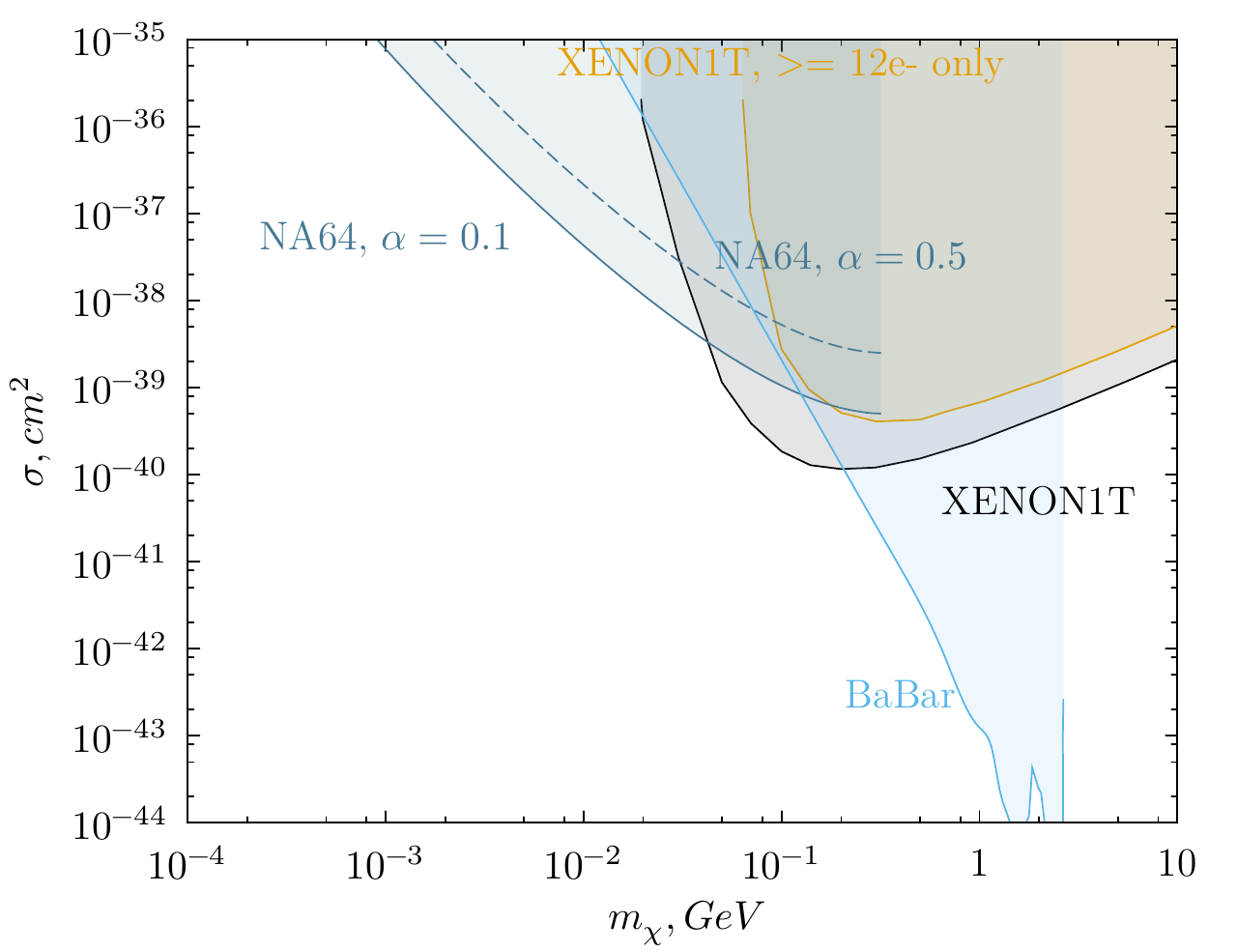}
\caption{Comparison of 90 \% C.L. upper limits on LDM-electron scattering cross-sections calculated by using NA64 \cite{NA64explast3} and BaBar constraints on kinetic-mixing from Fig. 2 with results of direct searches by XENON1T \cite{XENON1T}. The  blue curves are calculated for $\alpha_D=0.1$, while the dashed blue 
for $\alpha_D=0.5$. The  Yellow dashed line shows the XENON1T limit obtained without considering signals with $<$ 12 produced electrons. } 
\label{cross-sec}
\end{center}
\end{figure}

Recently XENON1T collaboration has published new record results \cite{XENON1T} on the search for direct electron LDM scattering. 
New bounds on elasic electron LDM cross sections  were obtained for $m_{\chi} \geq 30~MeV$. For the 
model with dark photon the use of the formula (\ref{53}) and the results of ref.\cite{XENON1T}  allows to  derive bound on $\epsilon^2\alpha_D$.
In Fig.\ref{cross-sec} the comparison of 90 \% C.L. upper limits on the cross-sections of LDM electron scattering
 transmitted by dark photon mediator $A'$  calculated by using NA64 \cite{NA64explast3} and BaBar bounds 
and  the  XENON1T \cite{XENON1T} bounds has been presented   for $\alpha_D = 0.1$. For $m_{\chi} \leq 50~MeV$ the NA64 bound 
is stronger than the XENON1T bound. 
For pseudo-Dirac fermions with not too small $\delta = \frac{m_{\chi_2} - m_{\chi_1}}{m_{\chi_1}}$ the reaction of $\chi_2$ electroproduction 
$\chi_1 ~e \rightarrow \chi_2~e$ for nonrelativistic LDM $\chi_1$  is prohibited due to energy 
conservation law, while elastic $\chi_1 ~e \rightarrow \chi_1~e$ scattering is absent at tree level that  
extremely complicates the direct LDM detection for pseudo-Dirac fermions.

\section{NA64 experiment}

\subsection{Invisible mode}

NA64 experiment\cite{NA64} at the CERN SPS employs the electron beam from the H4 beam line in the North Area (NA). 
The beam delivers $\approx 5 \times 10^{6}~e^-$ per SPS spill of $44.8~s$ produced by the primary $400~GeV$ proton beam 
with an intensity of a few $10^{12}$ protons on target. 
The NA64  experiment   is a fixed target experiment  searching for dark sector particles at 
the CERN Super Proton Synchrotron(SPS) by using active beam dump technique combined with missing energy approach
\cite{NA64, Segm, Crivelli,  Kirpich}. 
If new light boson  $A'$ exists it could be produced in the reaction of high energy electrons scattering off nuclei. 
Compared to the traditional beam dump experiment the main advantage of the NA64 experiment is that its sensitivity is 
proportional to the $\epsilon^2$. While for the classical beam dump experiments the sensitivity is 
proportional to the $\epsilon^2 \cdot \epsilon^2$, where one $\epsilon^2$ comes from   new particle production 
in the dump and another $\epsilon^2$ is from the LDM interaction in far detector. Another advantage 
of the NA64 experiment is that due to the higher energy of the incident beam, the centre-of-mass system is 
boosted relative to the laboratory system. This boost leads to enhanced hermeticity of the detector 
providing a nearly full solid angle coverage.    

The NA64 method of the search 
can be illustrated by considering the search for the dark photon $A'$ production 
for invisible $A'$ decays $A' \rightarrow \chi \bar{\chi}$ into LDM particles. 
A fraction $f$ of the primary beam energy $E_{A'} = fE_0$ is carried away by $\chi$ LDM particles, 
which penetrate the target and detector without interactions resulting in zero energy deposition. The remaining 
part of beam energy  $E_e  = (1-f)E_0$ is deposited in the target by the scattered electron. The 
occurrence of the $A'$ production via  the reaction   
$e Z \rightarrow e  Z A';  ~A' \rightarrow \chi \bar{\chi}$ would appear as an excess of events with a 
signature of a single isolated electromagnetic (e-m) shower in the active dump with energy $E_e$ 
accompanied  by a missing energy $E_{miss} = E_{A'} = E_0 - E_e$ above those expected from backgrounds. 
Here we assume that LDM  particles $\chi$ traverse the detector without decaying 
visibly. 
\begin{figure}[tbh!!]
\begin{center}
\includegraphics[width=0.9\textwidth]{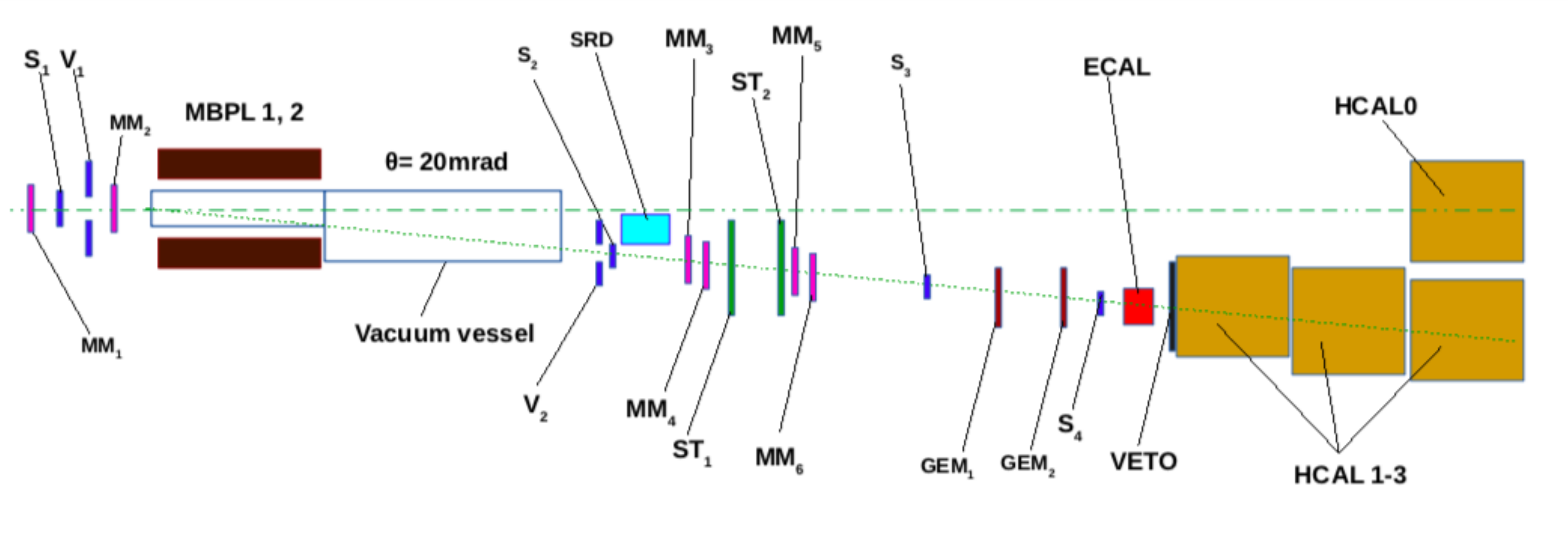}%
\vskip-0.cm{\caption{Schematic illustration of the setup to search for invisible  decays of the bremsstrahlung $A'$s 
produced in the reaction  $eZ \rightarrow eZ A'$ of 100 GeV e$^-$ incident  on the active ECAL target.}
\label{setup}}
\end{center}
\end{figure} 
Currently, the NA64 employs the $100~GeV$ electron beam from $H4$  beam line at the North  Area (NA) of the 
CERN SPS. The beam was optimized to transport the electrons with the maximal 
intensity $\geq 10^7$ per SPS spill with the momentum $100~GeV/c$.  The NA64 detector is schematically 
shown in Fig.\ref{setup}. The setup utilized the beam defining  scintillator  (Sc) counters $S1 -S3$  and veto V1, 
and  the spectrometer consisting of  two successive dipole magnets with the 
integral magnetic field of $\approx 7~T\cdot m$ and low-material-budget tracker. The tracker is a set 
of upstream Micromegas chambers $(T1, T2)$  and downstream Micromegas, GEM and Straw tube stations, 
measuring the beam $e^-$ momenta, $P_e$ with the precision $\delta P_e /P_e \approx 10^{-2}$ \cite{NA64}. 
The magnets also serve as an effective filter rejecting the low energy electrons present in the beam. 
The key feature of NA64 is the use of synchrotron radiation $(SR)$ from high energy electrons in the 
magnetic field to significantly enhance electron identification and suppress background from a hadron
 contamination in the beam. A 16 m long vacuum vessel was installed between the magnets and the ECAL to 
minimize absorption of the SR photons detected immediately at the downstream end of the vessel with a 
SRD, which is array of $PbSc$  sandwich counters  of a very fine longitudinal segmentation assembled 
from $80-100~\mu m$ $Pb$ and $1~mm~Sc$  plates with wave length shifting (WLS) fiber read-out. This 
allowed to additionally suppress background from hadrons, that could knock off electrons 
from the output vacuum window of the vessel producing a fake $e^- ~SRD$ tag, 
by about two orders of magnitude. The detector is also equipped with an active target, 
which is a hodoscopic electromagnetic calorimeter (ECAL) for the measurement of the electron energy 
deposition, $E_{ECAL}$, with the accuracy $\delta E_{ECAL}/E_{ECAL} \approx 0.1/\sqrt{E_{ECAL}[GeV]}$ 
as well as the $X$, $Y$ coordinates of the incoming electrons by using the transverse $e-m$ shower profile. The ECAL 
is a matrix of $6 \times 6$ Shashlik-type counters assembled with $Pb$ and $Sc$ plates with $WLS$ fiber read-out. 
Each model is $\approx 40$ radiation lengths $(X_0)$ and has an initial part $\approx4~X_0$ used as a 
preshower (PS) detector. By requiring the presence of in-time SR signal in all three SRD counters, and using  
the information of the longitudinal and lateral shower development in the ECAL, the initial level of 
the hadron contamination in the beam $\pi/e^- \leq 10^{-2} $ was further suppressed by more than 4 
orders of magnitude, while the electron ID at the level $\geq 95 \%$. A high-efficiency veto counter $ Veto $, 
and a massive, hermetic hadronic calorimeter (HCAL) of $\approx~30$ nuclear interaction lengths  $(\lambda_{int})$ 
were positioned after the ECAL.  The $Veto$ is a plane of scintillation counters used to veto charged 
secondaries incident on the HCAL detectors  
from upstream $e^-$ interactions. The HCAL which was an assembly of four modules $HCAL0-HCAL3$ served as an efficient veto to detect muons of 
hadronic secondaries produced of in the $e^-A$ interactions ECAL target. The $HCAL$ energy resolution is 
$\delta E_{HCAL}/E_{HCAL}       \approx 0.6/\sqrt{E_{HCAL}[GeV]}$.    
\subsection{Visible mode}
\begin{figure}[tbh!!]
\begin{center}
\includegraphics[width=0.9\textwidth]{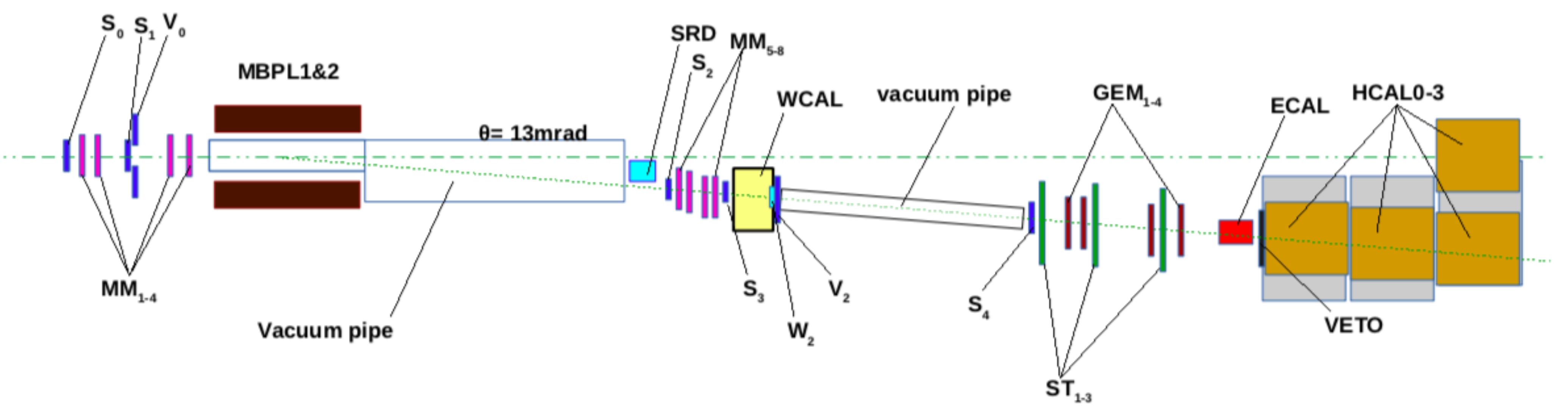}
\vskip-0.cm{\caption{Schematic illustration of the setup to search for visible  
$A', X \rightarrow e^+e^-$ decays 
decays of the bremsstrahlung $A',~X$ 
produced in the reaction  $eZ \rightarrow eZ A'$ of 100 GeV e$^-$ incident  on the active WCAL target.\label{setupvis}}}
\end{center}
\end{figure}

The NA64 setup designed for the searches for decays $X, A' \to e^+ e^- $ of the $X$ bosons, which could explain the $^8$Be anomaly (see below 5.1.2) and 
the  $A'$ is schematically shown in Fig.\ref{setupvis}. 
The NA64 experiment for visible $A' \rightarrow e^-e^+$ searches employs the optimized electron beam
from the H4 beam line in the North Area (NA) of the CERN SPS. The beam delivers $ 5 \times 10^6$ EOT per SPS  spill of $4.8 ~s$ produced by the primary $400~GeV$ proton
beam with an intensity of a few $10^{12}$ protons on target. Two scintillation  counters, $S1$ and $S2$ were used for the beam definition,
while the other two, $S3$ and $S4$, were used to detect the  $e ^+ e^-$ pairs. The detector is equipped with a magnetic
spectrometer consisting of two MPBL magnets and a low material budget tracker. The tracker was a set of four
upstream Micromegas (MM) chambers $(T1, T2)$ for the incoming e- angle selection and two sets of downstream
MM, GEM stations and scintillator hodoscopes $(T3, T4)$ allowing the measurement of the outgoing tracks  \cite{NA64}.
To enhance the electron identification the synchrotron radiation (SR) emitted by electrons was used for their effi-
cient tagging and for additional suppression of the initial hadron contamination in the beam $\pi/e$  $10^{-2}$ down to
the level $10^{-6}$ \cite{Crivelli}. The use of SR detectors (SRD) is a key point for the hadron background suppression and
improvement of the sensitivity compared to the previous electron beam dump searches \cite{NA64}. The dump is a
compact electromagnetic (e-m) calorimeter WCAL made as short as possible to maximize the sensitivity to short
 lifetimes while keeping the leakage of particles at a small
level. The WCAL was assembled from the tungsten and
plastic scintillator plates with wave lengths shifting fiber
read-out. The first (last) few layers of the WCAL were read
separately to form a signal from a preshower (veto $W_2$) counter. Immediately after the $W_2$
there is also one more veto counter $V2$, and several meters downstream the
signal counter $S4$ and tracking detectors. 
These detectors
are followed by another e-m calorimeter (ECAL), which
is a matrix of $6 \time 6$ shashlik-type lead - plastic scintillator
sandwich modules \cite{shashlyk}. Downstream the ECAL the detector 
was equipped with a high-efficiency veto counter,
and a thick hadron calorimeter (HCAL) \cite{NA64} used as a
hadron veto and muon identificator.
For the cuts selection, calculation of various efficiencies
and background estimation the package for the detailed
full simulation of the experiment based on Geant4 \cite{geant4}
is developed. It contains the subpackage for the simulation of various types of DM particles based on
the exact tree-level calculation of cross sections \cite{Exact, Exactkirpich}.
The method of the search for 
$A' \rightarrow e^-e^+$  decays is
described in \cite{NA64}. The application of all further
considerations to the case of the 
$X \rightarrow e^+e^-$ decay is
straightforward. If the $A'$  exists, it could be produced via
the coupling to electrons wherein high energy electrons
scatter off a nuclei of the active WCAL dump target,
followed by the decay into $e^+e^-$  pairs:
\begin{equation}
e^-   ~+~ Z  \rightarrow e^- ~+~ Z ~+~ A', ~~A' \rightarrow e^-e^+ \\.
\label{55}
\end{equation}
The reaction (\ref{55}) typically occurs within the first few
radiation lengths $(X_0 )$ of the WCAL. The downstream
part of the WCAL serves as a dump to absorb completely
the e-m shower tail. The bremsstrahlung $A'$ would penetrate the rest of the dump and the veto counter $V2$ without 
interactions and decay in flight into an $e^ + e^-$  pair
in the decay volume downstream the WCAL. A fraction (f ) of the primary beam energy $E_1 = f E_0$ is deposited 
in the WCAL by the recoil electron from the
reaction (\ref{53}). The remaining part of the primary electron energy 
$E_2=(1-f)E_0$ is transmitted through the 
dump by the $A^0$ , and deposited in the second downstream
calorimeter ECAL via the 
$A' \rightarrow e^+e^-$  decay in flight.
 The occurrence of  $A' \rightarrow e^+e^-$ decays produced in $eZ$
interactions would appear as an excess of events with
 two e-m-like showers in the detector: one shower in the
 WCAL and another one in the $ECAL$, with the total
 energy $E_{tot} = E_{WCAL} +E_{ECAL}$ equal to the beam energy
 $(E_0 )$, above those expected from the background sources. 

\section{Current  and future NA64  results}

In this section  we briefly discuss  last NA64  results and the perspectives of the 
NA64e(future NA64 experiment with electron beam) and NA64$\mu$(future NA64 experiment 
with muon beam).

\subsection{NA64e }

\subsubsection{Invisible mode. Dark photon bounds}

\begin{figure}[tbh!!]
\begin{center}
\includegraphics[width=0.45\textwidth,height=0.37\textwidth]{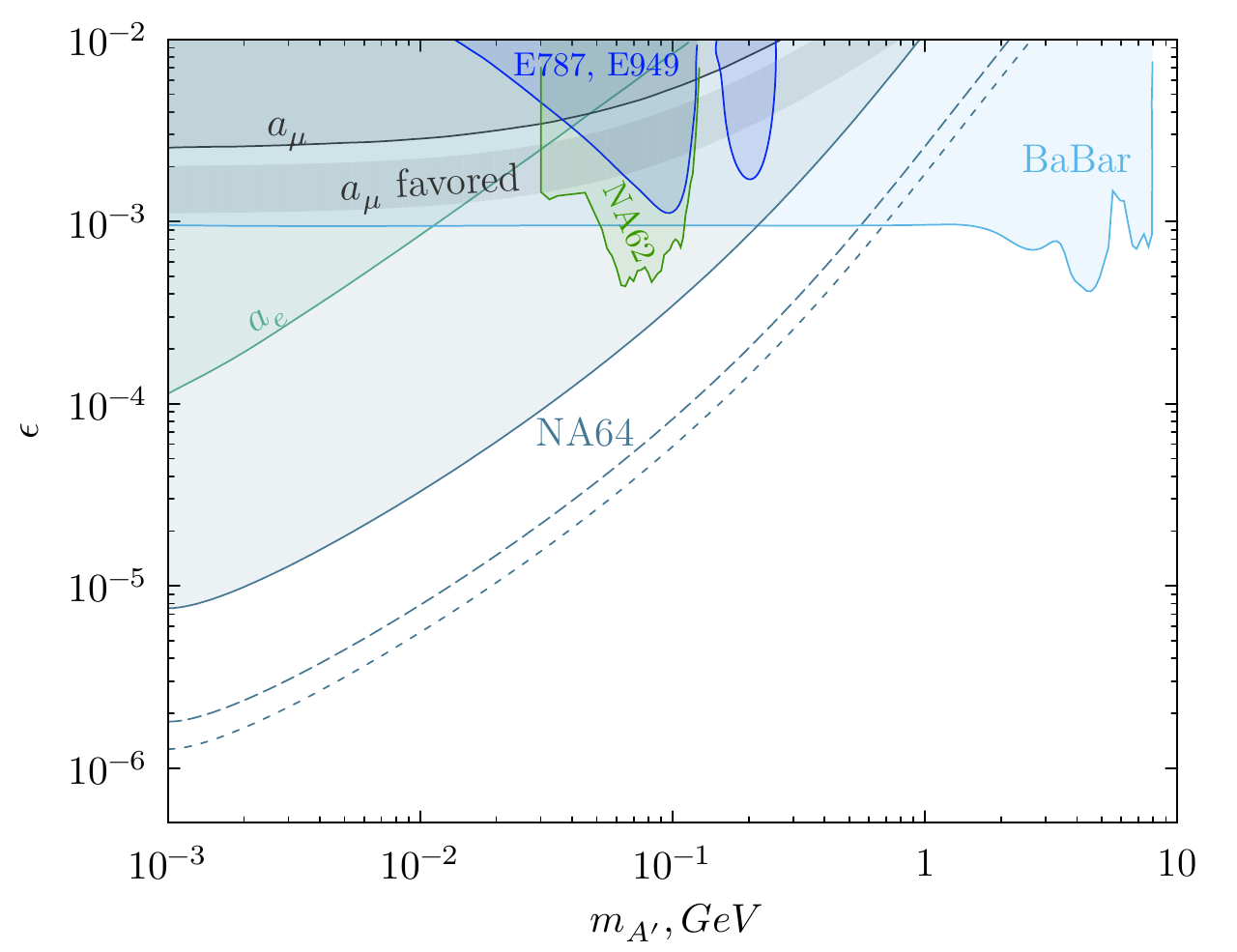}
\includegraphics[width=0.45\textwidth,height=0.4\textwidth]{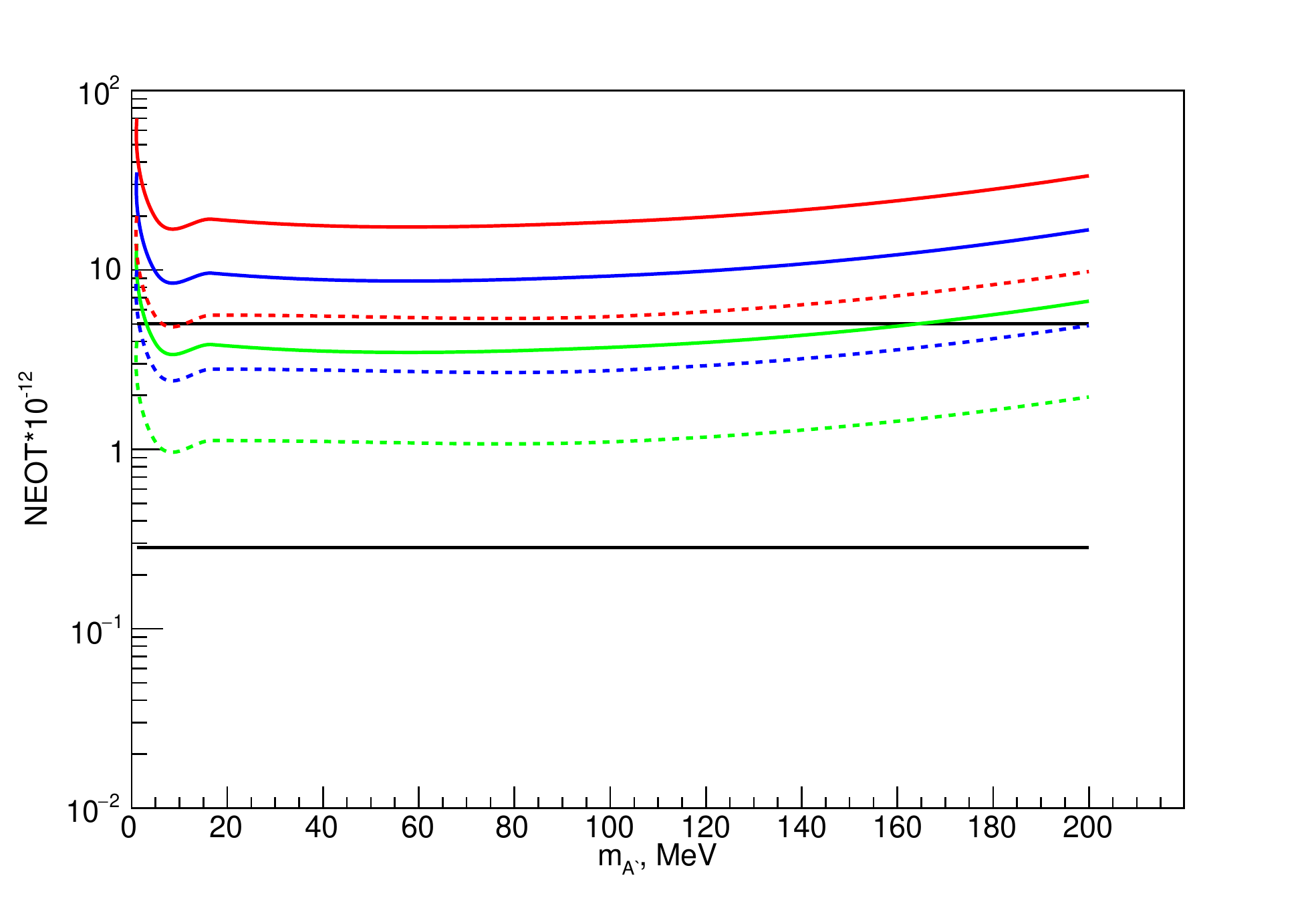}
\includegraphics[width=0.45\textwidth,height=0.4\textwidth]{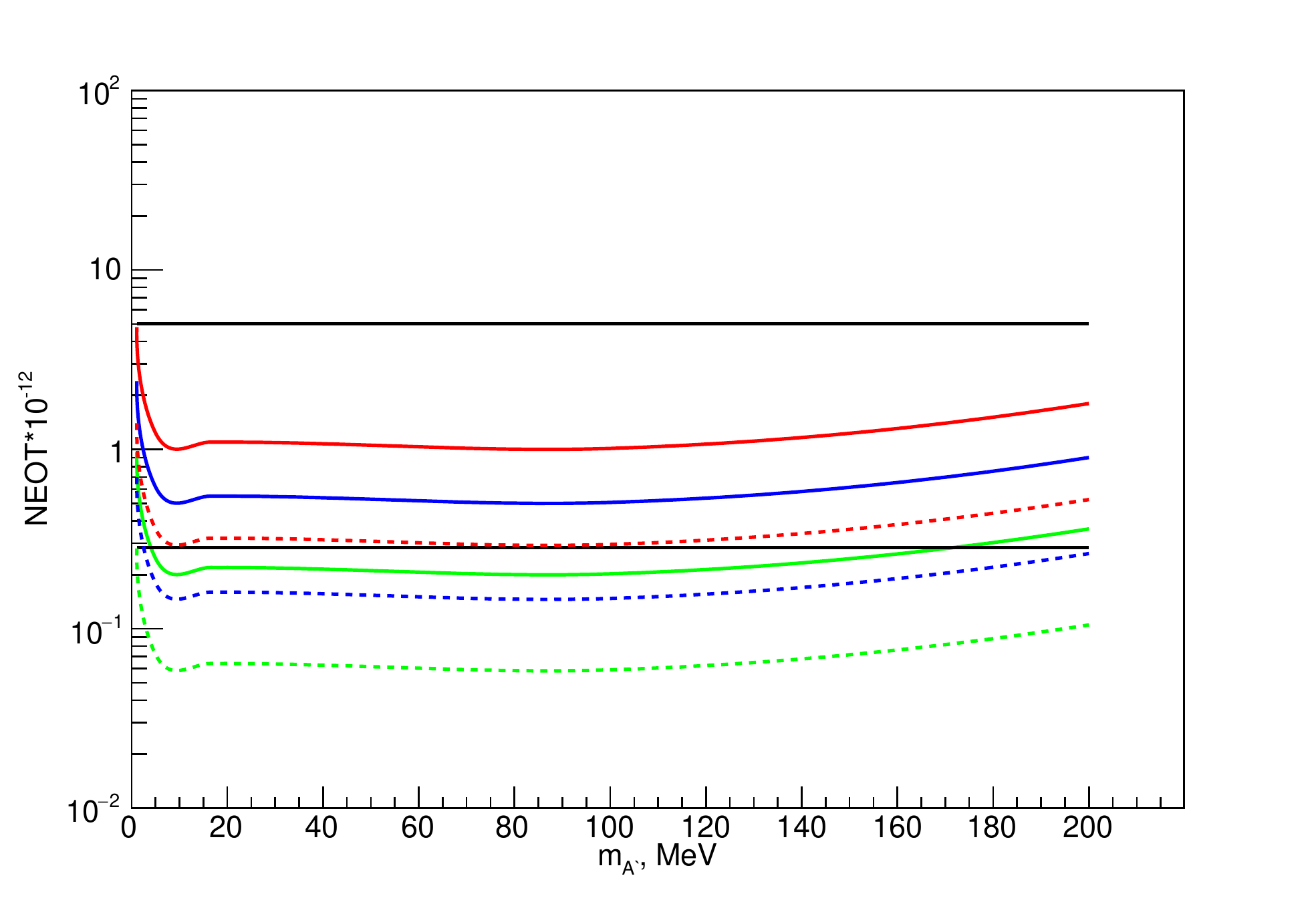}
\includegraphics[width=0.45\textwidth,height=0.4\textwidth]{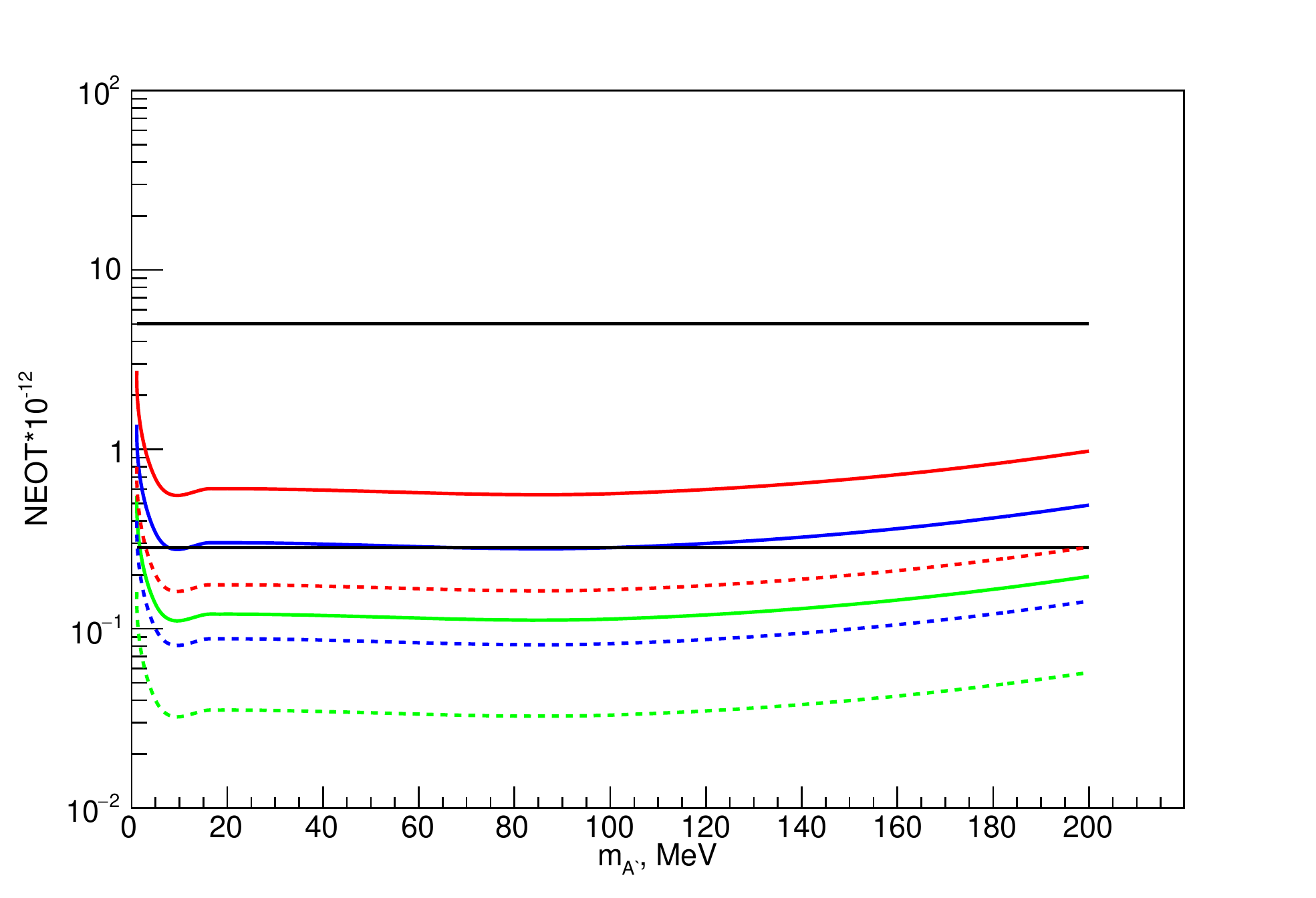}
\caption {The upper l.h.s. panel  shows the NA64 90\% C.L. current   bound (solid)  \cite{NA64explast3},  
and projected boundes for $5\times 10^{12}$(dashed)  and $10^{13}$(dotted)  in the ($m_{A'},  \epsilon^2$) plane.
The upper r.h.s plot and lower plots  show the required number of EOT for the 90\% C.L. 
exclusion of the $A'$ with a given mass 
$m_{A'}$ in the ($m_{A'}, n_{EOT}\times10^{-12}$ )  plane for pseudo-Dirac  with 
$\delta \ll1$(the upper r.h.s panel), Majorana (the lower l.h.s. panel), 
and scalar (the lower r.h.s. panel)   DM models for 
$\frac{m_{A'}}{m_{\chi}} = 2.5$ (solid),  and $= 3$ (dashed), and 
 $\alpha_{D} = $ 0.1 (red), 0.05 (blue),  
and  
0.02   (green). Upper(lower)  black lines correspond to $n_{EOT} = 5 \times 10^{12}(2.84 \times 10^{11})$. 
The curves under lower black line are excluded by last NA64 results  \cite{NA64explast3}.
\label{fig:excl-eps}}
\end{center}
\end{figure}

The NA64 collected $NEOT = 2.84\cdot10^{11}$ statistics in  the 2016-2018 years. Recently NA64 collaboration \cite{NA64explast3} has 
been analyzed 
these data and obtained new bounds on $\epsilon$ parameter\footnote{The assumption that  $Br(A' \rightarrow invisible) = 1$ has been used.}
  by factor $\sim 2.5$ stronger the previous bound \cite{NA64explast1}, see the upper l.h.s. panel 
in Fig.\ref{fig:excl-eps}. 
After the long shutdown (LS2)  at CERN the NA64 experiment plans to accumulate  
 $NEOT \gtrsim 5 \times 10^{12}$. The NA64e future expected 
limits on mixing strength $\epsilon$  after the LS2 period assuming the 
zero-background case   are shown in the upper l.h.s. panel in 
 Fig.\ref{fig:excl-eps}. 

To estimate NA64 LDM discovery potential we have used the formulae of Appendix A to calculate  the predicted value of $\epsilon^2$ as 
a function of $\alpha_D$, $m_{\chi}$ and $m_{A'}$ in the assumption that in the early Universe LDM was in thermo equilibrium. 
We used the values $\alpha_D = 0.02, ~0.05, ~0.1$ and $\frac{m_{A'}}{m_{\chi}} = 2.5, ~3$. We have made the calculations for 
the case of scalar, Majorana and pseudo-Dirac LDM with $(\delta \ll 1)$. Our results \cite{GKKK}   are presented in   Fig.\ref{fig:excl-eps}. 
The upper r.h.s plot and lower plots   in Fig.\ref{fig:excl-eps} show the required number of $EOT$ for the 
90\% C.L. exclusion of the $A'$ with a given mass 
$m_{A'}$ in the ($m_{A'}, n_{EOT}\times10^{-12}$ )  plane for pseudo-Dirac  with 
$\delta \ll1$ (the upper r.h.s panel), Majorana (the lower l.h.s. panel), 
and scalar
(the lower r.h.s. panel)   LDM  models 
for $\frac{m_{A'}}{m_{\chi}} = 2.5$ (solid),  and = 3 (dashed), and 
 $\alpha_{D} = $ 0.1 (red), 0.05  (blue),  
and 0.02 (green).
We see that NA64 experiment has already excluded scalar LDM model with $\alpha_D  \leq 0.1$, 
 $\frac{m_{A'}}{m_{\chi}} \geq  3$ and Majorana LDM with  $\alpha_D  = 0.02$,  $\frac{m_{A'}}{m_{\chi}} \geq  2.5$. 
     As one can see from  Fig.\ref{fig:excl-eps}    with $n_{EOT} = 5 \times10^{12}$  NA64e 
will be  able to 
exclude  the most interesting and natural  LDM scenarios in the $A'$ mass range  
$1~MeV \leq m_{A'} \leq 150$~MeV
except the most difficult case of  pseudo-Dirac LDM with 
$\alpha_D = 0.1$,  $\alpha_D = 0.05$ and  $\frac{m_{A'}}{m_{\chi}} = 2.5 $. 

\subsubsection{The problem with resonance region}

 The expressions  for the annihilation cross-sections   are  proportional to 
 the factor $K =\epsilon^2\alpha_D (\frac{m^4_{A'}}{m_{\chi}^2} -4)^{-2}$. From the assumption that in the early Universe the LDM 
was in equilibrium with the SM matter we can predict the dependence of $K $ on DM mass $m_{\chi}$, see Appendix A. 
In the resonance 
region $m_{A'} \approx 2m_{\chi}$  the  $\epsilon^2$ parameter   is proportional to $K^{-1}$ 
that can reduce the predicted $\epsilon^2$ value by (2 - 4) orders of magnitude \cite{Feng} in comparison with the 
often used reference point  $\frac{m_{A'}}{m_{\chi}} =3$. It means  that  NA64 experiment and probably 
 other future experiments 
will  not be able to test the region
 $m_{A'} \approx 2m_{\chi}$ completely. 
It should be mentioned that   the values of $m_{A'}$ and $m_{\chi}$ are  arbitrary, so the case
$m_{A'} = 2m_{\chi}$   could be considered  as some fine-tuning. It is 
natural to require the absence of significant fine-tuning. 
We   require   that  $(\frac{m_{A'}}{2m_{\chi}} -1) \geq 0.25$, i.e. $m_{A'} \geq 2.5 m_{\chi}$.
In our estimates (see Fig.\ref{fig:excl-eps}) 
we  used  two values   $\frac{m_{A'}}{m_{\chi}} = 2.5   $ and   $\frac{m_{A'}}{m_{\chi}} = 3 $. As it follows 
from  the previous subsection  the NA64 will be able to test the most interesting LDM models 
for the case of  significant fine-tuning absence.  

\subsubsection{Visible mode. The $^8$Be anomaly.}
The ATOMKI experiment of Krasznahorkay et al. \cite{17mev} has reported the observation of a 6.8 $\sigma$ excess of events
in the invariant mass distributions of $e^+ e^-$ pairs produced in the nuclear transitions of excited  $^8Be^*$ to its ground state via
internal pair creation.  This anomaly can be interpreted as the emission of a new 
protophobic gauge $X$ boson with a mass of 16.7 MeV followed by its $X\to e^+ e^-$   decay assuming that the $X$ has
non-universal couplings to quarks, coupling
to electrons in the range $2\times 10^{-4} \lesssim \epsilon_e \lesssim 1.4\times 10^{-3}$ and the lifetime 
$10^{-14}\lesssim \tau_X \lesssim 10^{-12}$~s \cite{FENG}. 
It has motivated worldwide theoretical and experimental 
efforts towards light  and weakly coupled  vector bosons,  see, e.g. \cite{jk}-\cite{pf}. 
Another strong motivation to the search for a new light boson decaying into $e^+ e^-$ pair is provided by the Dark Matter puzzle discussed previously. 
\begin{figure}[tbh!!]
\begin{center}
\includegraphics[width=0.6\textwidth]{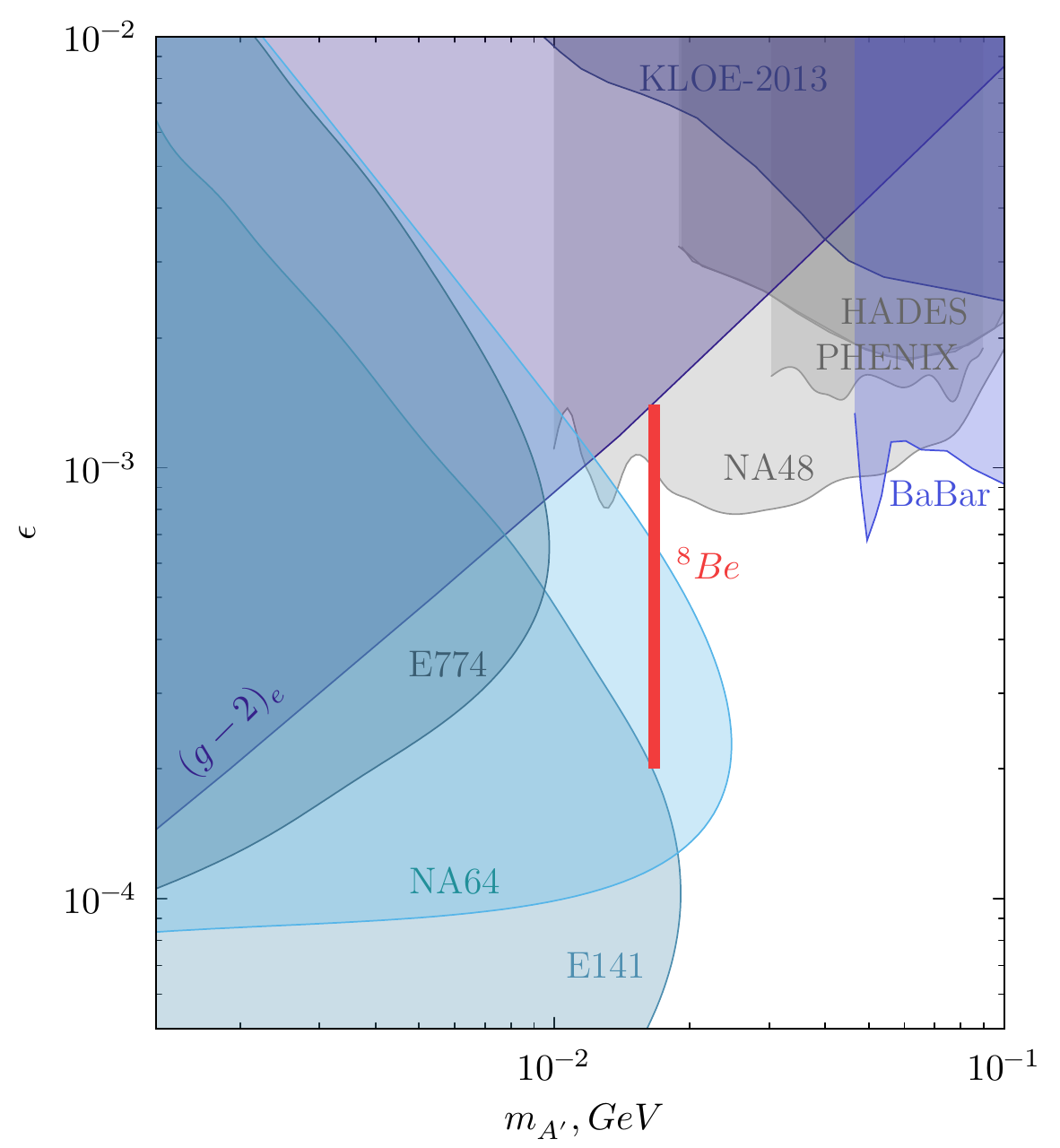}
\vskip-0.cm{\caption{The 90 \% C.L. exclusion area in the $(m_{\chi}; \epsilon)$ 
from the NA64 experiment (blue area).  For the mass of 
$16.7~MeV$, the $X - e$ coupling region excluded by NA64 is $1.2 \times10^{-4} < \epsilon_e < 6.8 \times 10^{-4}$.\label{exclvis}}}
\end{center}
\end{figure} 
\par The NA64e  combined 90\% C.L. exclusion limits on the mixing  $\epsilon$ as a function of the $A'$ mass are shown in Fig.~\ref{exclvis} together 
with the current constraints from other experiments \cite{na64-be-new}. The NA64  results exclude the X-boson as an explanation of the $^8$Be* anomaly
for the $X-e^-$ coupling  $ \epsilon_e  \lesssim 6.8 \times 10^{-4}$ and the mass value of 16.7 MeV, leaving the still unexplored  region
$6.8~\times  10^{-4} \lesssim  \epsilon_e  \lesssim 1.4 \times 10^{-3}$ for further searches. 
Note that in recent paper \cite{Krasnikovlast} the last NA64 data \cite{na64-be-new} has been analyzed. It was shown that at 90 \% C.L. 
 models with pure vector or  axial vector couplings of electron with $X(16.7)$ boson are excluded but the chiral couplings $V \pm A$ are still possible 
and moreover it is possible to explain both electron $g_e-2$ and muon $g_{\mu}-2$ anomalies  \cite{Krasnikovlast}.   
\par Very recently the ATOMKI group reported a similar excess of events at approximately the same invariant mass in the nuclear transitions of another nucleus, $^4$He \cite{atomki-new}. This dramatically  increases the importance of confirmation of the observed excess by another nuclear physics experiment, as well as independent  searches for the X in a particle physics experiment. Therefore, the NA64 experimental  approach based on the using two independent  electromagnetic calorimeters, one as an active dump (WCAL)  for the $X$ boson  production and another one (ECAL) for the $X\to e^+ e^-$  decay detection is extremely timely. 
To cover the remaining parameter space for the $X-e$ couplings, which corresponds to a very short-lived X boson case with a lifetime  $\tau_X \lesssim 10^{-13}$ s, is very challenging. A more accurate future measurement after LS2 should include also the $e^+ e^-$ pair invariant mass  reconstruction. This requires the use of a high-precision tracker with an excellent two-track resolution capability combined with a magnetic spectrometer for the accurate decay  electron and positron momenta measurements to finally reconstruct the invariant mass of the $X$ with a good precision. For this NA64e will need a substantial upgrade of the current setup with a new high-resolution trackers, e.g.  based on micromegas detectors, a new WCAL with a better optimised thickness,  and a new synchrotron radiation detector with higher granularity. 
This makes further searching  quite challenging but very exciting and important.

\subsubsection{NA64$e$ and the search for $Z'$ boson coupled with $L_{\mu} - L_{\tau}$ current}
Light $Z'$ boson which couples with   $L_{\mu} - L_{\tau}$ current will mix with ordinary photon 
at one-loop level \cite{gkm}. Namely, an account of one-loop  propagator diagrams with virtual $\mu$- and $\tau$-leptons 
leads to nonzero $\gamma -Z'$ kinetic mixing $-\frac{\epsilon}{2}F^{\mu\nu}Z^{`}_{\mu\nu}$ 
where $\epsilon$ is the finite  mixing strength given by \cite{Holdom} 
\begin{equation}
\epsilon_{1l} = \frac{8}{3} \frac{ee_{\mu}}{16\pi^2}{\rm ln}(\frac{m_{\tau}}{m_{\mu}}) = 1.4 \cdot 10^{-2} \cdot e_{\mu} \,.
\label{56}
\end{equation}  
Here $e$ is the electron charge, $e_{\mu}$ is electron $Z^`$ charge  and $m_\mu, ~ m_\tau$ are the muon and tau lepton masses respectively. 
It should be stressed that we assume 
 that possible tree level mixing  $-\frac{\epsilon_{tree}}{2}F^{\mu\nu}Z^`_{\mu\nu} $ 
is absent or much smaller than one-loop mixing $\frac{\epsilon_{1L}}{2}F^{\mu\nu}Z^{\mu\nu} $. To be precise, we 
assume that there is no essential cancellation between tree-level and one-loop mixing terms $|\epsilon_{tree} + \epsilon_{1l} | 
\geq  |\epsilon_{1l}|$ . 
For $m_{Z'} \ll m_{\mu}$ the value $e_{\mu} = (4.8 \pm 0.8) \cdot 10^{-4}$ from  Eq.(\ref{56})  leads to the prediction of the corresponding mixing value   
\begin{equation}
\epsilon_{1l} = (6.7 \pm 1.1) \cdot 10^{-6}
\label{57}
\end{equation}
Thus, one can see that  the $Z'$  interaction  with the $L_{\mu} - L_{\tau}$ current induces 
at one-loop level the $\gamma - Z' $ mixing of $Z'$  with ordinary photon 
which allows to probe $Z'$  not only in muon or tau induced reactions but also with intense electron beams. In particular, this loophole  opens up  the possibility 
of  searching  the new weak leptonic force mediated by the $Z'$  in experiments looking for dark photons ($A'$).
  The fact that  
 the $\gamma-Z'$ mixing of Eq.(\ref{57}) is at an experimentally interesting  level  is very exciting. We point out further  that a
 new intriguing possibilities for the complementary searches of the $Z'$ in the currently ongoing 
experiment  NA64 \cite{NA64,NA64explast3} exists. Indeed, the NA64 aimed at the direct search for invisible decay of sub-GeV dark photons in the reaction 
$e^- + Z \to e^- + Z + A'; ~ A' \to invisible$  of high energy electron scattering off heavy nuclei \cite{NA64}. The experimental signature 
of the invisible decay of $Z'$ produced in the reaction  $e^- + Z \to e^- + Z + Z'; ~ Z' \to invisible$  due to mixing of Eq.(\ref{56})
 is  the same - it is an event with a large missing energy carried away by the $Z'$. Thus, by using Eq.(\ref{57})  and  bounds on the $\gamma - A'$ mixing 
 the NA64 can also set constraints on   coupling  $e_\mu$.
 
 The current  NA64 bounds on the $\epsilon$ parameter for the  dark photon mass  region  $1\lesssim m_{Z'} \lesssim 10$ MeV are in the range 
 $ 0.7 \cdot 10^{-5} \lesssim  \epsilon \lesssim 3 \cdot 10^{-5}$  
\cite{NA64explast3}. 
  Taking into account that the sensitivity of the experiment   scales as $\epsilon \sim 1/\sqrt{n_{EOT}}$,   
 results in  required  increase of statistics by a factor $\simeq$ 30  in order  to improve  sensitivity up to the  mixing value of 
 Eq.(\ref{57}) for this $Z'$ mass   region. This would allow  either to discover  the  
$Z'$ or exclude it as an explanation of the $g_{\mu} -2 $ anomaly for the  substantial part of the  mass range $m_{Z'} \ll m_{\mu}$  by  using the electron beam. 
The direct search for the $Z'$ in missing-energy  events in the reaction  $\mu Z \rightarrow \mu Z Z'; Z'\to invisible$  in the 
dedicated experiment   with the muon beam at CERN  would  then be an important cross check  of results obtained with the  electron beam. 
Let us note that the mixing given by the Eq.(\ref{57})  would also lead to an extra contribution to the elastic $\nu e \to \nu e$  scattering signal 
 in the solar neutrino measurement at the Borexino experiment \cite{takeshi1}.
 The BOREXINO data  on the elastic $\nu_{\mu}e$ scattering \cite{BOREXINO} lead to lower bound 
on $m_{Z'} \geq (5-10)~MeV$ by assuming that muon anomaly is explained due to 
existence of light $Z'$ boson interacting with $L_{\mu} - L_{\tau}$ current and 
there is no tree level mixing between photon and $Z'$, i.e.  $\epsilon_{tree} = 0$.
 The measurement of $\nu - e$  elastic scattering  in the LSND experiment \cite{lsnd} set  a similar bound to the $e_\mu$ coupling for $m_{Z'} \lesssim 10$ MeV \cite{takeshi1}.
The expected 90\% C.L.  NA64 exclusion regions   in the ($m_{Z'}, e_\mu$) 
plane (dashed curves) from the measurements with the electron beam for  $\simeq 4\times 10^{12}$ and   $\simeq 4\times 10^{13}$ 
EOT  and muon beams for 
$\simeq 10^{12}$  muons on target (MOT)  \cite{gkm} 
are shown in Fig.\ref{Fig11}. Constraints from the BOREXINO  \cite{takeshi1}, CCFR   \cite{ccfr}, and BABAR \cite{BABAR1} experiments,  
as well as the BBN excluded area \cite{ takeshi1, kamada}  are also shown.
The parameter space shown in Fig.{\ref{Fig11} could  also be probed by other electron experiments such as Belle II \cite{takeshi2},  
BDX \cite{BDX1, BDX2},  and LDMX \cite{LDMX},  which  would provide  important complementary results.

\subsection{The experiment  NA64$\mu$}   
Recently,  the  NA64 collaboration proposed to carry out further searches  for dark sector and other rare processes in missing energy events from high energy  muon interactions in a hermetic detector  at the CERN SPS  \cite{NA64mu, NA64mu1}.
\par A dark sector of particles predominantly weakly-coupled to the second and possibly third generations 
of the SM  is  motivated by several theoretically interesting models.  Additional to gravity this new very  weak interaction between the visible and dark sector could be mediated either by a scalar ($S_\mu$) or  $U'(1)$ gauge  bosons ($Z_\mu$)  interacting  with  ordinary muons. In a class of $L_\mu - L_\tau$ models the corresponding $Z_\mu$ could be light and have the  coupling strength lying  in the experimentally accessible region. If such $Z_\mu$ mediator exists it  could also explain the muon $g_{\mu}-2$ anomaly - the discrepancy between the predicted and measured values of the muon anomalous magnetic moment \cite{NA64mu}.
\par The proposed  extension of the NA64 experiment called NA64$\mu$ 
aiming mainly at searching for invisible  decays of the $\zm$ either to neutrinos or LDM particles \cite{NA64mu1}. 
The primary goal of the  experiment in the 2021 pilot run with the $\simeq 100-160$ GeV M2 beam is to commission the NA64$\mu$ detector and  to probe for the first time the still unexplored area of the coupling  strengths 
and masses $M_{\zm} \lesssim 200$ MeV that could explain the muon $g_{\mu}- 2 $ anomaly. Another strong point of NA64$\mu$ is its  capability for  a sensitive search for 
dark photon mediator ($A'$)  of DM production in invisible decay mode in the  mass range $m_{A'} \gtrsim m_\mu$, thus making the experiment extremely 
complementary to the ongoing NA64e and greatly increases the discovery potential of sub-GeV dark matter. Other searches for $S_\mu$'s decaying invisibly to dark sector particles, and  millicharged particles  will probe a still unexplored  parameter areas \cite{NA64mu1}.

\subsubsection{Searching  for the $\mu ~+~Z \rightarrow \mu ~+~Z ~+~Z_{\mu}, ~Z_{\mu} \rightarrow \nu \bar{\nu}$   }
The reaction  of the $Z_{\mu}$ production is a rare  event. For the previously mentioned parameter space, it is  expected to occur 
with the rate $\lesssim  \alpha_\mu/\alpha \sim 10^{-6}$ with respect to the  ordinary photon production rate. Hence, 
its observation presents a challenge for the detector design and performance. 
The  experimental setup  specifically designed to search for the $Z_{\mu}$ is schematically shown in Fig. \ref{setup-na64mu}.

\begin{figure}[tbh!]
\begin{center}
\includegraphics[width =0.6\textwidth]{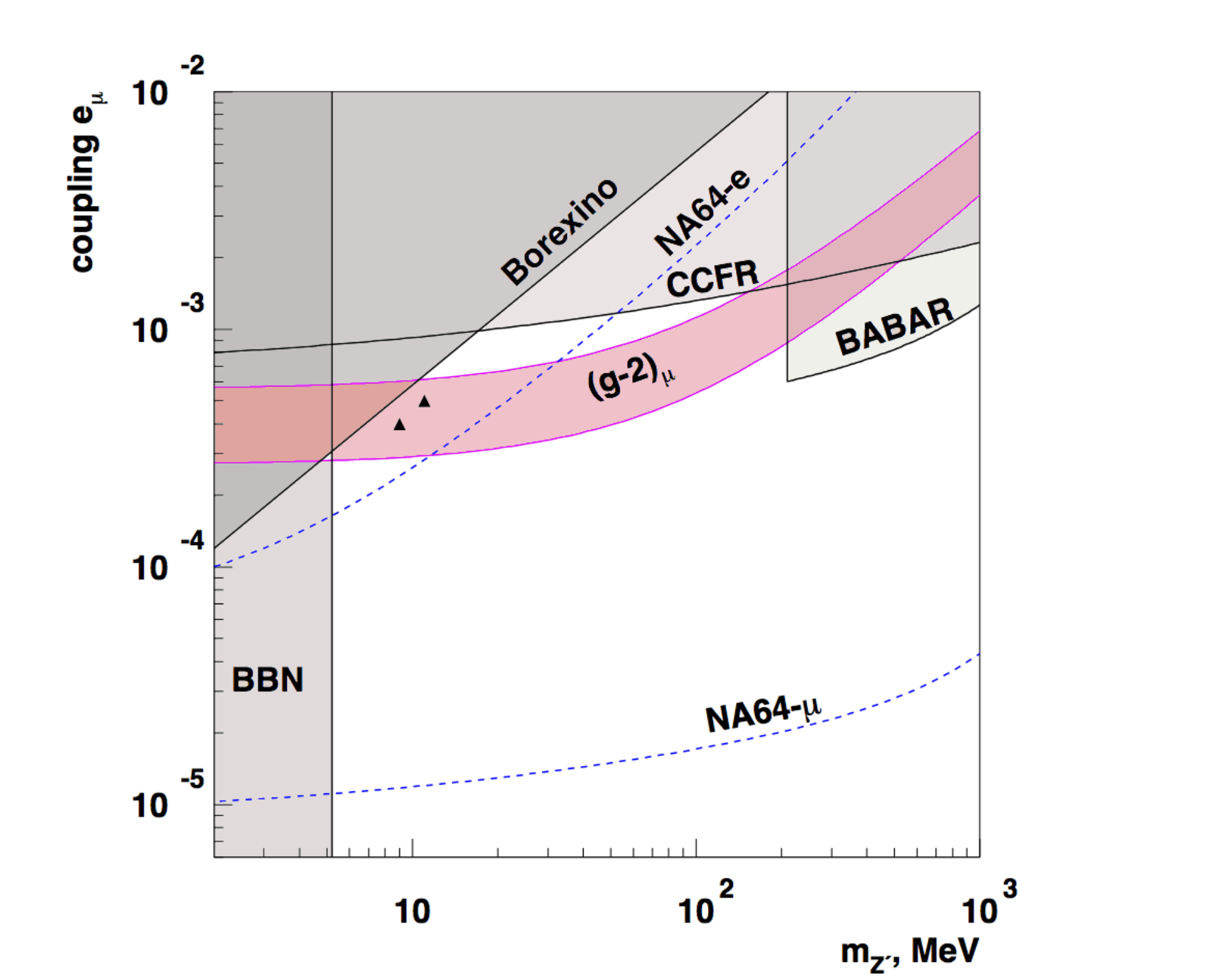}
\caption{The NA64 90\% C.L. expected exclusion regions in the ($m_{Z'}, e_\mu$) 
plane (dashed curves) from the measurements with the electron (NA64$e$, $\simeq 4\times 10^{12}$ EOT  and  muon (NA64$\mu$, $\simeq 10^{12}$ MOT) beams,
taken from ref. \cite{NA64mu, NA64mu1}.  
  Two triangles indicate   reference points  corresponding to 
 the mass $m_{Z'} = 9$ and 11 MeV, and coupling $e_{\mu} = 4\times10^{-4}$ and $5\times10^{-4}$, respectively, 
 which are used to explain the IceCube results, see ref.\cite{takeshi1} for details.}
\label{Fig11}
\end{center}
\end{figure}

\begin{figure}[tbh!]
\begin{center}
\includegraphics[width =0.95\textwidth]{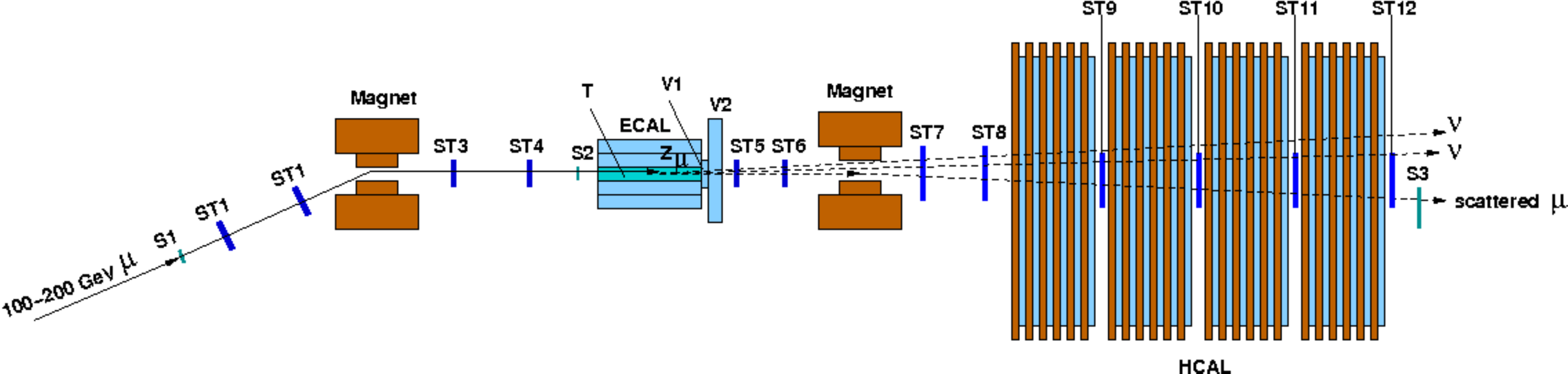}
\caption{Schematic illustration of the NA64$\mu$ setup to search for invisible $Z_{\mu}$ decays in the reaction $\mu Z \rightarrow \mu ZZ_{\mu}$ 
\cite{NA64mu}. \label{setup-na64mu}}
\end{center}
\end{figure}


The experiment could employ the upgraded muon beam at the CERN SPS. 
The  beam was designed to transport high fluxes of muons of the maximum momenta 
in the range between 100 and 225 GeV/c that could be derived from a primary proton beam of 450 GeV/c
with the intensity between 10$^{12}$ and 10$^{13}$ protons per SPS spill.
The detector shown in Fig.\ref{setup-na64mu} utilizes  two, upstream and downstream,  magnetic spectrometer sections
consisting of  dipole magnets and a set of low-material budget  straw tubes chambers, ST1-ST4 and ST5-ST6, respectively,   
allowed reconstruction and  precise measurements of incident and scattered in a target muons. It also uses 
 scintillating fiber hodoscopes  S1,S2, defining the primary muon beam, and S3, S4, and S5 defining the scattered muons,  
the active target $T$  surrounded by a  high efficiency  electromagnetic calorimeter (ECAL) serving  as a veto 
 against photons and other secondaries emitted from the target  at large angles. Downstream the target the detector
 is equipped with  high efficiency forward veto counters V1 and V2 and a massive, completely hermetic hadronic calorimeter 
(HCAL) located at the end of the setup to detect energy deposited by secondaries from the $\mu^- A \to anything$ primary muon  
interactions  with nuclei $A$  in the target. The HCAL has lateral and longitudinal 
 segmentation, and also serves for the final state muon identification. For searches at low  energies,  Cherenkov counters  
to enhance the incoming muon tagging efficiency can be used.

The method of the search is the following.
The bremsstrahlung $Z_{\mu}$s are produced in the reaction 

\begin{equation}    
\mu ~+~Z \rightarrow \mu ~+~Z ~+~Z_{\mu}, ~Z_{\mu} \rightarrow \nu \bar{\nu}
\label{58}
\end{equation}
   from the high energy muon scattering off nuclei in the target. 
 The reaction (\ref{58}) is  typically occurred uniformly over the length of the target.
The $Z_{\mu}$ is either stable or decaying invisibly if its mass $M_{Z_{\mu}}\leq 2 m_\mu$, or, as shown,  it could  
subsequently decay  into a $\mu^+\mu^-$ pair if $M_{Z_{\mu}} > 2 m_\mu$. In the former case,  
 the $Z_{\mu}$  penetrates  the T,  veto V1, V2 and the massive HCAL without interaction. In the later case, it could   decays 
in flight into a  $\mu^+\mu^-$ pair, resulting in the di-muon track  signature in the detector.  
The bremsstrahlung $Z_{\mu}$ then either penetrates the rest of the detector without interactions, resulting in 
zero-energy deposition in the V1, V2 and HCAL , or it could  decay in flight into a $\mu^+\mu^-$ pair 
if its mass is greater than the mass of two muons.  A fraction ($f\lesssim 0.3$) of the primary beam energy 
$E_\mu = f E_0$  is carried away by the scattered muon which is detected by the second magnetic spectrometer arm. 
For the radiation length 
$X_0 \lesssim$ 1 cm, and the total thickness of the target  $\simeq 30$ cm  the energy leak  from the target 
into the V1 is  negligibly small. The remained part of the primary muon energy $E_2 = (1-f)E_0$ is transmitted through the "HCAL wall" 
by the $Z_{\mu}$ , or  deposited partly in the HCAL  via the  $Z_{\mu}$ decay in flight
$Z_{\mu} \rightarrow \mu^+\mu^-$.  At $Z_{\mu}$ energies $E_{Z_{\mu}}\lesssim 50$ GeV,  the opening
 angle  $\Theta_{\mu^+\mu^-} \simeq M_{Z_{\mu}}/E_{Z_{\mu}}$ of the decay $\mu^+\mu^-$ pair is big enough to be resolved in 
two separated tracks in the M1 and M2 so the pairs are mostly  detected as a double track event.  
  The HACL  is served  as a dump to absorb completely the energy  of secondary particles produced in 
the primary pion or kaon interaction in the target.  In order to suppress  background due to the detection 
inefficiency,  the detector must be longitudinally completely hermetic. To enhance detector hermeticity, 
the hadronic calorimeter has the total thickness of $\simeq 28 ~\lambda_{int}$ (nuclear interaction lengths) 
and  placed behind the DV. 

The signature of the reaction (\ref{58}) is 
\begin{itemize}
\item the presence of incoming muon with energy around 150 GeV,
\item the presence of scattered muon with energy $\lesssim 80$ GeV,
\item no energy deposition in the HCAL 
\item no energy deposition in the HCAL EE
\end{itemize}
 
The occurrence of $Z_{\mu}$  produced in $\mu^- Z $ interactions would appear as an excess of 
events with a single low energy muon accompanied by zero-energy deposition  in the detector. 
The backgrounds for the reaction (\ref{58}) have been analyzed in ref. \cite{NA64mu, NA64mu1}. The main backgrounds are due to $\mu$ 
low-energy tail, HCAL nonhermeticity, $\mu$ induced photonuclear reactions and $\mu$ trident events \cite{NA64mu, NA64mu1}. 
These backgrounds were estimated in ref.\cite{NA64mu, NA64mu1} and they are rather small $\lesssim 10^{-12}$. 

The expected sensitivity of this experiment for $\alpha_{\mu}$ for different $Z_{\mu}$ masses and for 
$10^{12}$ muons on target  is shown in Fig. \ref{Fig12}. Note that in refs.\cite{posp,krnj,chen} the possibility 
to use muon beam for the search for light scalar particles has been discussed.

\begin{figure}[tbh!]
\begin{center}
\includegraphics[width =0.5\textwidth]{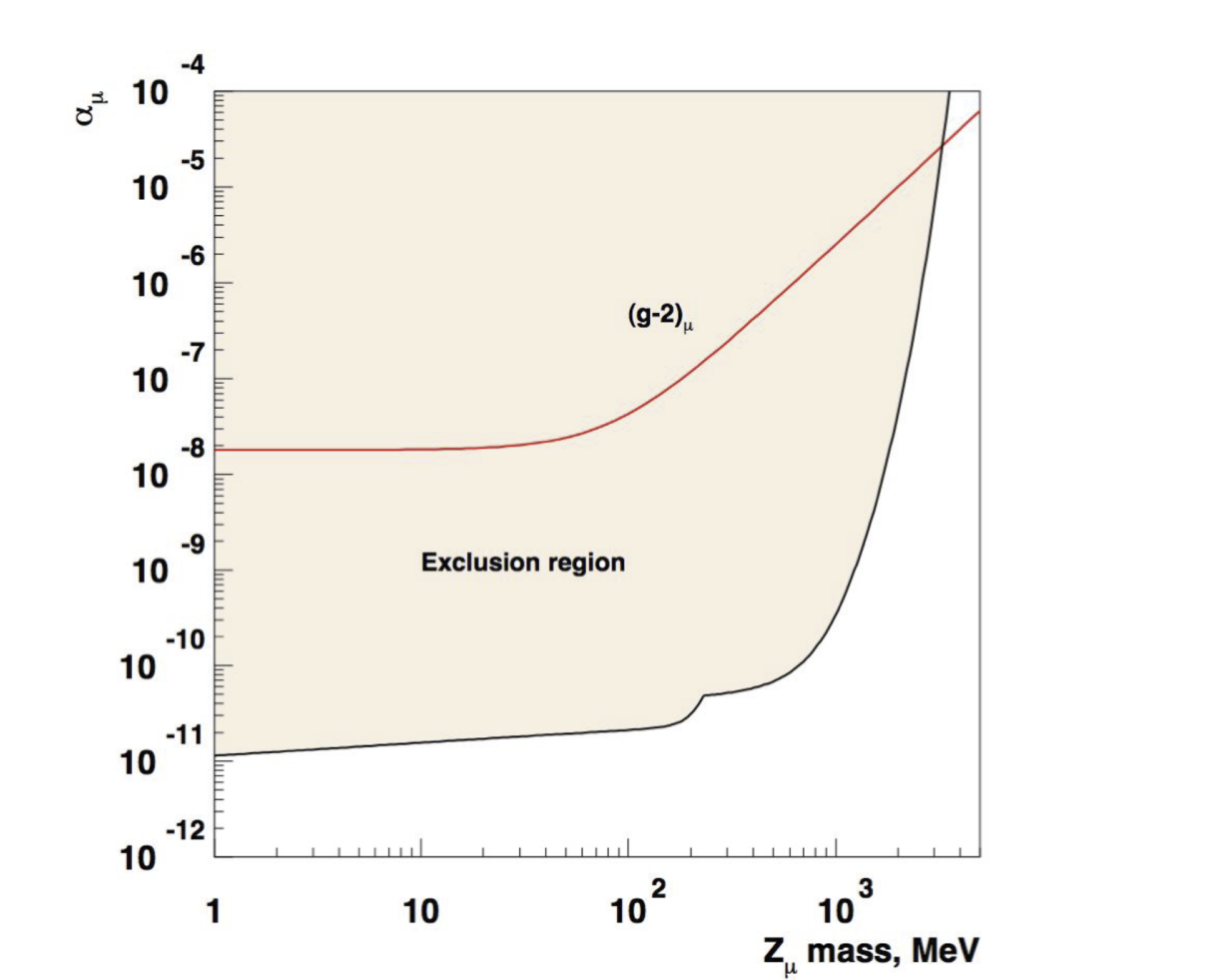}
\caption{
Expected constraints on the $\alpha_\mu$ coupling constant as a function of the $Z_{\mu}$ mass
 for $10^{12}$ $\mu$ at energy $E_{\mu}=150$ GeV \cite{NA64mu, NA64mu1}.
   }
\label{Fig12}
\end{center}
\end{figure}


 In the $A'$ dark photon model muons and electrons interact with the dark photon with the same coupling constant.
Hence, similar to the reaction of Eq.(\ref{55}), the dark photons will be also produced in the
  reaction of  
 Eq.(\ref{58}) with the same experimental signature of the missing energy.  
\begin{figure}[tbh!!]
\begin{center}
\includegraphics[width=0.5\textwidth,height=0.5\textwidth]{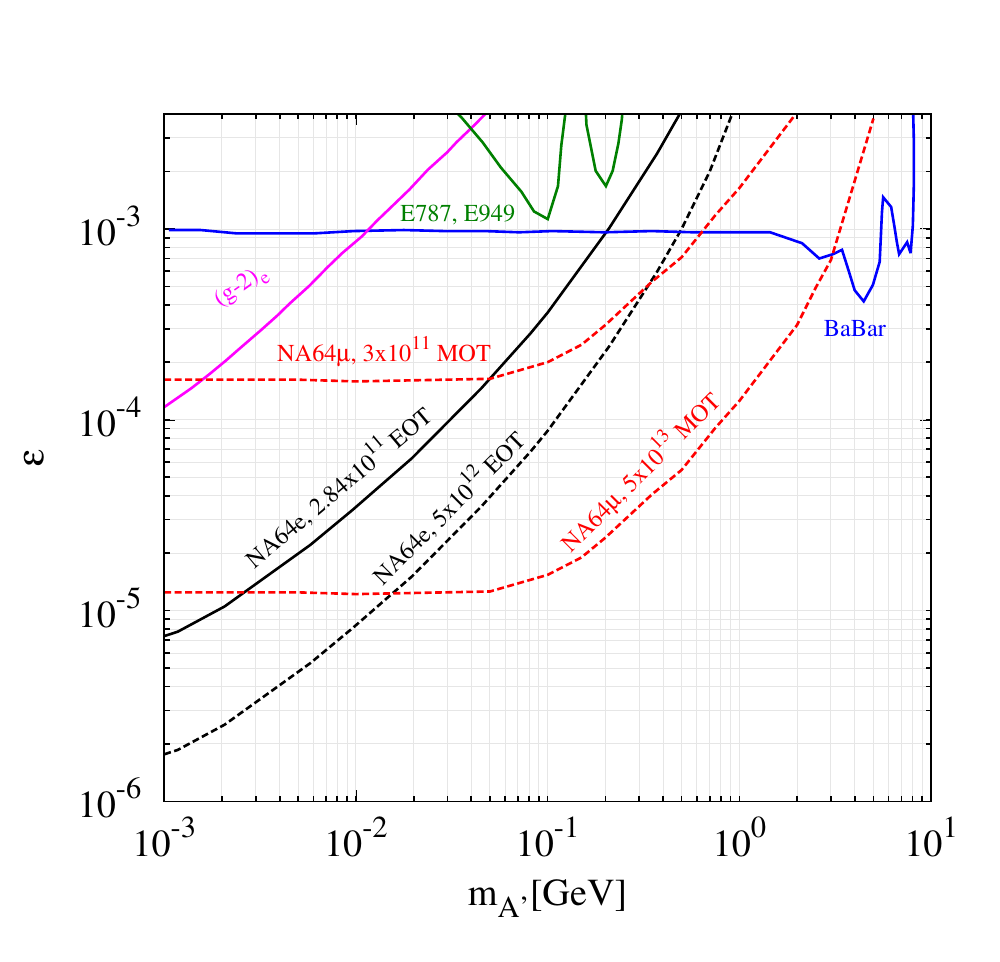}
\caption {The NA64e 90\% C.L. current \cite{NA64explast3} and expected exclusion bounds  obtained with 
$2.84\times10^{11}$ EOT and  $5\cdot 10^{12}$ EOT, respectively,  in the ($m_{A'}, \epsilon$) 
plane.  The NA64$\mu$ projected  bounds calculated  for  $n_{MOT} = 5 \cdot 10^{12}$  and 
  $5\cdot 10^{13}$  are also shown.
 \label{fig:excl-muon}} 
\end{center}
\end{figure} 
 For the $A'$ mass region  $m_{A'} \gg m_e$,  the total cross-section of the dark photon electroproduction $e Z \rightarrow e ZA'$  scales as    
$
\sigma^{e}_{A'}\sim \epsilon_{e}^2 / m_{A'}^2 
$. 
On the other hand,  for the dark photon masses, $m_{A'} \lesssim m_\mu$, the similar $\mu Z \rightarrow \mu ZA'$ cross-section can be approximated  in the bremsstrahlung-like limit as
$
\sigma^{\mu}_{A'}\sim \epsilon_{\mu}^2 /  m_\mu^2
$.
Let us now  compare expected sensitivities of the $A'$ searches with NA64e and NA64$\mu$  experiments for the
same number $\simeq 5\times 10^{12}$   particles on target.  Assuming the same signal efficiency 
 the number of $A'$ produced by the 100 GeV electron and muon beam 
 can  approximated, respectively,   as follows
\begin{equation}
N^{e}_{A'} \approx   \frac{\rho N_{av}}{A} \cdot n_{EOT} L^{e} \sigma_{A'}^{e}, \qquad 
N^{\mu}_{A'} \approx   \frac{\rho N_{av}}{A} \cdot n_{MOT} L^{\mu} \sigma_{A'}^{\mu}, 
\label{59}
\end{equation}
where  $L^{e} \simeq X_0$  and $L^{\mu}\simeq 40 X_0$ are  the 
typical distances that are  passed by an electron  and muon, respectively,   before producing
the $A'$ with the energy $E_{A'} \gtrsim 50$ GeV   in the NA64 active Pb target of the total thickness of $\simeq 40$ radiation length ($X_0$) 
\cite{NA64mu}.  
The detailed comparison of the calculated $A'$ sensitivities of NA64e and NA64$\mu$ 
 is shown  in Fig.\ref{fig:excl-muon},  where the 90\% C.L. limits on the mixing $\epsilon$ are shown for 
a different number of particles on target for both the NA64e and NA64$\mu$ experiments. 
The limits were obtained for the background free case by using exact-tree-level 
(ETL) cross-sections rather than the improved Weizsacker-Williams (IWW) ones calculated  for NA64e  in ref.\cite{Exactkirpich}, 
 and for  the NA64$\mu$ case in this work. The later are shown in  Fig.~\ref{fig:etlCSmuon}  as a function of 
$E_{A'}/E_\mu$ for the Pb target and mixing value $\epsilon=1$.  One can see that in a wide 
range of masses, $20 \mbox{ MeV} \lesssim m_{A'} \lesssim 1 \mbox{ GeV}$, the  
total IWW cross-sections  are larger by a factor $\simeq 2$ compared to the ETL ones. 
 As the result, the typical  limits on $\epsilon$ 
 for the ETL case are worse by about  a factor $\simeq 1.4$  compared to the IWW case. 
 \begin{figure}[tbh!!]
\begin{center}
\includegraphics[width=0.75\textwidth]{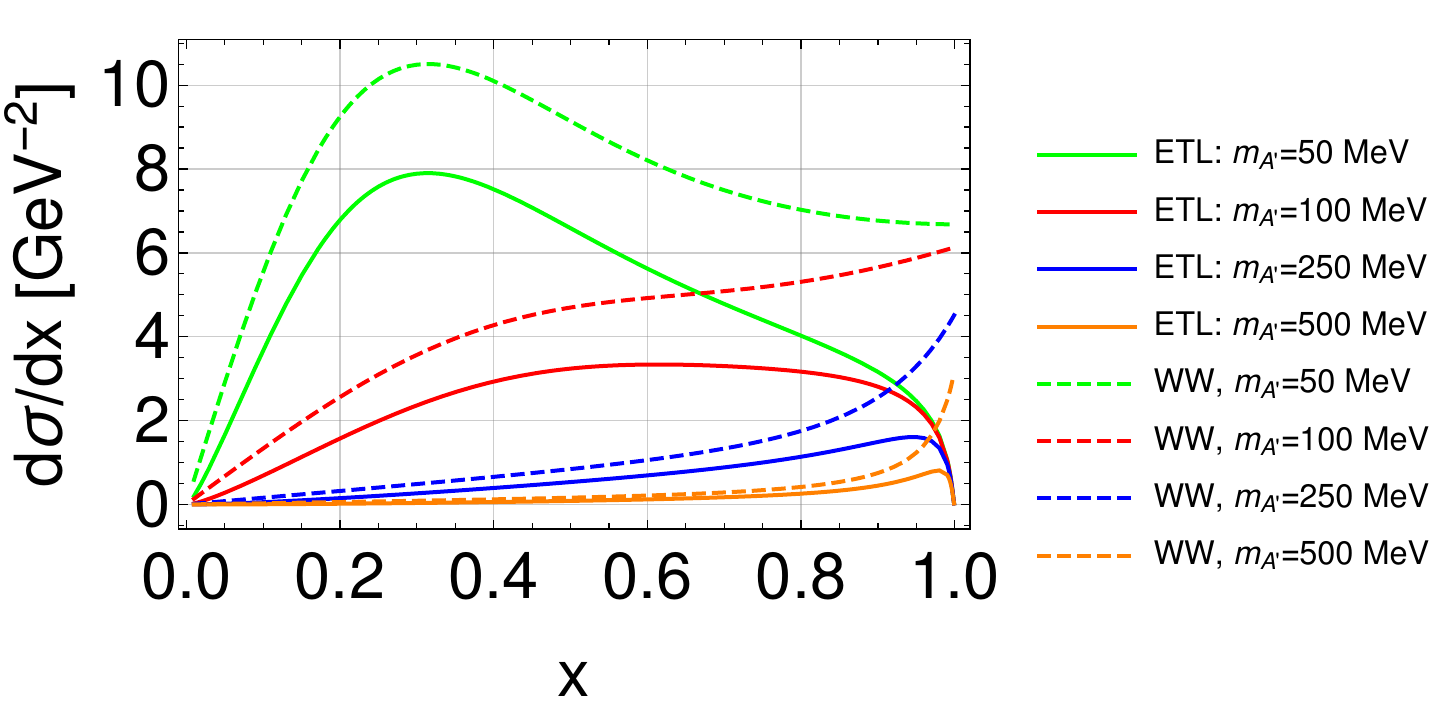}
\caption {Cross-section of dark photon 
production by muons as a function of $x=E_{A'}/E_\mu$ for various masses $m_{A'}$ and $\epsilon =1$. Solid lines represent ETL cross-sections and
dashed lines show the cross-sections calculated in IWW approach.
\label{fig:etlCSmuon}} 
\end{center}
\end{figure}  
For  $n_{EOT} = n_{MOT}=5 \cdot 10^{12}$ the sensitivity of NA64e 
  is enhanced for  the mass range $m_e \ll  m_{A'} \simeq 100$ MeV 
while for the $A'$ masses  $ m_{A'} \gtrsim 100$ MeV NA64$\mu$  allows 
to obtain  more stringent limits  on $\epsilon$  
in comparison with   NA64e. 

\subsection{Combined LDM sensitivity of NA64e and NA64$\mu$ \cite{GKKK}}
The                 estimated  NA64e and NA64$\mu$  limits on the $\gamma-A'$ mixing strength, allow us to set the 
combined NA64e and NA64$\mu$ constraints  on the LDM models, which are shown 
in the $(y;~m_\chi$)  plane in Fig.\ref{fig:comb-limit}.
\begin{figure}[tbh!!]
\begin{center}
\hspace{-0.5cm}{\includegraphics[width=.45\textwidth,height=0.4\textwidth]{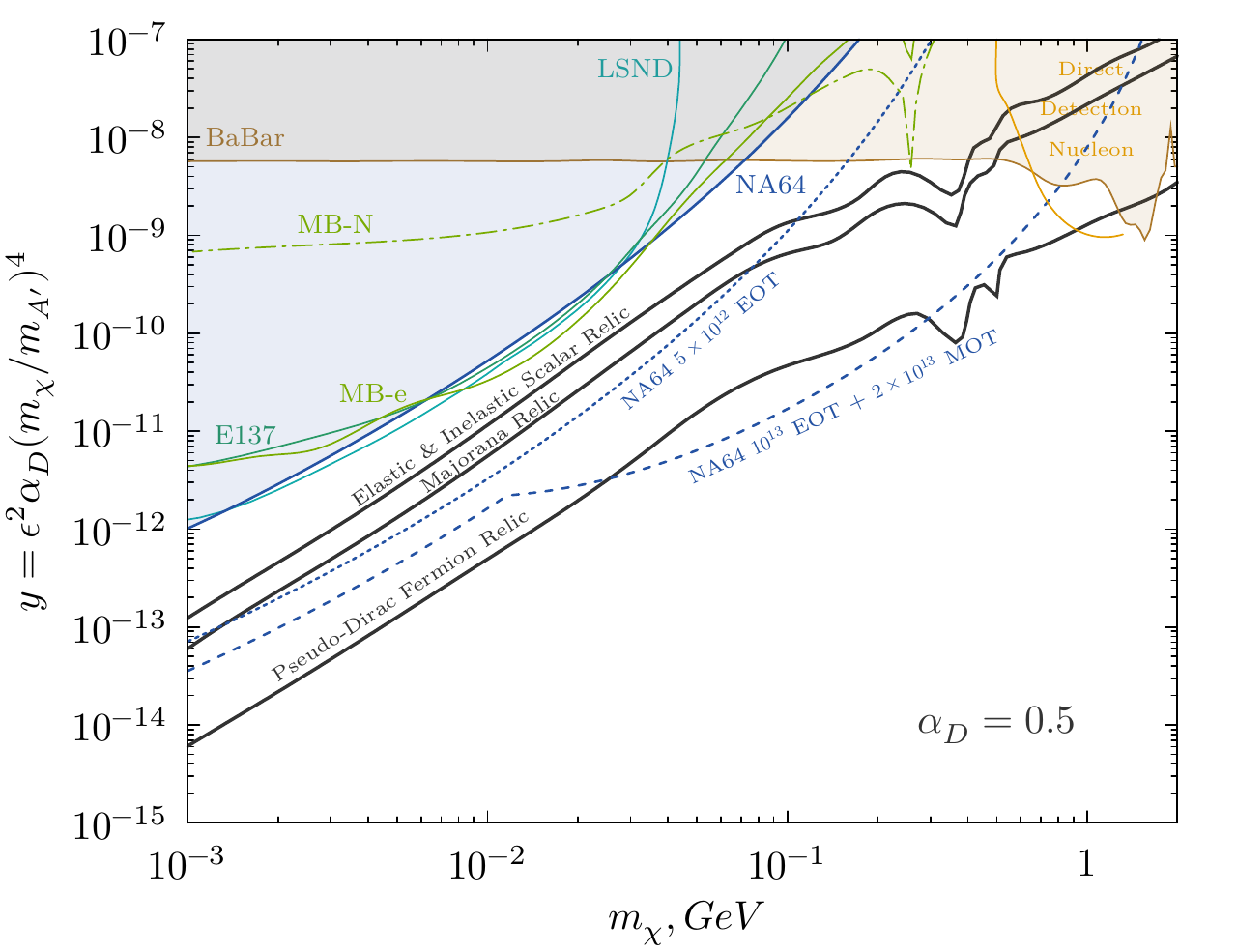}
\includegraphics[width=.45\textwidth,height=0.4\textwidth]{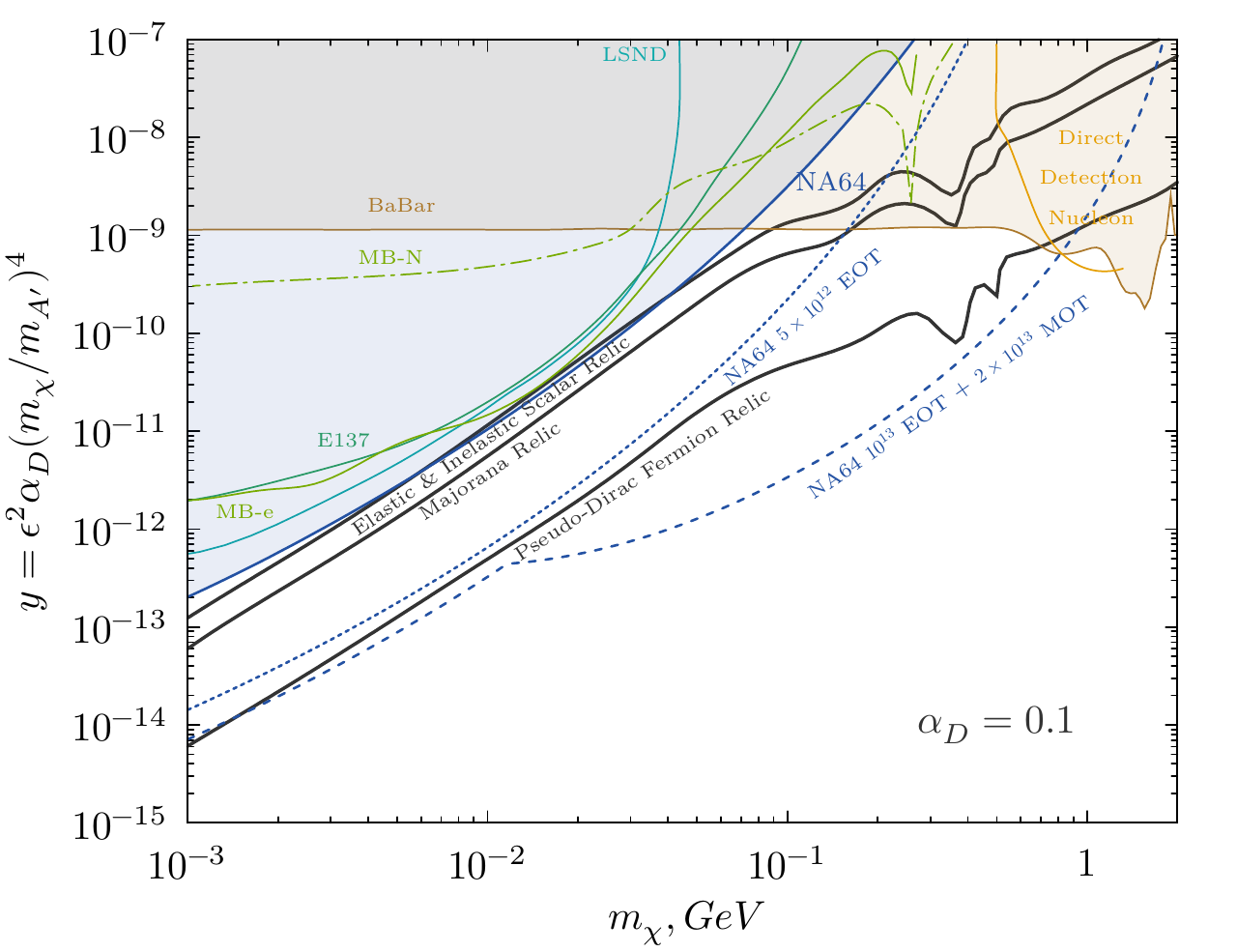}}
\includegraphics[width=.48\textwidth,height=0.4\textwidth]{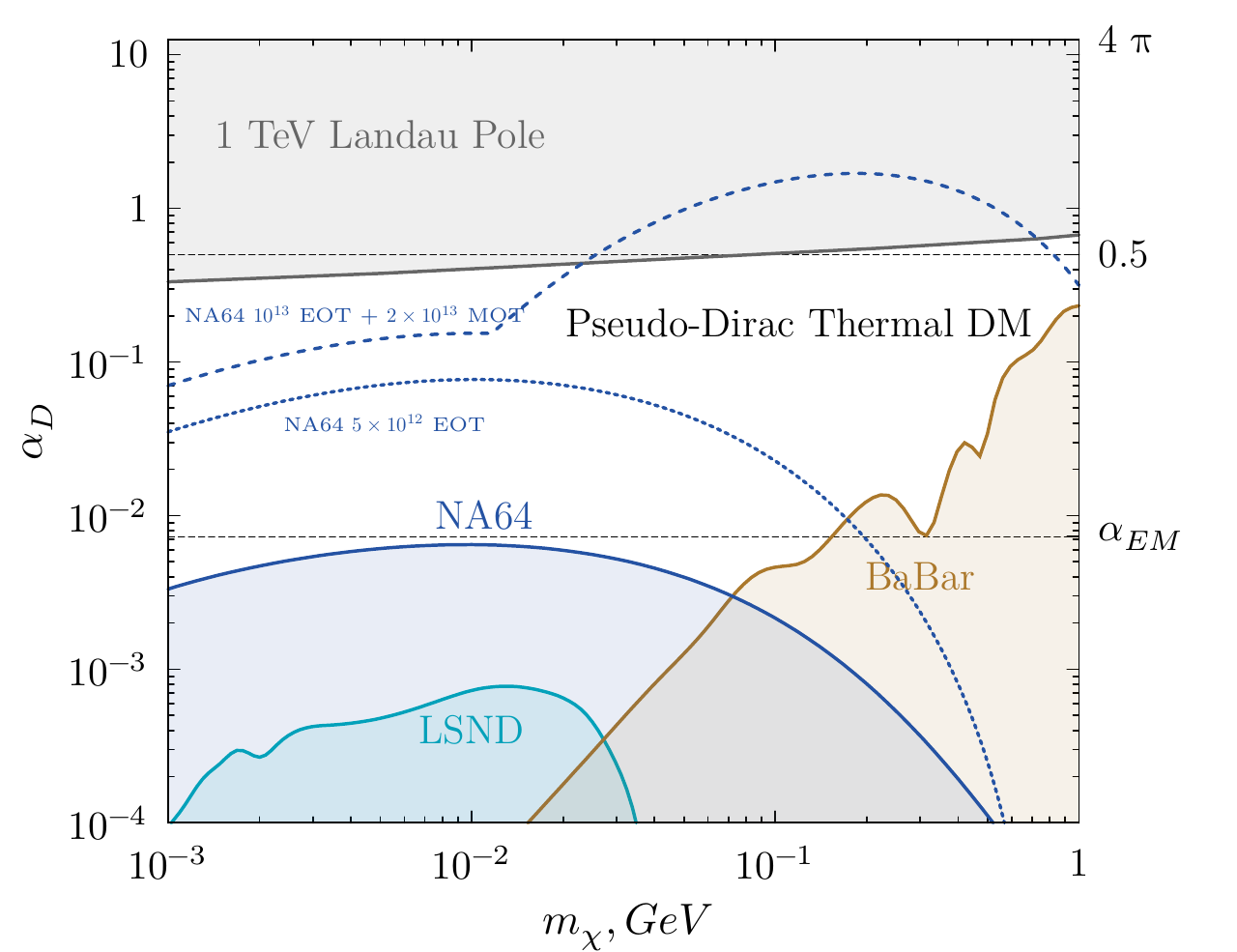}
\includegraphics[width=.48\textwidth,height=0.4\textwidth]{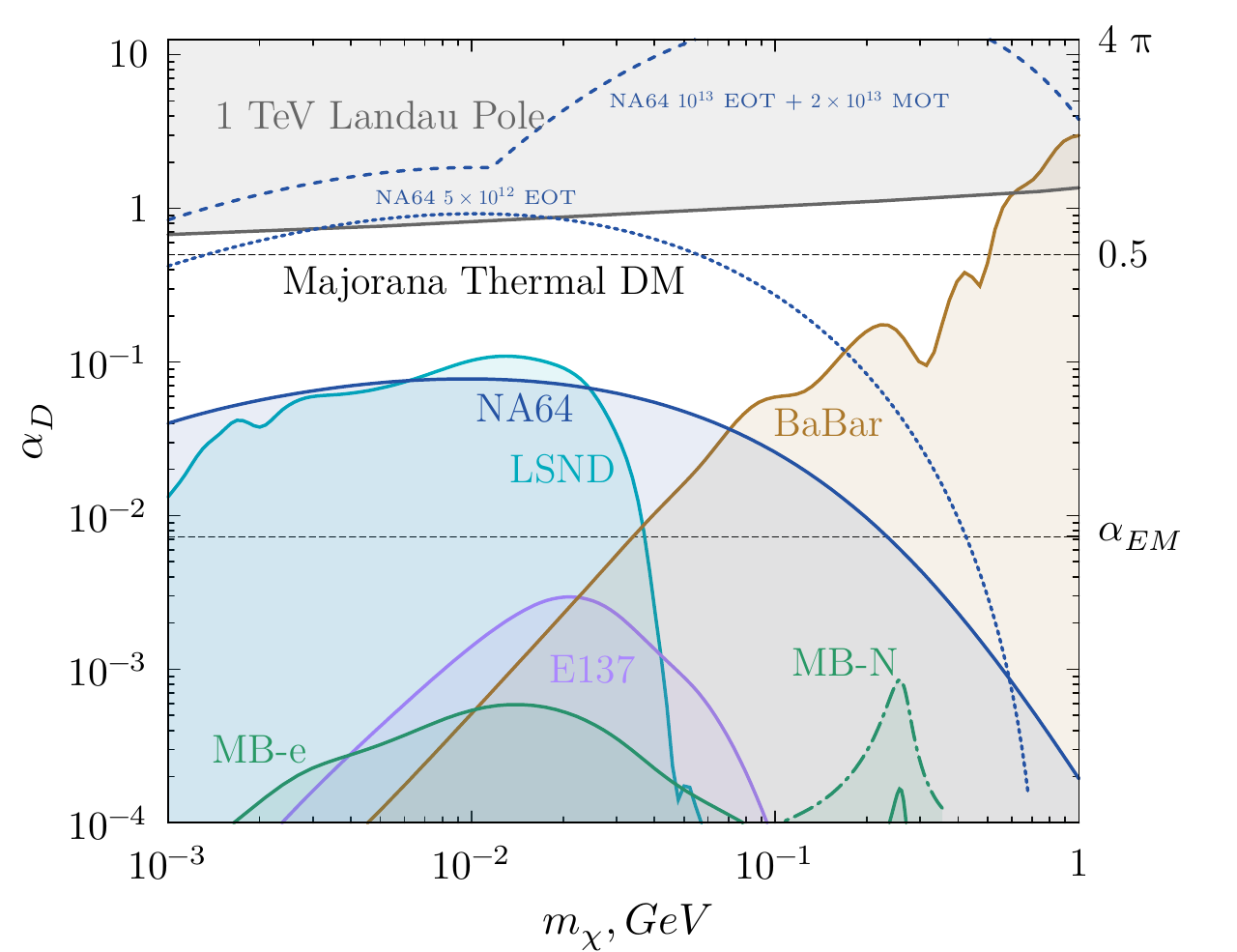}
\caption {The  NA64 90\% C.L. current (solid) \cite{NA64explast3} and  expected (dotted light blue)  exclusion bounds  for 
 $ 5\times10^{12}$ EOT  in the ($m_{\chi}, y$) and ($m_{\chi}, \alpha_D$)  planes. The combined limits from NA64e and NA64$\mu$ are also shown 
 for  $10^{13}$ EOT plus $2\times 10^{13}$ MOT (dashed  blue). The limits are  calculated for $\alpha_D=0.1$ and  0.5,  and   $m_{A'}=3m_{\chi}$.
The results are also  shown in comparison with bounds obtained from  the results of the 
LSND~\cite{lsnd},   E137 \cite{E137}, BaBar \cite{BABAR1}  and MiniBooNE \cite{MINIboon} experiments.
  \label{fig:comb-limit}} 
\end{center}
\end{figure} 
 As discussed in Appendix {\bf A}, as a result of  
the $\gamma - A'$ mixing  the cross-section of the DM particles annihilation into the SM particles is proportional to $\epsilon^2$. Hence using  
constraints on  the DM annihilation  cross-section 
one  can derive constraints 
 in the ($y \equiv \epsilon^2 \alpha_D (m_\chi/m_{A'})^4; ~m_{\chi}$) plane and restrict the LDM models with the masses  $m_{\chi} \lesssim 1$~GeV. 
\par The combined limits\cite{NA64explast3} obtained  from the 
data sample of the 2016, 2017 and  2018 runs and expected from the run after the LS2   are shown in  the top panels of 
Fig.~\ref{fig:comb-limit} together with combined limits from NA64e and NA64$\mu$ for  
 $10^{13}$ EOT and $2\times 10^{13}$ MOT, respectively.
The plots show also the comparison of our results with the limits of other experiments.  
It  should be 
 noted  that   the $\chi$-yield in the NA64 case scales as $\epsilon^2$ rather than  $\epsilon^4 \alpha_D$ as in beam dump experiments. Therefore, 
for sufficiently small values of $\alpha_D$  the NA64 limits  will be much stronger. This is 
illustrated in the upper right panel of  Fig.~\ref{fig:comb-limit},  where the NA64 limits are shown for  
 $\alpha_D = 0.1$. One can see  that  for this or smaller values of   $\alpha_D$  the direct search for LDM  at NA64e with $5\times10^{12}$ EOT
  excludes  the scalar and Majorana  models of  the LDM production via vector mediator with $\frac{m_{A^{'}}}{m_{\chi}} = 3$
   for the full    mass region  up to $m_\chi \lesssim 0.2$ GeV.  
While being combined with the NA64$\mu$  limit, the NA64 will  exclude the models with $\alpha_D \leq 0.1$ 
for the  entire mass region up to $m_\chi \lesssim 1$ GeV. 
 So we see that for the full mass range $m_{\chi} \lesssim 1$ GeV 
 the obtained combined NA64e and NA64$\mu$ bounds are more stringent than the limits obtained from the results of  NA64e 
 that allows probing the full sub-GeV DM parameter space.

\section{Other future experiments}

There are a lot of planned experiments   devoted to the search for 
both visible and invisible $A'$ decay modes. Here we briefly describe 
the most interesting future experiments.

\subsection{SHiP at CERN}

The proposed experiment SHiP \cite{SHiP1} at CERN is intended  to look for visible decays 
$A' \rightarrow   e^+e^-, \mu^+ \mu^- , \pi^+\pi^-$ of long lived 
$A'$ boson.
Also SHiP  can search for LDM by detection of 
the LDM scattering in neutrino detector at the 400 GeV SPS beam line at CERN. The 
detector consists of OPERA-like bricks of lead and emulsions placed in magnetic field. 
The LDM detection occurs via electon LDM elastic  scattering. The dominant backgrounds are 
expected related with neutrino scattering processes and can be reduced using several 
cuts. For $N = 10^{20}~POT$\footnote{POT $\equiv$ protons on target} the sensitivity is
$y \equiv \epsilon^2 \alpha_D(\frac{m_{\chi}}{m_{A'}})^4  \geq 10^{-12}$ for $m_{\chi} \leq O(1)~GeV$ 
\cite{lightdark1}. 

\subsection{Belle-II at KEK}

Belle-II \cite{Belle-2} is a multi-purpose detector with sensitivity to invisible $A'$ decays via mono-photon in the range 
$M_{A'} \leq 9.5~GeV$ can look for $A'$ invisible decays using the reaction 
$e^+e^- \rightarrow \gamma (A' \rightarrow invisible) $. Belle-II also can search for 
visible $A'$ decays. First data  with full luminosity 
$L_t = 50 ~ab^{-1}$ are expected   in 2025. The future sensitivity is $\epsilon^2 \geq 10^{-9}$ for 
$m_{A'} < 9.5~GeV$. 

\subsection{MAGIX at MESA}
 Visible dark photon decay searches with dipole spectrometer MAGIX at the $105~MeV$ polarized 
electron beam are planned at MESA accelerator complex \cite{MESA2}. The electroproduction  reaction $eZ \rightarrow eZA'$ 
and 
visible decay mode $A' \rightarrow e^+e^-$  will be  used     to identify the $A'$ as di-electron 
resonance. The expected sensitivity to the  $\epsilon^2$ parameter is up to $10^{-9}$ for $10~MeV < m_{A'} < 60~MeV$. 

\subsection{PADME at LNF}
The reaction $e^+e^- \rightarrow \gamma (A' \rightarrow invisible)$ is used for the search for  dark photon. 
For $10^{13}$ positron on target the expected sensitivity is $\epsilon^2 \geq 10^{-7} $
for $m_{A'} < 24~MeV$ \cite{PADME1}. The collection of data started at the end of 2018.

\subsection{VEPP3 at BINP} 
The proposed experiment at BINP\cite{VEPP3} is similar to PADME experiment . The expected sensitivity is planned to be 
$\epsilon^2 \geq 10^{-8}$ in the range $5 < m_{A'} < 22~MeV$. 



\subsection{BDX at JLab}

BDX  ar JLab is an electron beam -dump experiment \cite{BDX1, BDX2}.  The experiment is sensitive to elastic DM scattering $e\chi \rightarrow e\chi$ 
in the far  detector after  electron nuclei production in $eZ \rightarrow eZ(A' \rightarrow \chi \bar{\chi})$. 
The expected sensitivity is $y \geq 10^{-13}$ for $1~MeV < m_{\chi} < 100~MeV$.

\subsection{DarkLight at JLab}

In this experiment dark photons are produced in the reaction $ep \rightarrow epA'$ colliding the $100~MeV$ electron beam on a gaseous hydrogen target 
\cite{DarkLight, BDX2}.  The main peculiarity of this experiment is the possibility to detect the scattered electron  and recoil proton, enabling the 
reconstruction of invisible $A'$ decays. Also the search for visible $A' \rightarrow e^+e^-$ decays is possible. The expected 
sensitivity is $\epsilon^2 \geq 10^{-6}$ for $10~MeV < m_{A'} < 80~MeV$.

\subsection{LDMX}
This experiment is similar to NA64 experiment and will use the electroproduction reaction $eZ \rightarrow eZ(A' \rightarrow \chi \bar{\chi})$ 
for the dark photon search \cite{LDMX}. The LDMX(Light Dark Matter Experiment) will measure both missing energy and 
missing momentum that  is extremely important for background suppression. The expected sensitivity for the $\epsilon$ 
parameter is 
up to $10^{-6}$ for $m_{A'} = 1~MeV$ \cite{LDMX}. The extended LDMX will be able to increase sensitivity to 
the $\epsilon$ parameter by factor 
$~10$.

\section{Conclusion}
Active beam-dump searches for dark sector physics in missing energy events  have been proven by the NA64 experiment to be very powerful and sensitive via both 
invisible and visible decays of dark vector mediator.
 The future combined sensitivity of searches with both electron and muon beams has a great potential to probe a large region of the remaining 
 LDM parameter space, especially towards the higher LDM masses. 
Remarkably,  that with the statistics accumulated during years 2016-2018 NA64 already starts probing  the sub-GeV DM parameter space. 
While  with $ 5\times10^{12}$  EOT   NA64 with electron beam is able to test the scalar and Majorana  
LDM scenarios for $\frac{m_{A'}}{m_{\chi}} \geq 2.5$.  
The  combined NA64 results  with electron and muon beams and with 
$\gtrsim 10^{13}$ EOT,   $2\times 10^{13}$ MOT, respectively,  will  allow to fully explore   the parameter space of 
other  interesting LDM models like pseudo-Dirac DM model or the model with new light vector boson $Z_{\mu}$.
 This makes NA64e and NA64$\mu$  extremely complementary to each other, as well as 
 to the planned LDMX experiment \cite{LDMX}, and greatly increases the  NA64  discovery potential of  sub-GeV DM.  
 
 There are several alternatives \cite{Rev2018} to  the dark photon model based 
on the use of gauge symmetries like $U(1)_{B-L}$  or  $U(1)_{B - 3e}$. As in the dark photon model 
the observed value of the LDM density allows to estimate the 
coupling constant $\epsilon$ of new light $Z'$ boson with electron. The value of the $\epsilon$ parameter for such models 
coincides with the $\epsilon$ value for dark photon model up to some factor $k \leq 3$ \cite{Rev2018}, 
so NA64e can also test such  models. For instance, 
for the model with $(B - L)$ vector interaction
 NA64e is able to exclude  scalar and Majorana LDM scenarios in full analogy with the case of dark photon model.    

However it should be stressed that for $m_{A'} \approx 2m_{\chi}$ the DM annihilation cross-section is proportional to
$(m^2_{A'} - 4 m^2_{\chi})^{-2}$. 
As a consequence the predicted value of the $\epsilon^2$ parameter is proportional to 
$(\frac{m^2_{A'}}{4m^2_{\chi}} -4)^2 $ 
that can reduce the  $\epsilon^2$ value by (2 - 4) orders of magnitude in comparison with the 
reference point  $\frac{m_{A'}}{m_{\chi}} =3$ \cite{Feng}. It means  that  NA64 experiment as other future experiments like LDMX \cite{LDMX} 
are  not able to test the region
 $m_{A'} \approx 2m_{\chi}$ completely\footnote{The values of $m_{A'}$ and $m_{\chi}$ are  arbitrary, so the case
$m_{A'} \approx  2m_{\chi}$   could be considered  as some fine-tuning. It is 
natural to assume  the absence of significant fine-tuning. 
In this paper we  require   that  $|\frac{m_{A'}}{2m_{\chi}} -1| \geq 0.25$.}.

Current  accelerator experimental data\footnote{The review of nonaccelerator bounds 
can be found in ref.\cite{Jeong}.}  restrict rather strongly the 
explanation of the $g_{\mu} - 2 $ muon anomaly due to existence of new light gauge boson 
but not completely eliminate it. The most popular  model where dark photon $ A' $ interacts with the SM electromagnetic current 
due to mixing $ \frac{\epsilon}{2} F_{\mu \nu}F^{`\mu\nu} $ term is excluded.    
The Borexino data on neutrino electron elastic scattering
exclude  the models where   $Z'$ interacts with both leptonic  and $B - L$ currents.
The interaction of the $Z^{`}$ boson with $L_{\mu} - L_{\tau}$ 
current  is  excluded    for   $m_{Z'} \geq 214~MeV$ while still leaving the region of lower masses 
unconstrained.  NA64$\mu$ is able to test the model with $L_{\mu} - L_{\tau}$ interaction 
at  $m_{Z'} \leq 214~MeV$ as a model explaining muon $g_{\mu}-2$ anomaly.

\section*{Acknowledgments}
We are indebted  to our colleagues from the NA64 Collaborations, in particular to P. Crivelli, D.V. Kirpichnikov, M.M. Kirsanov and V. Lyubovitsky, for many useful discussions and comments. We would also like to thank R. Dusaev for his help in designing and preparing  several figures.
\newpage

 {\bf Appendix A: DM density calculations}

 $$  $$

The observed homogeneity and isotropy of the Universe enable us to describe the overall geometry
and evolution of the Universe in terms of two cosmological parameters accounting for
the spatial curvature and the overall expansion (or contraction) of the Universe 
that is realized in the Freedman-Robertson-Walker metric\footnote{As a review, see for example \cite{Universe00, Universe0}.}
\begin{equation} 
ds^2 = dt^2 - R^2(t)[\frac{dr^2}{1 - kr^2} +r^2(d\theta^2 + \sin^2 \theta d\Phi^2)] \,.
\label{60}
\end{equation}
The curvature constant $k$ takes three values $k = 1, -1, 0$ 
that corresponds to closed, open and spationally flat geometries.   
The cosmological equations are derived from Einstein's equations
\begin{equation}
R_{\mu\nu} - \frac{1}{2}g_{\mu\nu} = 8 \pi G_N T_{\mu\nu} + \Lambda g_{\mu\nu} \,.
\label{61}
\end{equation}
We shall use  the standard assumption that an effective energy-momentum tensor
$T_{\mu\nu}$ is a perfect fluid, for which
\begin{equation}
T{\mu\nu} = -pg_{\mu\nu} +(p+\rho)u_{\mu}u_{\nu} \,,
\label{62}
\end{equation}
where $p$ is the pressure, $\rho$ is the energy-density and $u = (1,0,0,0)$ is the velocity 
vector for the isotropic fluid in co-moving coordinates. For the metric (\ref{60}) and 
the energy-momentum tensor (\ref{62}) the Einstein equations (\ref{61}) lead to 
Friedman-Lemaitre equations 
\begin{equation}
H^2  = \frac{8\pi G_N\rho}{3} - \frac{k}{R^2} + 
\frac{\Lambda}{3} \,,
\label{63}
\end{equation}
\begin{equation}
\frac{1}{a^2(t)}\cdot \frac{d^2a}{ dt^2} = \frac{\Lambda}{3} - \frac{4\pi G_N}{3}(\rho + 3p) \,,
\label{64}
\end{equation}
\begin{equation}
H(t)  \equiv  \frac{1}{R(t)}\frac{dR}{dt} \,,
\label{65}
\end{equation} 
where $H(t)$ is the Hubble parameter and $\Lambda$ is cosmological constant.
Energy conservation $T^{\mu\nu}_{;\nu} = 0$ leads to the equation
\begin{equation}
\frac{d\rho}{dt} = -3H(\rho + p) \,.
\label{66}
\end{equation}
The equation (\ref{66}) allows to determine today critical density $\rho_c$ that corresponds to 
flat Universe with $k =0$ and $\Lambda = 0$ in the equations (\ref{63},\ref{64}), namely
\begin{equation}
\rho_c = \frac{3 H^2}{8\pi G_N} = 1.05 \cdot 10^{-5} ~h^2 ~GeV ~cm^{-3}
\label{67}
\end{equation}
Here the parameter $h$ is defined by 
\begin{equation}
H \equiv 100 ~h ~km ~s^{-1} Mpc^{-1}
\label{68}
\end{equation}
and its experimental value is $h =  0.72 \pm 0.03$ \cite{particledata}.
The cosmological density parameter $\Omega_{tot}$ is defined as the energy density 
relative to the critical density 
\begin{equation}
\Omega_{tot} = \rho/\rho_c \,.
\label{69}
\end{equation}
One can rewrite the equation (\ref{63}) in the form
\begin{equation}
\frac{k}{R^2} = H^2(\Omega_{tot} - 1)
\label{70}
\end{equation}
As a consequence of the equation (\ref{70}) we see that for $ \Omega_{tot} > 1$ the Universe is closed, 
for $ \Omega_{tot} < 1$ the Universe is open and for  $ \Omega_{tot} = 1$ the Universe 
is spatially flat.
It is often necessary to distinguish different contributions 
to the density $\Omega_{tot}$. It is convenient to define present-day density parameters 
for pressureless matter $\Omega_m$ and relativistic particles $\Omega_r$ plus 
the vacuum dark energy density $\Omega_{V}    $ and the dark matter density $\Omega_{d}$.
Current data  give \cite{particledata}
\begin{equation}
\Omega_V = 0.73 \pm 0.01 \,,
\label{71}
\end{equation}
\begin{equation}
\Omega_{d}       = 0.23 \pm 0.01 \,.
\label{72}
\end{equation}

It is expected  that the early Universe can be described by a radiation-dominated 
equation of state. In addition it is assumed that through much of the radiation-dominated period, 
thermal equillibrium is established by the rapid rate of particle interactions relative to 
the expansion rate of the Universe. In equilibrium thermodynamic quantities like 
energy density, pressure and entropy are calculable quantities in the 
ideal gas approximation.  The density of states for particle $i$ is given by 
\begin{equation} 
dn_i = \frac{g_i d^3\vec{p}}{(2\pi)^3}(\exp[\frac{E_i - \mu_i}{T_i}] \pm 1)^{-1} \,.
\label{73}
\end{equation} 
Here $g_i$ counts the number of degrees of freedom of particle $i$, 
$E^2_i =  \vec{p}^2 + m^2_i$, $\pm$ corresponds to either Fermi or Bose statistics, 
$\mu_i$ is the chemical potential\footnote{For the Universe the effects of nonzero chemical potential are small 
so we shall use the approximation with zero chemical potentials $\mu_i = 0$} and $ T_i$ is the temperature. 
The  energy density, the pressure, the number density  and the entropy density 
are given by the formulae
\begin{equation}
\rho_i = \int E_i dn_i \,,
\label{74}
\end{equation}
\begin{equation}
p_i = \frac{1}{3}\int \frac{{\vec p}^2_i}{E_i} dn_i \,,
\label{75}
\end{equation}
\begin{equation}
n_i = \int dn_i \,,
\label{76}
\end{equation}
\begin{equation}
s_i = \frac{\rho_i + p_i - \mu_i n_i}{T_i} \,.
\label{77}
\end{equation}

For instance, for photons with $g_{\gamma} = 2$ polarization states 
the energy density, pressure, density of the number of photons and the entropy density 
are given by the formulae
\begin{equation}
\rho_{\gamma} = \frac{\pi^2}{15}T^4 \,,
\label{78}
\end{equation} 
\begin{equation}
p_{\gamma} = \frac{1}{3} \rho_{\gamma}  \,,
\label{79}
\end{equation} 
\begin{equation}
s_{\gamma} = \frac{4\rho_{\gamma}}{3T}  \,,
\label{80}
\end{equation} 
\begin{equation}
n_{\gamma}  = 0.243 T^3 \,.
\label{81}
\end{equation} 
The number density of nonrelativistic particles is given by the formula 
\begin{equation}
n_{nonrel} = g \frac{1}{ (2\pi)^{3/2}}(mT)^{3/2}exp(-\frac{m}{T})  \,,
\label{82}
\end{equation}
where $g$ is the number of polarizations.
As a consequence of the equations (\ref{63}), (\ref{64}) and the definition (\ref{77})  of the entropy density 
one can find that the total entropy is conserved, namely 
\begin{equation}
\frac{d(sR^3)}{dt} = 0\,.
\label{83}
\end{equation}

At the very high temperatures associated with the early Universe, massive 
particles are pair produced, and are part of the thermal bath. At 
high temperature $T \gg m_i$ we can neglect masses and approximate the energy density 
by including those particles with $m_i \ll T$, namely
\begin{equation}
\rho = (\sum_{B} g_B + \frac{7}{8}\sum_{F}g_F)\frac{\pi^2}{30}T^4 \equiv g_{\rho}T^4 \,,
\label{84}
\end{equation}
where $g_{B(F)}$ is the number of degrees of freedom of each boson (fermion) and the sum runs 
over all bosons and fermions with $m \ll T$. The factor $7/8$ is due to the difference between 
the Fermi and Bose integrals (\ref{73}-\ref{77}). The equation (\ref{84}) defines the effective number of degrees of 
freedom. For instance, for temperature $ m_e < T < m_{\mu}$ the effective number $g_{\rho} = 
43/4$.

To obtain   estimate of dark matter density  we have to solve the Boltzmann equation 
\begin{equation}
\frac{dn_{d}}{dt} + 3H(T)n_{d} = - <\sigma v_{rel}>(n^2_{d} - n^2_{d,eq})\,.
\label{85}
\end{equation}
Here 
\begin{equation}
n_{d}(T) = \int \frac{d^3p}{2\pi^3} f_{d}(p,T) \,
\label{86}
\end{equation}
and $f_{d}(p,T)$ is DM distribution function. 
The equilibrium nonrelativistic DM density is

\begin{equation}
n_{d,eq} = g_{d} \frac{1}{ (2\pi)^{3/2}}(m_{\chi} T)^{3/2}exp(-\frac{m_{\chi}}{T})  \,,
\label{87}
\end{equation}
where $m_{\chi}$ is the mass of DM particle.    The  $<\sigma v>$ is thermally pair averaged 
cross section \cite{Universe00, Go}
\begin{equation}
<\sigma v> = \frac{1}{8m^4_{\chi}TK_2(\frac{m_{\chi}}{T})^2}\int^{\infty}_{4m^2_{\chi}} ds 
\sigma(s)\sqrt{s}(s - 4m^2_{\chi})K_1(\frac{\sqrt{s}}{T}) \,.
\label{88}
\end{equation}
In nonrelativistic approximation $<\frac{m_{\chi}\vec{v}^2}{2}> =  \frac{3T}{2}$.

The DM relative density parameter $\Omega_{d}$ is represented in the form
\begin{equation}
\Omega_{d} = \frac{m_{\chi} s_oY_0}{\rho_c} \,,
\label{89}
\end{equation}
where $s_0 \equiv s(T_0)$ is today dark 
entropy density and  $ Y \equiv \frac{n_{d}}{s}$ is approximately constant for iso-entropic Universe
$(Y(t_{d}) \approx Y(t_0))$. 
The evolution equation for $Y(t)$ reads 
\begin{equation}
\frac{dY}{dt} = -s<\sigma_{rel}>(Y^2 - Y^2_{eq}) \,.
\label{90}
\end{equation}
The equation (\ref{90}) can be rewritten in the form
\begin{equation}
\frac{dY}{dx} =   \frac{1}{3H} \frac{ds}{dx}  <\sigma_{rel}>(Y^2 - Y^2_{eq}) \,.
\label{91}
\end{equation}
Here $x = \frac{m_{\chi}}{T}$ and $T$ is photon temperature.  
Note that for the flat Universe the Hubble parameter $H = (\frac{8}{3}\pi G\rho)^{1/2} $.
The effective degrees of freedom for the energy and entropy densities are defined by
\begin{equation}
\rho  = g_{eff}(T)\frac{\pi^2}{30}T^4 \,,
\label{92}
\end{equation}
\begin{equation}
s  = h_{eff}(T)\frac{2\pi^2}{45}T^3 \,
\label{93}
\end{equation}
respectively, in such a way that the $g_{eff}(T) = h_{eff}(T) = 1$ 
for a relativistic species with one internal or spin degree of freedom. 
Taking into account (\ref{93}) equation (\ref{91}) takes the form
\begin{equation}
\frac{dY}{dx} = -(\frac{45}{\pi} G)^{-1/2} \frac{g^{1/2}_*m_{\chi}}{x^2}  <\sigma v_{rel}>(Y^2 - Y^2_{eq}) \,,
\label{94}
\end{equation}
where
\begin{equation}
g^{1/2}_* = \frac{h_{eff}}{g^{1/2}_{eff}}(1 + \frac{1}{3}\frac{T}{h_{eff}}\frac{dh_{eff}}{dT}) \,.
\label{95}
\end{equation}
    The equilibrium density $Y_{eq}$ is given by 
\begin{equation}
Y_{eq} = \frac{45g}{4\pi^2}\frac{x^2K_2(x)}{h_{eff}(\frac{m_{\chi}}{x})} \,.
\label{96}
\end{equation}
The  solution of the equation (\ref{94}) allows to determine the freeze-out temperature $T_d$. 
The decoupling temperature $T_d$ is usually defined by the equation 
\begin{equation}
\Delta \equiv  Y - Y_{eq} = \delta Y_{eq} \,.
\label{97}
\end{equation}
In the approximation $\frac{d\Delta}{dx} =0$
the  equation
\begin{equation}
(\frac{45}{\pi} G)^{-1/2} \frac{g^{1/2}_*m_{\chi}}{x^2}  <\sigma v_{rel}>Y_{eq}\delta(\delta + 2) = 
-\frac{d\ln Y_{eq}}{dx} \,
\label{98}
\end{equation}
allows to determine the decoupling temperature $T_{d}$. 
 The parameter $\delta$ is usually taken to be $\delta = 1.5$.
After the decoupling we can neglect $Y_{eq}$ in the equation (\ref{85}) and the integration 
from $T_d$ to $T_0$ gives \cite{Universe00, Go}
\begin{equation}
\frac{1}{Y_0} = \frac{1}{Y_d} + (\sqrt{\frac{\pi}{45}}M_{PL})^{-1} \cdot [ \int^{T_{d}}_{T_0}( g^{1/2}_*(<\sigma v >)dT ] \,.
\label{99}
\end{equation} 
Numerically $Y_d \gg Y_o$ and we can neglect it, so we obtain  \cite{Universe00, Go}
\begin{equation}
Y_o \approx \sqrt{\frac{\pi}{45}}M_{PL}[ \int^{T_{dec}}_{T_0}( g^{1/2}_*(<\sigma v >)dT ]^{-1} \,.
\label{100}
\end{equation}

The DM relic density can be numerically estimated as
\begin{equation}
\Omega_{d}h^2 =8.76 \times 10^{-11} GeV^{-2}[ \int^{T_{d}}_{T_0}( g^{1/2}_*<\sigma v >)\frac{dT}{m_{\chi}} ]^{-1} \,.
\label{101}
\end{equation}
In nonrelativistic approximation with  $<\sigma v_{rel}> = \sigma_o x^{-n}_f$
one can find that the previous formula takes the form \cite{Universe00, Go}
\footnote{Here $n =0$ corresponds to $s$-wave annihilation and $n=1$ 
corresponds to $p$-wave annihilation}  
\begin{equation}
\Omega_{DM}h^2 = 0.1\Bigl(\frac{(n+1)x_f^{n+1}}{(g_{*s}/g^{1/2}_*)}\Bigr)\frac{0.876\cdot 10^{-9}GeV^{-2}}{\sigma_0} \,,
\label{102}
\end{equation}
where $x_f = \frac{m_{\chi}}{T_d}$.
The following approximate formula \cite{Universe00} takes place for $x_f$: 
\begin{equation}
x_f = c - (n + \frac{1}{2}){\rm ln}(c) \,,
\label{103}
\end{equation}
\begin{equation}
c = {\rm ln}(0.038(n+1)\frac{g}{\sqrt{g_*}}M_{Pl}m_{\chi}\sigma_0) \,.
\label{104}
\end{equation}
Here $g_*$, $g_{*s}$ are 
the effective relativistic energy and entropy degrees of freedom and g is an internal number of freedom degree.  
If DM particles differ from DM antiparticles $\sigma_o = \frac{\sigma_{an}}{2}$.

For s-wave annihilation cross-section with $n = 0$
\begin{equation}
<\sigma v_{rel}> =  7.3 \cdot 10^{-10}GeV^{-2}\cdot \frac{1}{g^{1/2}_{*,av}}(\frac{m_{\chi}}{T_d}) \,.
\label{105}
\end{equation}
Here $g^{1/2}_{*,av} = \frac{1}{T_d}\int^{T_d}_{T_o} (g_{*s}/g^{1/2}_*) dT$.  
The calculations show that  $1 \leq c_s \equiv \frac{m_{\chi}}{10T_d}   \leq  1.5$ 
at $1~MeV \leq m_{\chi} \leq 100~MeV$. So we find that 
\begin{equation}
<\sigma v_{rel} > =  7.3 \cdot 10^{-9}GeV^{-2}\cdot \frac{1}{g^{1/2}_{*,av}} c_S \,.
\label{106}
\end{equation}
For the Dirac fermion DM $\chi$ with dark photon as a messenger between DM and SM sectors the 
nonrelativistic annihilation
cross-section into electron positron pair is\footnote{Here we assume that $m_{\chi} \gg m_e$ }
\begin{equation}
\sigma_{an}(\chi\bar{\chi} \rightarrow e^-e^+) v_{rel} = \frac{16\pi \epsilon^2 \alpha_D m^2_{\chi}}{(m^2_{A'}- 4 m^2_{\chi})^2} \,.
\label{107}
\end{equation}

\begin{equation}
\epsilon^2\alpha_D = 2.0\cdot 10^{-8}GeV^{-2}\cdot \frac{(m^2_{A'} - 4 m^2_{\chi})^2}{m^2_{\chi}}  \cdot \frac{2c_s}{g^{1/2}_{*,av}} \,.
\label{108}
\end{equation}
For  $m_{A'} = 3 m_{\chi}$ we find
\begin{equation}
\epsilon^2\alpha_{D} =  0.5\cdot 10^{-12} \cdot (\frac{m_{\chi}}{MeV})^2 \cdot \frac{2c_s}{g^{1/2}_{*,av}} \,.
\label{109}
\end{equation}
At $20~MeV \leq m_{\chi} \leq 200~MeV$ 
 and  $   T \leq 100~ MeV$ the effective value $g^{1/2}_{*,av} \approx  3.3$, so  we find that 
\begin{equation}
\epsilon^2\alpha_D \sim 0.4 \cdot 10^{-12} \cdot (\frac{m_{\chi}}{MeV})^2  \,.
\label{110}
\end{equation}
Note that for pseudo-Dirac DM the predicted value for $\epsilon^2\alpha_D$ is bigger than the corresponding value 
for fermion DM.
For the p-wave cross-section in nonrelativistic approximation 
$<\sigma v_{rel}> = <B v^2_{rel}> =  6 B \cdot (\frac{T}{m}_{\chi})$.
An analog of the formula (\ref{105}) is
\begin{equation}
    6 B =  14.6 \cdot 10^{-10}GeV^{-2}\cdot \frac{1}{g^{1/2}_{*,av}}(\frac{m_{\chi}}{T_d})^2 \,.
\label{111}
\end{equation}
Here $g^{1/2}_{*,av} = \frac{2}{T^2_d}\int^{T_d}_{T_o} T (g_{*s}/g^{1/2}_*) dT$.  
For the p-wave annihilations the  estimates are  similar to the Dirac fermion case,
namely for $1~MeV \leq m_{\chi} \leq 200~MeV$ we find that $\frac{m_{\chi}}{T_d} = 10 \cdot c_p$ with $1 \leq c_p \leq 2$. 

For the charged scalar DM   the  nonrelativistic annihilation cross-section into electron-positron pair is 
\begin{equation}
\sigma v_{rel} = \frac{8\pi}{3} \frac{\epsilon^2\alpha\alpha_D m^2_{\chi}v^2_{rel}}
{(m^2_{A'} - 4 m^2_{\chi})^2 } \,,
\label{112}
\end{equation} 
An analog of the formula (\ref{108}) is
\begin{equation}
\epsilon^2\alpha_{D} = 4.0\cdot 10^{-7}GeV^{-2} \frac{(m^2_{A'} - 4 m^2_{\chi})^2}{m^2_{\chi}}  \frac{2c^2_p}{g^{1/2}_{*,av}} \,.
\label{113}
\end{equation}
For $m_{A'} = 3 m_{\chi}$ we find
\begin{equation}
\epsilon^2\alpha_{D} =  10^{-11} \cdot (\frac{m_{\chi}}{MeV})^2  \frac{2c_p}{g^{1/2}_{*,av}} \,.
\label{114}
\end{equation}


As a reasonable estimate we take 
\begin{equation}
\epsilon^2\alpha_D \sim  10^{-11} \cdot (\frac{m_{\chi}}{MeV})^2  \,.
\label{115}
\end{equation}
For Majorana fermions the typical estimate for $\epsilon^2\alpha_D$ 
has additional factor $\approx 2$.  

\newpage
 {\bf Appendix B:    Detection of long lived particles at NA64}

$$  $$


In pseudo-Dirac scenario \cite{lightdark1} the  Majorana particles $\chi_1$ and $\chi_{2}$ 
are produced   in the reactions
\begin{equation}
eZ \rightarrow eZ A'  \,,
\label{116}
\end{equation}
\begin{equation}
A' \rightarrow \chi_1 \chi_2 \,.
\label{117}
\end{equation}
Here we assume that $m_{\chi_2} > m_{\chi_1}$.  In pseudo-Dirac model the decay 
\begin{equation}
\chi_2 \rightarrow \chi_1 e^+e^- \,
\label{118}
\end{equation} 
allows to avoid GMB restrictions \cite{Planck} on the $s$-wave DM annihilation cross-section.
The decay width $\chi_2 \rightarrow \chi_1 e^+e^-$  is given by the formula \cite{Mohlabeng}
\begin{equation}
\Gamma(\chi_2 \rightarrow \chi_1 e^+e^-) \approx \frac{4\epsilon^2\alpha\alpha_D\Delta^5}{15\pi m^4_{A'}} \,,
\label{119}
\end{equation} 
where $\Delta = m_{\chi_2} - m_{\chi_1}$.
For the  case of dominant  $A' \rightarrow e^+e^-$ decay 
the dark photon decay length is given by the formula \cite{Bjorken}
\begin{equation}
l_{A'} = \gamma_{A'} c \tau \approx 0.8 ~mm~ \cdot (\frac{\gamma_{A^`}}{10})\cdot (\frac{10^{-4}}{\epsilon})^2 \cdot 
\frac{100~MeV}{m_{A'}} \,,
\label{120}
\end{equation}
where $\gamma_{A'} = \frac{E_{A'}}{m_{A'}}$.
The analogous formula for $\chi_2 \rightarrow \chi_1e^{+}e^-$ decay length is
\begin{equation}
l_{\chi_2} = \gamma_{\chi_2} c \tau \approx 0.8 ~mm~ \cdot (\frac{\gamma_{\chi_2}}{10})\cdot(\frac{10^{-4}}{\epsilon})^2 
\cdot \kappa^{-1}
\cdot 
\frac{100~MeV}{m_{A'}} \,.
\label{121}
\end{equation}
Here $\gamma_{\chi_2} = \frac{E_{\chi_2}}{m_{\chi_2}}$ and 
$\kappa =  \frac{4\alpha_D\Delta^5}{5\pi m^5_{A}}$. 
As a numerical  example we use the point \cite{Mohlabeng} 
$m_{A'} = 3 m_{\chi_1}$,  $\Delta = 0.4 m_{\chi_1}$ and $\alpha_D = 0.1$. 
For this point we find that 
\begin{equation}
l_{\chi_2} = \gamma_{\chi_2} c \tau \approx 0.8 ~mm~ \cdot (\frac{\gamma_{\chi_2}}{10})\cdot (\frac{10^{-4}}{\epsilon})^2 
\cdot 
\frac{100~MeV}{m_{A'}}\cdot  0.43 \cdot 10^6 \,.
\label{122}
\end{equation}
For NA64 experiment with $100~GeV$ electron beam the $A'$  energy is $\sim 100~GeV$ and 
approximately $E_{\chi_2} \sim \frac{E_{A'}}{2} \approx 50~GeV $. As a crude  estimate we shall use  
 $\gamma_{\chi_2} = \frac{50~GeV}{m_{\chi_2}}$.
As a result we find
\begin{equation}
l_{\chi_2} \equiv \gamma_{\chi_2} c \tau \approx 
7.6 ~cm~ (\frac{10^{-1}}{\epsilon})^2 
\cdot 
(\frac{100~MeV}{m_{A'}})^2  \,.
\label{123}
\end{equation}
For instance, for $m_{A'} = 100~MeV$  and 
$\epsilon = 10^{-2}$  
\begin{equation}
l_{\chi_2} \approx 8 ~m \,.  
\label{124}
\end{equation}

So the problem arises  - is it possible  to derive  bounds on $\epsilon^2$ at finite $l_{\chi_2}$ from NA64 data?
The NA64 experiment for the search for invisible $A'$ decays consists of ECAL with the length $60~cm$.
 Also we have 3 HCAL modules each with $l_{HCAL} = 170~cm$ 
and the distance between the end of the ECAL and the begining of the HCAL modules is $80~cm$. So 
the distance between the begining of ECAL and the end of the last HCAL section(the end of NA64 experiment) 
is $l_{exp} = 6.5~m$. The active zone of ECAL is $l_{ECAL,act}  \approx 45~cm$. 
Suppose the  $A'$ is produced in ECAL and immediately decays into $\chi_2 \chi_1$( this assumption is correct since 
$\alpha_D = 0.1$ and  $\Gamma(A^` \rightarrow \chi_2\chi_1)$  is not small) and   $\chi_2$ decays into $\chi_1e^+e^-$  
with the decay length $l_{\chi_2}$. 
The probability that $\chi_2$ does not decay within NA64, i.e. between the ECAL and the HCAL, is 
\begin{equation}
P_{\chi_2}(outside~decay) = exp(-\frac{l_{exp}}{l_{\chi_2}})\,,
\label{125}
\end{equation}
where $l_{\chi_2}$ is  the $\chi_2$ decay length. 
We can use 
the NA64 results on the search for invisible dark photon decays.
The bound on mixing parameter is
\begin{equation}
\epsilon^2 \leq \epsilon^2_{NA64, up} \cdot (P_{\chi_2}(outside~decay))^{-1} \,,
\label{126}
\end{equation}
where    $\epsilon^2_{NA64, up}$ is the NA64 upper bound \cite{NA64explast3} obtained in the assumption that 
$Br(A' \rightarrow invisible) = $100\%. 
Also the situation with  
$\chi_2$ decaying withing the ECAL is possible. In this case we have missing 
energy due to decay chain $A' \rightarrow \chi_1 \chi_2 \rightarrow \chi_1 \chi_1 e^+e^-$ 
and nonobservation of 2 $\chi_1$ particles.
The average missing energy in this decay is $ E_{miss}  \approx 0.5 E_{A'} + \frac{1}{3}E_{A'} $
and it is bigger than the used in  NA64 missing energy cut $E_{miss} > E_{miss, cut} =  50~GeV$. So 
we can detect the events related with the $\chi_2$ decay within ECAL by the measurement of missing energy. 
The probability that $\chi_2$ decays within ECAL active zone is
\begin{equation}
P_{\chi_2}( decays~ in~ ECAL) = 1 - exp(-\frac{l_{ECAL, act}}{l_{\chi_2}})  \,.
\label{127}
\end{equation}
So total probability of the $\chi_2$ detection with the use of energy missing cut is 
\begin{equation}
P_{\chi_2} =  
P_{\chi_2}( decays~in~ ECAL)  + P_{\chi_2}(outside ~decay) =
(1 - exp(-\frac{l_{ECAL, act}}{l_{\chi_2}})) +  exp(-\frac{l_{exp}}{l_{\chi_2}}) \,.
\label{128}
\end{equation}

For arbitrary $l_{\chi_2}$ 
the expression (\ref{128}) for $P_{\chi_2}$ has minimal value at 
\begin{equation}
l_{exp} exp(-\frac{l_{exp}}{l_{\chi_2, min}}) =l_{ECAL, act}exp(-\frac{l_{ECAL, act}}{l_{\chi_2, min}}) \,
\label{129}
\end{equation}
or 
\begin{equation}
l_{\chi_2, min} = \frac{l_{exp} - l_{ECAL,act}}{\ln(\frac{l_{exp}}{l_{ECAL,act}})}  \,
\label{130}
\end{equation}
and $P_{\chi_2, min}$ is
\begin{equation}
P_{\chi_2, min} = (1 - \exp(-\frac{l_{ECAL,act}}{l_{exp}- l_{ECAL,act}}\ln\frac{l_{exp}}{l_{ECAL,act}}) +
\exp(-\frac{l_{exp}}{l_{exp}- l_{ECAL,act}}\ln\frac{l_{exp}}{l_{ECAL,act}})  \,.
\label{131}
\end{equation}
Numerically for $l_{exp} = 650~cm$ and $l_{ECAL,exp} = 40~cm$ we find 
\begin{equation}
P_{\chi_2, min}\approx 0.22 \,.
\label{132}
\end{equation}
The bound on $\epsilon^2$ reads
\begin{equation}
\epsilon^2 \leq \epsilon^2_{NA64, up} \cdot (P_{\chi_2,min})^{-1}
\approx 4.5 \cdot \epsilon^2_{NA64, up} \,.
\label{133}
\end{equation}
Here  $\epsilon^2_{NA64, up}$ is the NA64 bound for the case of invisible $A^`$ decay.
So we see that  NA64 is able to obtain upper bound on $\epsilon^2$ parameter 
for the case of visible $A^`$ decay with large missing energy in a model independent way.
 The 
knowledge  of $l_{\chi_2}$  allows to improve the bound (\ref{133}). 


\end{document}